\documentclass[useAMS,usenatbib]{mn2e}
\usepackage{epsfig}
\usepackage{lscape}
\usepackage{graphicx}
\usepackage{rotating}
\usepackage{times}
\def\cm3{cm$^{-3}$}
\def\kms{km~s$^{-1}$}

\def\lsun{L$_{\odot}$}
\def\rsun{R$_{\odot}$}

\def\msun{M$_{\odot}$}

\let\ts=\thinspace

\def\one{\ts {\,\sc i}}
\def\two{\ts {\,\sc ii}}
\def\three{\ts {\,\sc iii}}
\def\four{\ts {\,\sc iv}}

\def\beq{\begin{equation}}
\def\eeq{\end{equation}}

\def\lesssim{\mathrel{\hbox{\rlap{\hbox{\lower4pt\hbox{$\sim$}}}\hbox{$<$}}}}
\def\gtrsim{\mathrel{\hbox{\rlap{\hbox{\lower4pt\hbox{$\sim$}}}\hbox{$>$}}}}
\def\taur{$\tau_{\rm Rosseland}$}

\def\aj{AJ}
\def\pasp{PASP}
\def\apj{ApJ}
\def\apjs{ApJS}
\def\apjl{ApJL}
\def\aap{A\&A}
\def\araa{ARA\&A}

\def\mnras{MNRAS}
\def\nat{Nature}

\def\physrep{Phys.~Rep.}

\title[Type II-P SN spectroscopic modelling]{Non-LTE time-dependent
spectroscopic modelling of type II-plateau supernovae from the photospheric
to the nebular phase: case study for 15 and 25\,\msun\ progenitor stars}

\author[Luc Dessart and D. John Hillier]
{
Luc Dessart$^{1}$\thanks{E-mail: Luc.Dessart@oamp.fr},
D. John Hillier$^{2}$\\
$^{1}$ Laboratoire dÕAstrophysique de Marseille, Universit\'e de Provence,
CNRS, 38 rue Fr\'ed\'eric Joliot-Curie, F-13388 Marseille Cedex 13, France \\
$^{2}$ Department of Physics and Astronomy, University of Pittsburgh,
3941 O'Hara Street, Pittsburgh, PA, 15260, USA
}

\begin{document}

\date{Accepted 2010 August 17.  Received 2010 August 13; in original form 2010 July 5}

\pagerange{\pageref{firstpage}--\pageref{lastpage}} \pubyear{2009}

\maketitle

\label{firstpage}

\begin{abstract}
We present the first non-LTE time-dependent radiative-transfer simulations of supernovae (SNe)
II-Plateau (II-P) covering both the photospheric and nebular phases, from $\sim$10 to $\gtrsim$1000\,d after the explosion, and based on
1.2\,B piston-driven ejecta produced from a 15\,\msun\ and a 25\,\msun\ non-rotating
solar-metallicity star. The radial expansion of the gradually cooling photosphere gives rise to a near-constant luminosity
up to $\gtrsim$100\,d after the explosion.  The photosphere remains in the outer 0.5\,\msun\ of the
ejecta for up to $\sim$50\,d after the explosion.
As the photosphere reaches the edge of the helium core, the SN luminosity drops
by an amount mitigated by the progenitor radius and the $^{56}$Ni mass.
Synthetic light-curves exhibit a bell-shape morphology,
evolving faster for more compact progenitors, and with an earlier peak
and narrower width in bluer filters.
UV and $U$-band fluxes are very sensitive to line-blanketing, the metallicity, and the adopted model atoms.
During the recombination epoch synthetic spectra are dominated by H\,{\sc i} and metal lines, and are largely insensitive
to the differing H/He/C/N/O composition of our two progenitor stars.
In contrast, synthetic nebular-phase spectra reveal a broader/stronger O\,{\sc i} doublet line
in the higher-mass progenitor model, reflecting the larger masses of oxygen and nickel that are ejected.
Our simulations overestimate the typical luminosity and the visual rise time
of standard SNe II-P, most likely a consequence of our progenitor stars being too big and/or too hydrogen rich.
Comparison of our simulations with photospheric-phase observations of SN1999em of the same color are satisfactory.
Our neglect of non-thermal excitation/ionisation leads to a fast disappearance of continuum radiation
and Balmer-line emission at the end of the plateau phase.
With the exception of H\,{\sc i} lines, our nebular spectra show a striking similarity to contemporaneous observations
of SN1999em.
\end{abstract}


\section{Introduction}

Supernovae (SNe) II-Plateau (II-P) are the most represented SN type \citep{cappellaro_etal_97,smartt_etal_09}.
The theoretical prediction that they stem from the explosion of Red Supergiant (RSG) stars
 \citep{colgate_white_66,falk_arnett_77} is supported by direct identification of some SNe II-P's with their progenitor on pre-explosion images. The progenitors indicate main-sequence masses in the range 8-17\,\msun\ (see \citealt{smartt_09} for a review). After their death,  SN II-P produce a stellar-mass compact object, either a neutron star or a black hole \citep{heger_etal_03}. They are a key driver for the chemical evolution of galaxies through the core-embedded material they eject (in particular oxygen; \citealt{woosley_weaver_95}), as well as efficient dust factories \citep{sugerman_etal_06}.

Owing to their large luminosity, SNe II-P offer a means to constrain distances in the Universe
through a variety of methods known as the Expanding Photosphere Method \citep{KK74_EPM,E96,
DH05_epm,DH06_SN1999em}, the Spectral-fitting Expanding Atmosphere Method
\citep{MBB02_87A}, and the Standard Candle Method \citep{hamuy_pinto_02}.
Studying SNe II-P is thus of relevance for a wide range of topics in modern astrophysics.

Inferences on SNe II-P ejecta, as well as their progenitor stars, can be made  using radiation hydrodynamics and radiative transfer. For multi-band light-curve calculations, radiation hydrodynamics
assuming the gas in Local Thermodynamic Equilibrium (LTE), frequency-mean or frequency-dependent opacities
(including lines or not) and time-dependent
radiation transport are used with some success to constrain the explosion energy, the mass of the ejecta, the radius
of the progenitor star, and the amount of $^{56}$Ni produced in the explosion
\citep[e.g.,][]{falk_arnett_77,litvinova_nadezhin_85,blinnikov_etal_00,baklanov_etal_05,utrobin_07,tominaga_etal_09,DLW10a,DLW10b}. Calibrations drawn from such studies have been used to infer SNe II-P properties \citep{hamuy_03}.

However, an attractive and more physically-consistent alternative is to use radiation hydrodynamics
up to a few days after shock-breakout, when acceleration terms are sizeable, and then switch to a more sophisticated radiative-transfer treatment on a non-accelerating ejecta \citep{eastman_etal_94,KW09,DH10}. This then permits a more detailed treatment of line-blanketing, departures from LTE, and the dependence of the radiation field on frequency,
angle, and time. Recently, we sketched an approach that treats all these important aspects \citep{DH10} and applied it
to the early evolution of SN1987A. The method retains the sophistication
of non-LTE steady-state simulations focusing on the decoupling layers of SN ejecta \citep{DH05_qs_SN,baron_etal_07,dessart_etal_08},
but also treats important time-dependent terms appearing in the energy, statistical-equilibrium, and radiative-transfer equations
(see, for example, \citealt{UC05_time_dep,DH08}, for a partial treatment of these terms).

This initial-value problem is constrained at the start by employing a hydrodynamical input model of the explosion,
itself generated from a physically-consistent pre-SN model. Confrontation of synthetic spectra and light curves with observations
can thus be connected to the known ejecta properties, yielding constraints on the explosion and the progenitor properties.
Because we explicitly treat all processes controlling level populations, the effects of line emission, absorption, and overlap are accurately computed, and thus line profiles can be used to extract information on the ejecta.
Accurate spectral modelling is critical for the determination of abundances, excitation/ionisation sources, kinematics,
and ultimately the explosion and progenitor properties. By contrast, most SN studies treat bound-bound transitions approximately. These studies allow  inferences based on colour and light-curve evolution but little reliable information is extracted from lines.

In this paper, building upon our recent analysis of the Type II-peculiar SN1987A \citep{DH10},
we present the first non-LTE time-dependent radiative-transfer simulations of SNe II-P using {\sc cmfgen}
\citep{HM98_lb,DH05_qs_SN,DH08}, covering both the photospheric and nebular phases.
Because of the huge computational expense of this endeavour, we focus on only two ejecta
produced from the explosion of stars with main-sequence masses of 15\,\msun\ and 25\,\msun, and characterised
by an explosion energy of 1.2\,B (Woosley, private communication; \citealt{WH07}). We use these simulations to describe fundamental  properties of the SN gas and light, with special emphasis on spectra which have not previously been computed with this level of physical consistency in this context. Due to the limited set of inputs, we make no attempt to reproduce specific observations, but rather we seek to make a qualitative assessment of the general adequacy of our models.

In the next section, we present the pre-SN models that were used to generate the
hydrodynamical input ejecta from which our radiative-transfer simulations start.   The adopted model atoms, atomic data,
and the importance of non-LTE, are discussed in \S~\ref{model_atoms}. In \S~\ref{sect_gas},
we discuss the evolution of the gas properties throughout the photospheric and nebular phases,
including the photospheric properties and the ejecta ionisation.
We then  turn to our theoretical results for the radiation field.
In \S~\ref{sect_LC}, we first describe the bolometric-light evolution and multi-band light-curves,
before moving on to synthetic spectra, the sources of blanketing, and
line identifications in \S~\ref{sect_spec}. We also document the spectral differences predicted in our simulations for
the low and the high-mass progenitor models.
In \S~\ref{sect_comp_obs}, we compare our results to photometric and spectroscopic observations
of SNe II-P in general, and of SN1999em in particular, to identify the successes and failures of our simulations.
In \S~\ref{sect_concl}, we present our conclusions and plans for a forthcoming parameter
study of SNe II-P.

\begin{table}
\begin{center}
\caption[]{Properties of the pre-SN models s15 and s25 \citep{WH07}. The age is the time elapsed between the
main sequence and core collapse.}
\label{tab_presn}
\begin{tabular}{lcccccc}
\hline
Model &   $M_{\rm i}$  &  $M_{\rm f}$      &   $R_{\star}$      &    Age          &    $M_{\rm ejecta}$  &    $M_{\rm remnant}$ \\
             &   \msun            &  \msun             &        \rsun             &    Myr          &      \msun                    &    \msun    \\
\hline
s15      &      15               &   12.79              &       810               &    13.24         &      10.93                  &     1.83    \\
s25      &      25               &   15.84              &      1350              &    7.47          &        13.94                 &     1.90    \\
\hline
\end{tabular}
\end{center}
\end{table}

\begin{table*}
\begin{center}
\caption[]{Ejecta cumulative yields (upper table) and surface mass fraction (lower table)
for the most abundant species in models s15e12 and s25e12. Metal abundances not shown here
are taken at the solar-metallicity value. The total ejecta mass is 10.93\,\msun\ (13.94\,\msun) for
model s15e12 (s25e12). Model s15e12iso has the same ejecta characteristics as model s15e12
and differs only in composition of under-abundant species, e.g., Sc or Ba.
Numbers in parentheses correspond to powers of ten.}
\label{tab_comp}
\begin{tabular}{@{}l@{\hspace{1.6mm}}c@{\hspace{1.6mm}}c@{\hspace{1.6mm}}c@{\hspace{1.6mm}}
c@{\hspace{1.6mm}}c@{\hspace{1.6mm}}c@{\hspace{1.6mm}}
c@{\hspace{1.6mm}}c@{\hspace{1.6mm}}c@{\hspace{1.6mm}}
c@{\hspace{1.6mm}}c@{\hspace{1.6mm}}c@{\hspace{1.6mm}}
c@{\hspace{1.6mm}}c@{\hspace{1.6mm}}c@{}}
\hline
Model & \multicolumn{15}{c}{Ejecta Cumulative Yields} \\
\hline
   & $M_{\rm H}$ &$M_{\rm He}$ & $M_{\rm C}$  & $M_{\rm N}$  & $M_{\rm O}$ & $M_{\rm Ne}$& $M_{\rm Mg}$& $M_{\rm Si}$& $M_{\rm S}$ & $M_{\rm Ar}$ & $M_{\rm Ca}$ & $M_{\rm Ti}$& $M_{\rm Cr}$& $M_{\rm Fe}$
   & $ M_{^{56}{\rm Ni}}$    \\
   &    [\msun]  & [\msun]  &[\msun]    &   [\msun] &   [\msun]   &    [\msun]    &   [\msun]  &   [\msun]  &   [\msun]            &    [\msun]            &[\msun]             &          [\msun]    & [\msun]           &    [\msun]           & [\msun]            \\
\hline
s15e12            &     5.49(0)  &  4.00(0)  &  1.59(-1)  &  3.22(-2)  &  8.16(-1)  &  1.52(-1)  &  5.41(-2)  &  8.10(-2)  &  3.17(-2)  &  6.43(-3)  &  5.52(-3)  &  1.26(-4)  &  1.29(-3)  &  1.60(-2)  &  8.66(-2) \\
s25e12            &     4.01(0)  &  4.77(0)  &  3.53(-1)  &  4.48(-2)  &  3.32(0)  &  5.30(-1)  &  2.24(-1)  &  3.16(-1)  &  1.49(-1)  &  2.58(-2)  &  1.85(-2)  &  1.54(-4)  &  2.48(-3)  &  2.04(-2)  &  1.63(-1) \\
\hline
Model & \multicolumn{15}{c}{Ejecta Surface Mass Fractions} \\
\hline
  & $X_{\rm H}$    &$X_{\rm He}$ & $ X_{\rm C}$ &$ X_{\rm N}$ & $X_{\rm O}$& $X_{\rm Ne}$& $X_{\rm Mg}$& $X_{\rm Si}$& $X_{\rm S}$& $X_{\rm Ar}$& $X_{\rm Ca}$& $X_{\rm Ti}$ & $X_{\rm Cr}$& $X_{\rm Fe}$& $X_{^{56}{\rm Ni}}$    \\
\hline
s15e12  & 6.50(-1)  &  3.35(-1)  &  1.42(-3)  &  3.11(-3)  &  5.41(-3)  &  1.31(-3)  &  7.91(-4)  &  8.29(-4)  &  4.23(-4)  &  1.13(-4)  &  7.38(-5)  &  3.83(-6)  &  3.43(-5)  &  1.46(-3)  &  1.14(-9) \\
s25e12  & 5.28(-1)  &  4.56(-1)  &  9.17(-4)  &  5.54(-3)  &  3.89(-3)  &  1.37(-3)  &  8.03(-4)  &  8.33(-4)  &  4.25(-4)  &  1.13(-4)  &  7.40(-5)  &  3.83(-6)  &  3.43(-5)  &  1.46(-3)  &  1.58(-6) \\
\hline
\end{tabular}
\end{center}
\end{table*}

\section{Initial hydrodynamical model and setup}
\label{sect_setup}

   The approach we follow is similar to that described in \citet{DH10},
but we now concentrate on the SN ejecta resulting from the explosion of RSG stars.
Starting from a hydrodynamical model that describes the depth variation of
element mass fraction, density, temperature, radius and velocity, we model the non-LTE time-dependent radiative transfer of SNe II-P ejecta with {\sc cmfgen} \citep{HM98_lb,DH05_qs_SN,DH08,DH10}.
We assume the SN is spherically symmetric and smooth (i.e., un-clumped),
and free from external disturbances, such as interaction with the circumstellar environment or irradiation
from the newly-born neutron star.
The hydrodynamical models were computed by \citet{WH07} using the code {\sc kepler} \citep{weaver_etal_78}
by first evolving non-rotating, solar-metallicity, 15\,\msun\ and 25\,\msun\ stars from the main sequence
to the formation and collapse of their degenerate iron core, and then exploding
these by means of a piston.

The input model s15e12 (s25e12) was generated from model s15A (s25) of \citet{WH07} and corresponds to a SN ejecta with
a 1.2\,B energy and post-explosion time of 8$\times$10$^5$\,s (2$\times$10$^6$\,s; see Table~\ref{tab_presn} for details).
For the same ejecta kinetic energy, the model s25e12 yields a higher mass of
$^{56}$Ni (0.163 compared to 0.0866\,\msun) because it possesses a larger helium-core mass with more mass at larger density,
smaller radii and higher temperatures.
All models were artificially mixed by sweeping a ``box'' four times through the ejecta and averaging the composition
within that box. The box used in model s15e12 (s25e12) had a width of 0.427\,\msun\ (0.819\,\msun), corresponding
in each case to a tenth of the progenitor helium-core mass.
This moderate mixing softens the composition gradients but preserves the strong ejecta stratification,
hydrogen being absent in the inner few solar masses of the ejecta (below 1500\,\kms).
The $^{56}$Ni mass fraction is sizeable only below 1000--2000\,\kms\ but owing to mixing, shows a non-zero
mass fraction even at the outer edge of the ejecta.

\begin{figure*}
\epsfig{file=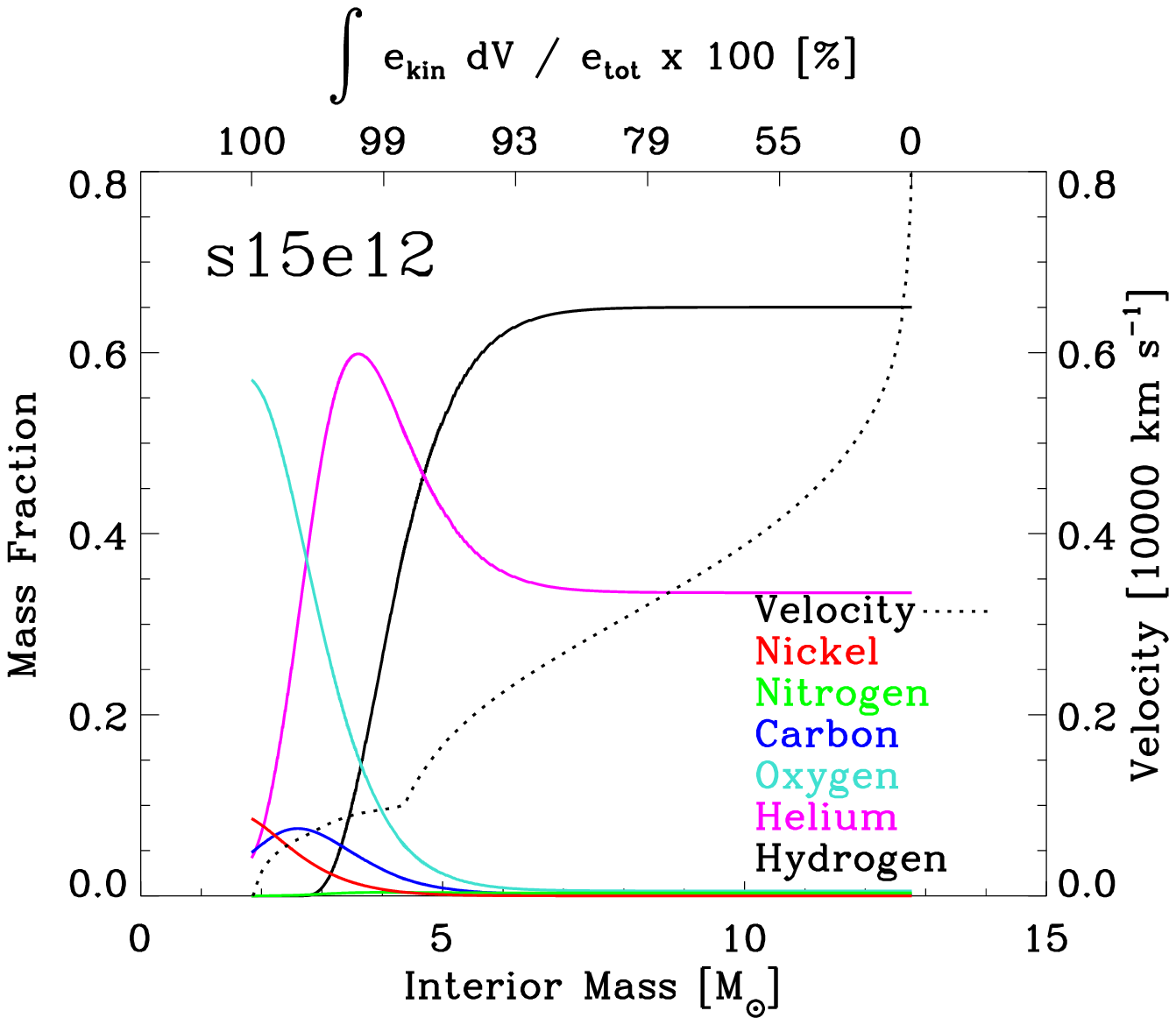,width=8.5cm}
\epsfig{file=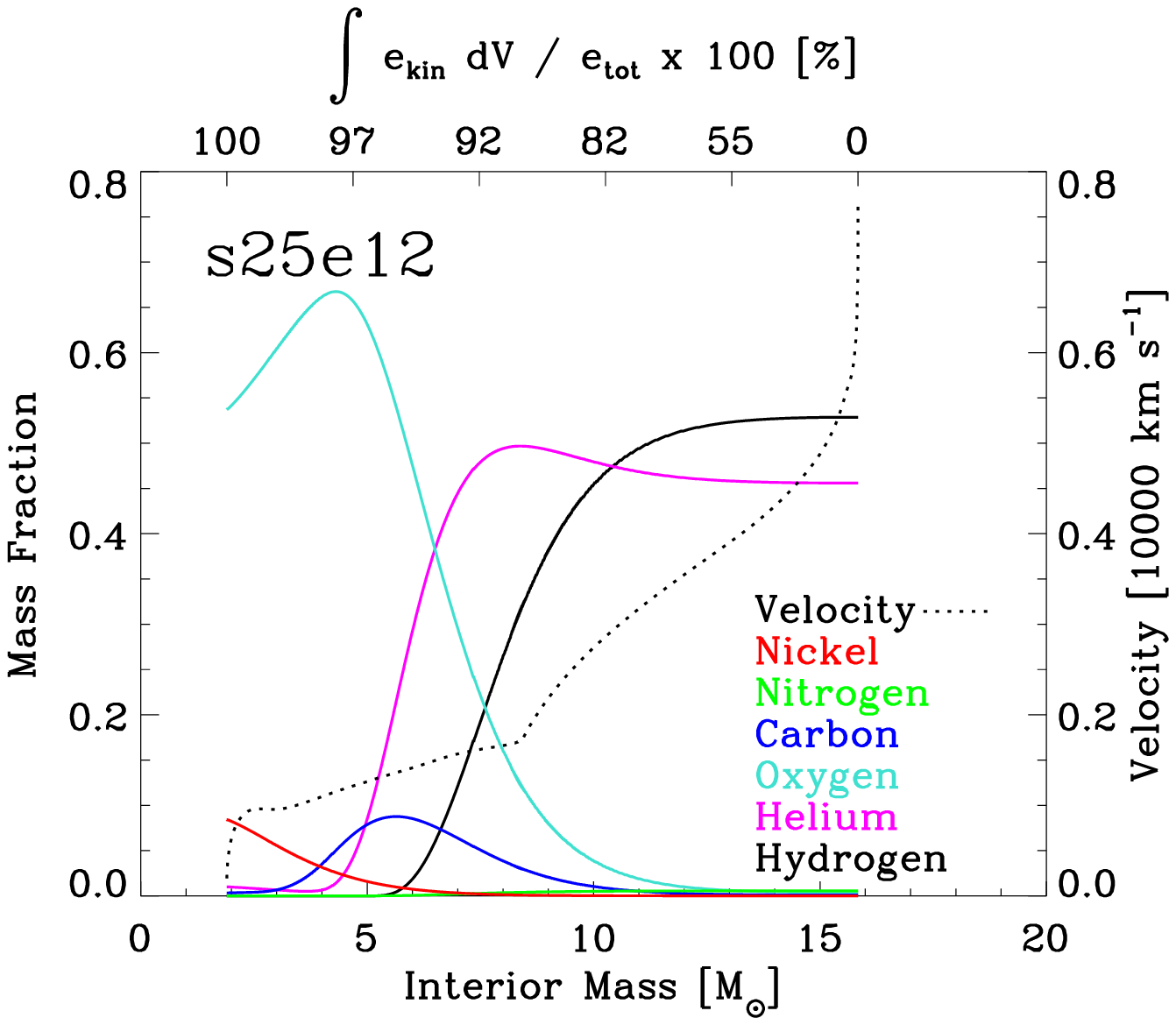,width=8.5cm}
\caption{
{\it Left:} Element mass fractions as a function of Lagrangian mass for the hydrodynamical
input model s15e12.
We show hydrogen (black), helium (pink), oxygen (turquoise), carbon (blue), nitrogen (green),
and $^{56}$Ni (red). We also overplot the velocity as a dotted black line, with the corresponding axis
given on the right. We include an axis at the top showing the distribution
of the mass-integrated specific kinetic energy (the integral bounds being the Lagrangian mass
$M_r$ and $M_{\rm tot}$). The bulk of the
ejecta kinetic energy is contained within what used to be the loosely-bound hydrogen-rich region of the progenitor
envelope. In contrast, the more gravitationally bound helium core is ultimately endowed with only a few percent
of the total ejecta kinetic energy.
{\it Right:} Same as left, but now for the s25e12 model. The helium core is endowed with a larger
fraction of the total ejecta kinetic energy than seen in the s15e12 model, although the largest share
is still contained in the hydrogen-rich region of the ejecta. For both models, the total ejecta kinetic energy is 1.2\,B.
\label{fig_comp_sn}
}
\end{figure*}

In the (standard) {\sc kepler} models s15e12 and s25e12, the species included are
H, $^3$He, $^4$He, $^{12}$C, $^{14}$N, $^{16}$O, $^{20}$O, $^{24}$Mg, $^{28}$Si,
$^{32}$S, $^{36}$Ar, $^{40}$Ca, $^{44}$Ti, $^{48}$Cr, $^{52}$Fe, $^{54}$Fe, $^{56}$Ni,
$^{56}$Fe, and ``Fe''. In this nomenclature, some species approximate certain nuclei.
For example, $^4$He approximates nuclei with $1<A<6$, where $A$ is the atomic mass,
and $^{48}$Cr stands for nuclei with $47<A<52$ [see Woosley, Heger, \& Weaver (2002) for details].
In such inputs, numerous species are not described individually (e.g., Na, although Na\one\ lines
are unambiguously see in SNe II-P spectra and require the knowledge of the sodium abundance
for a proper modelling). Some, like Ba, are altogether ignored.
We give a summary of the chemical composition for the dominant species in Table~\ref{tab_comp},
with both the cumulative ejecta yields (upper table) and the corresponding mass fractions
at the progenitor surface (lower table).  We also illustrate the distribution of the main elements
through the ejecta in Fig.~\ref{fig_comp_sn}.
Surface abundances of both models show evidence for CNO processing, with the higher-mass model exhibiting larger enhancements of both He and N, and a greater depletion of C and O.
The presence of CNO processed material in RSG atmospheres is corroborated by spectroscopic analyses of RSG stars
\citep{lancon_etal_07}, and understood to stem, in the absence of rotation, from convective dredge-up.
Stellar rotation is expected to aid the appearance of CNO processed material at the stellar surface \citep{meynet_maeder_00, meynet_etal_06}.  In the outer regions of the ejecta, mass fractions for elements
heavier than oxygen are identical for both models and correspond to the solar metallicity
(apart from the nucleosynthesised $^{56}$Ni).

The bulk of the kinetic energy is contained within the hydrogen-rich regions
of the ejecta, while the region corresponding to the helium core contains $\lesssim$10\% of the total
kinetic energy in model s25e12, and an even lower fraction in the model s15e12 (see top axis in Fig.~\ref{fig_comp_sn}).
The higher mass progenitor (which suffered enhanced mass
loss) has a lower mass hydrogen-rich envelope and a larger helium/oxygen ejected mass (Fig.~\ref{fig_comp_sn}).
As we discuss in this paper in relation to the degeneracy of photospheric-phase
SNe II-P spectra and light curves, the enhanced mass of ejected oxygen may be most
easily identified in SNe II-P observations taken at nebular times. Because it pertains to the systematic increase in
helium-core mass with main-sequence, it can help constrain the progenitor identity \citep{DLW10b}.

\begin{figure*}
\epsfig{file=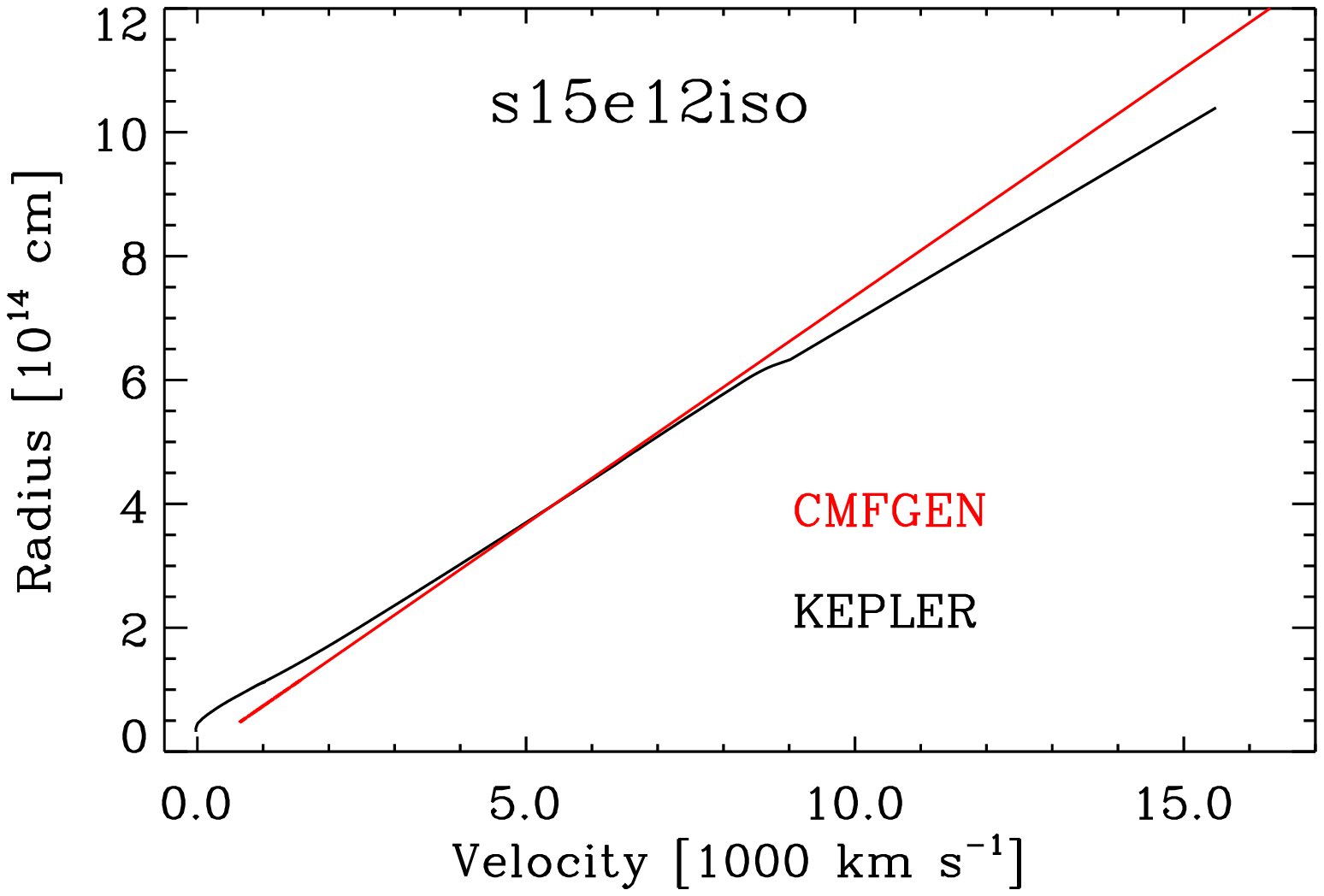,width=5.5cm}
\epsfig{file=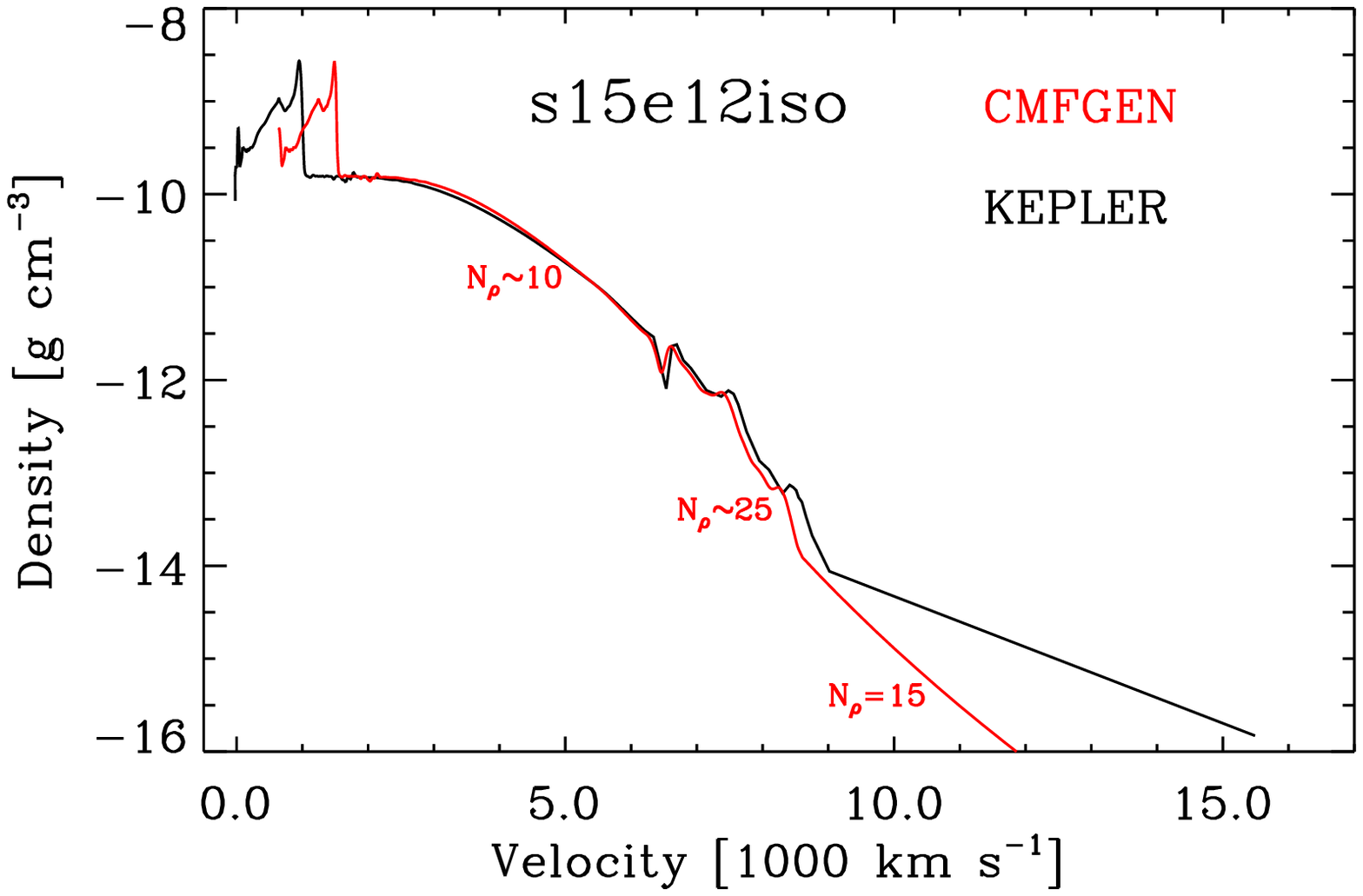,width=5.5cm}
\epsfig{file=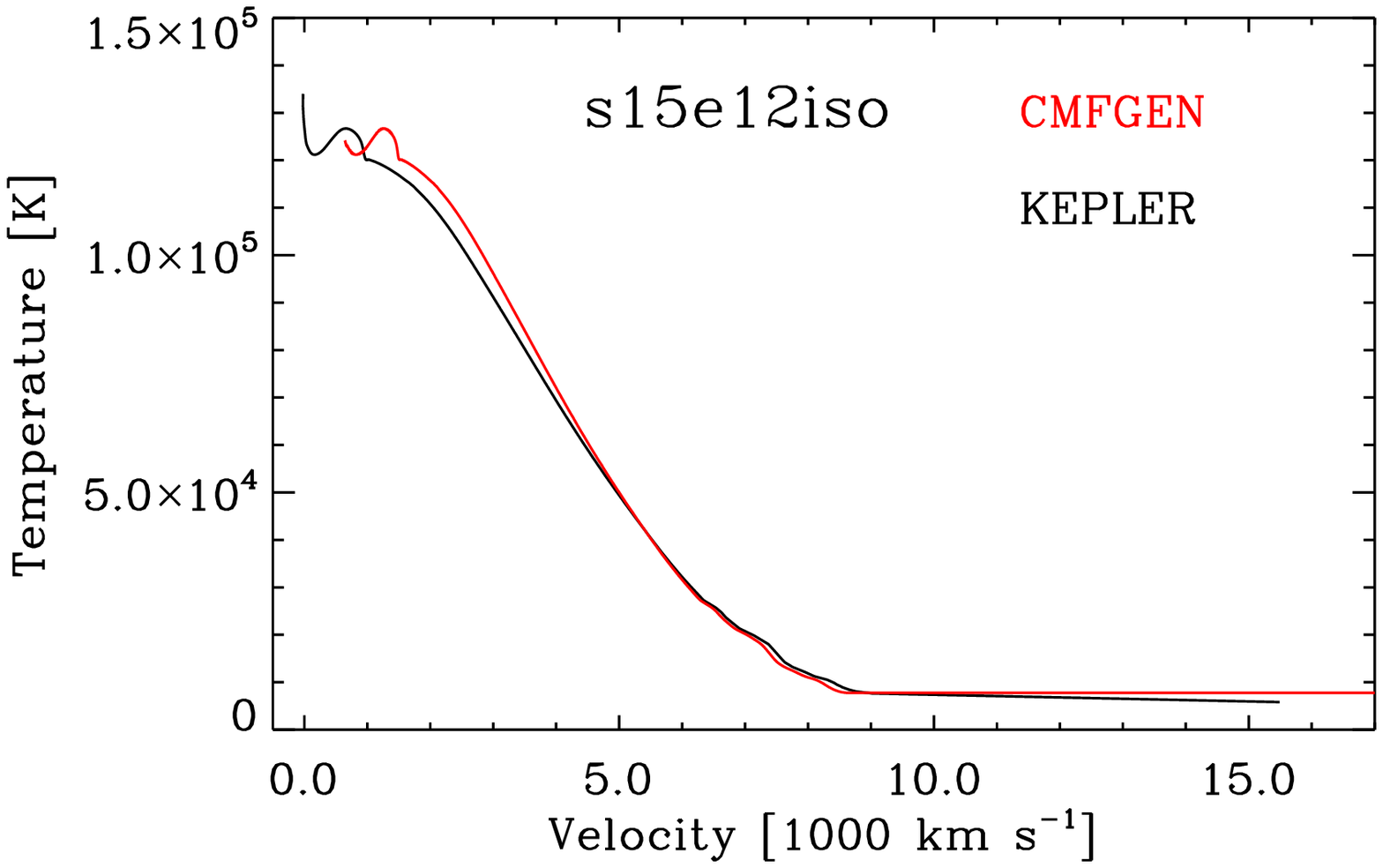,width=5.5cm}
\epsfig{file=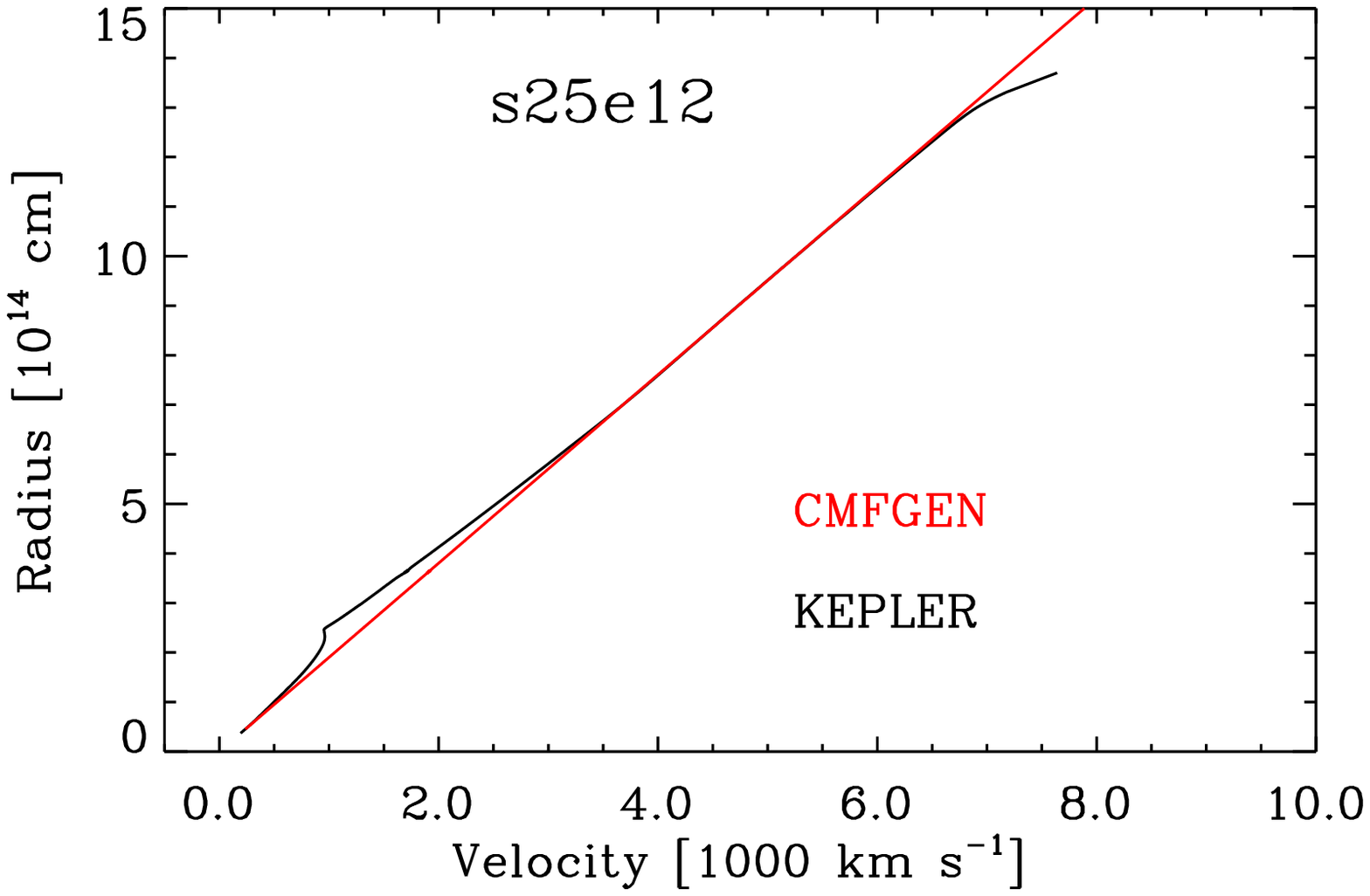,width=5.5cm}
\epsfig{file=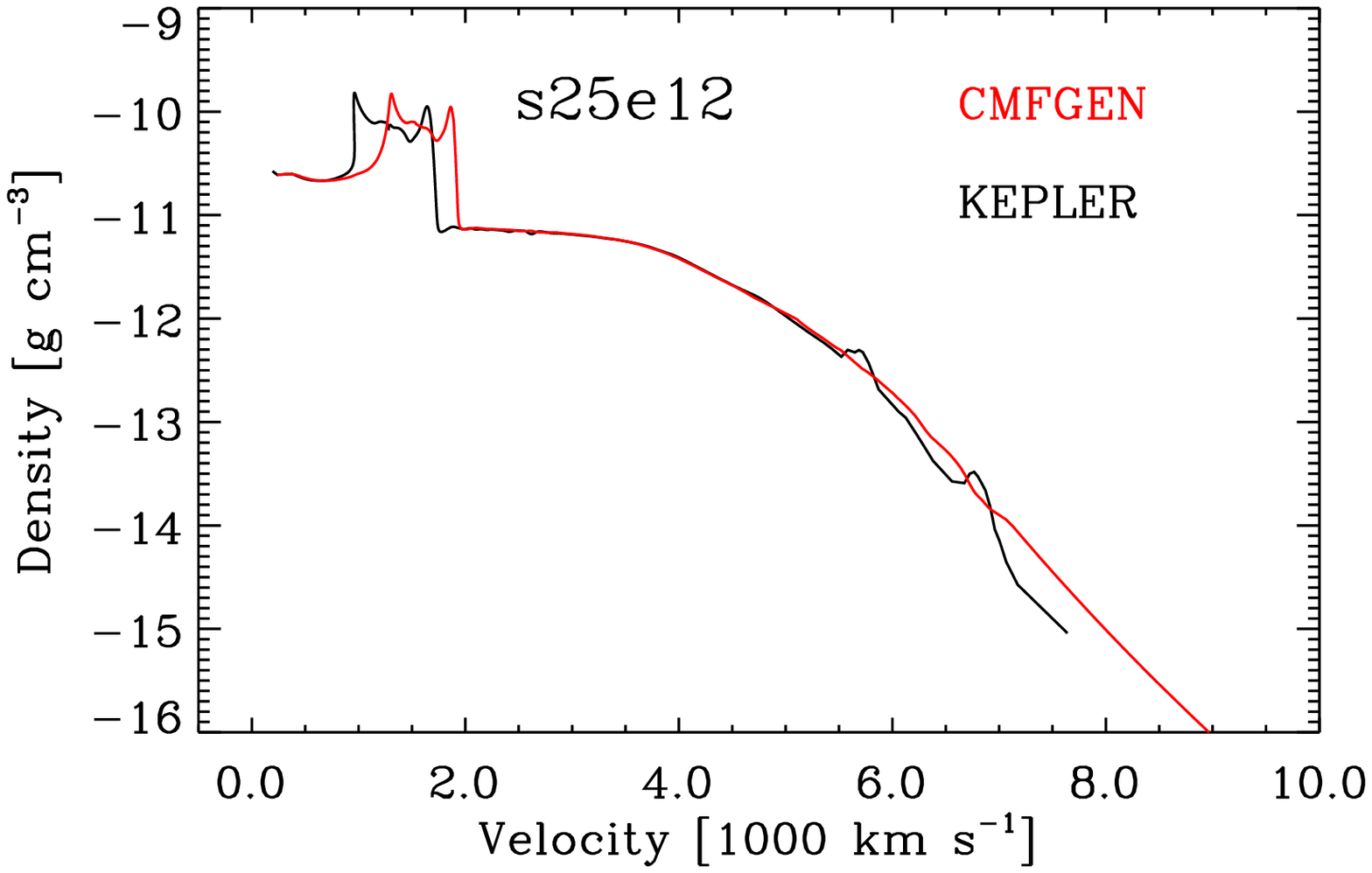,width=5.5cm}
\epsfig{file=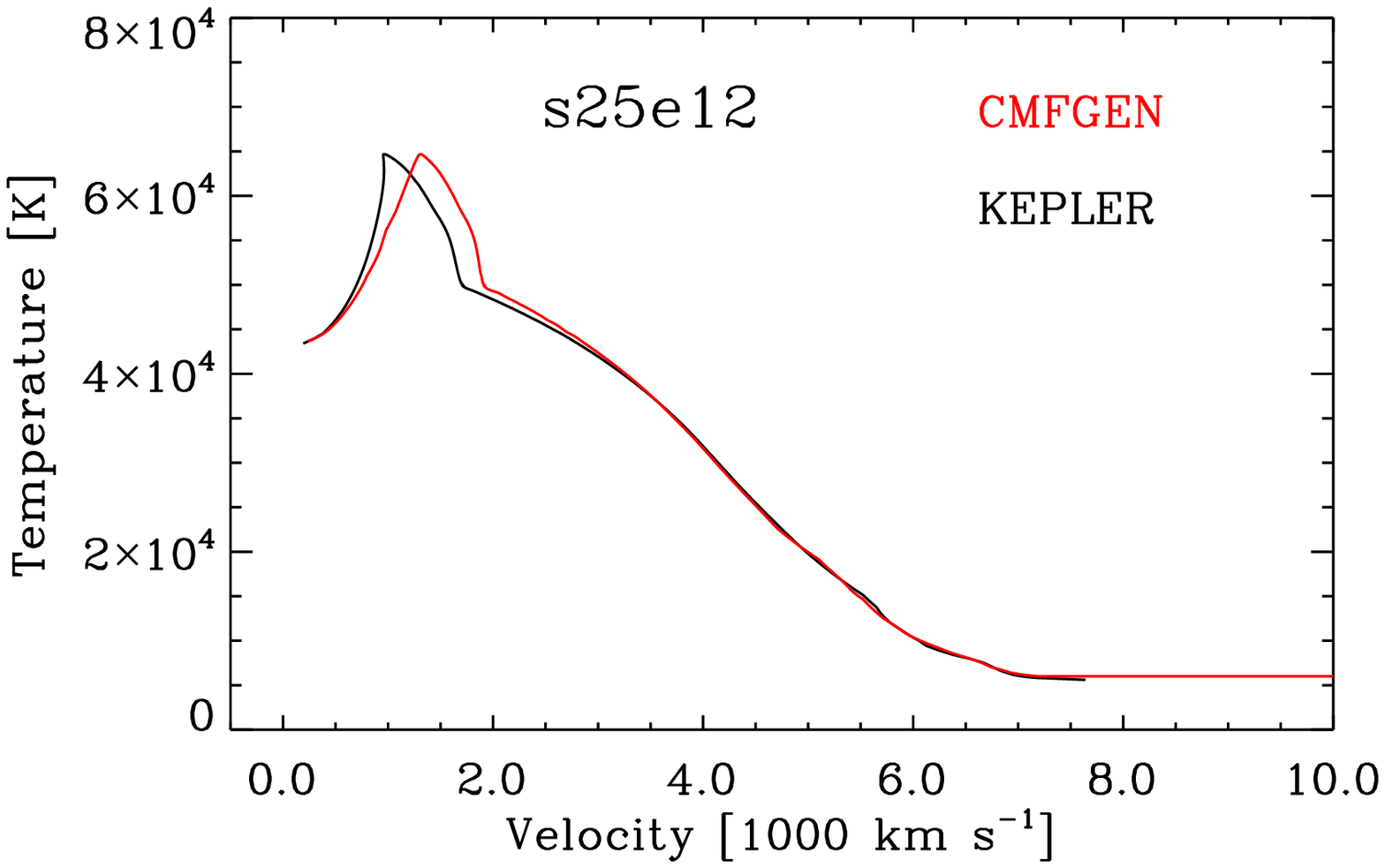,width=5.5cm}
\caption{{\it Top row:} Comparison between the initial {\sc kepler} hydrodynamical input s15e12iso (black) and the corresponding
ejecta structure we start with for the radiative-transfer simulations (red). The differences between the two
arise from our need of strict-homologous expansion ($r/v = {\rm constant}$) and of an outer optically-thin region where
photons are freely-streaming (which requires we stitch this outer region, which is lacking, to the {\sc kepler} model).
In the process, we artificially speed up regions with $v<2000$\,\kms\ by $\lesssim$50\%
and smooth the density structure of the SN ejecta.
We indicate in the middle panel the density exponent $N_{\rho} = -d \log \rho/d\log v$
for the inner, intermediate, and outer ejecta regions. As discussed in Section~\ref{sect_gas},
the photosphere resides in the region of steep density decline with $N_{\rho} \sim 20-25$
and $7.5<V_{\rm phot}/(1000$\,\kms)$<9$ for up to three weeks after explosion in model s15e12iso.
{\it Bottom row:} Same as top row, but now for model s25e12. The speed-up of the
inner regions is smaller, approximately 10\%, because we start with a model that is closer to homologous expansion.
\label{fig_comp_kepler_cmfgen}}
\end{figure*}

In parallel to the simulation based on model s15e12, we also perform a {\sc cmfgen} simulation
based on the detailed {\sc kepler} model s15e12iso, which has exactly the same structure in density/radius/
velocity/temperature, but was evolved with a more complete nuclear network.
The {\sc kepler} model s15e12iso includes every element of the periodic table (as well as all its
associated isotopes) up to Astatine (symbol At; $Z=$85). This detailed input permits the specific modelling
of any species (and associated ions).
In practice, apart from unstable isotopes, we use the sum of the individual isotopic mass fractions
to determine the abundance of the associated species.
Compared to models s15e12 and s25e12, our simulation of model s15e12iso treats in addition
the elements Ne, Na (no longer assumed to have an abundance at the solar-metallicity value and constant with depth),
Al, Ar, Ti, Sc, Cr, and Ba, adopting the abundance distributions computed by {\sc kepler}.
With this choice, we include all species that have a mass fraction larger than $\sim$10$^{-5}$ at any depth in the s15e12iso ejecta,
as well as any species with identified features in SNe II-P spectra (e.g., Sc and Ba).
When discussing general ejecta and radiation properties, we employ our simulations based on the s15e12 and s25e12 models,
but when comparing to observations we employ our simulations based on model s15e12iso as it describes more completely the
ejecta composition and the effects of line blanketing.

 Before starting our radiative-transfer calculations, we made a number of adjustments
to the hydrodynamical inputs. In both models, the innermost regions are not in homologous expansion
(despite the ejecta age of 9.26 and 23.15\,d for models s15e12 and s25e12, respectively)
so we enforce $r/v = {\rm constant}$ (adopting a midpoint in the ejecta) for compatibility with the radiative-transfer algorithm
which assumes $dv/dr=v/r$. This artificially enhances the small velocity of
the inner regions by $\lesssim$50\% in model s15e12 and by $\sim$10\% in model s25e12.
To ensure the radiation is free streaming at all wavelengths at the outer boundary, and to correct for the
poorly-resolved outer region in the input model, we trim and stitch an outer region to the input model
(characterized by a power-law density with an exponent of 15), extending it to larger radii/velocities.
The photosphere is always located inside this outer region, and is thus not directly affected
by this manipulation (see Section~\ref{sect_gas}).
At the new outer boundary, densities are on the order of 10$^{-20}$\,g\,cm$^{-3}$
and velocities on the order of 20\,000\,\kms.
To illustrate these adjustments, we show in Fig.~\ref{fig_comp_kepler_cmfgen} the ejecta properties
provided by {\sc kepler} and those we  start with in our {\sc cmfgen} simulations.

\begin{figure*}
\epsfig{file=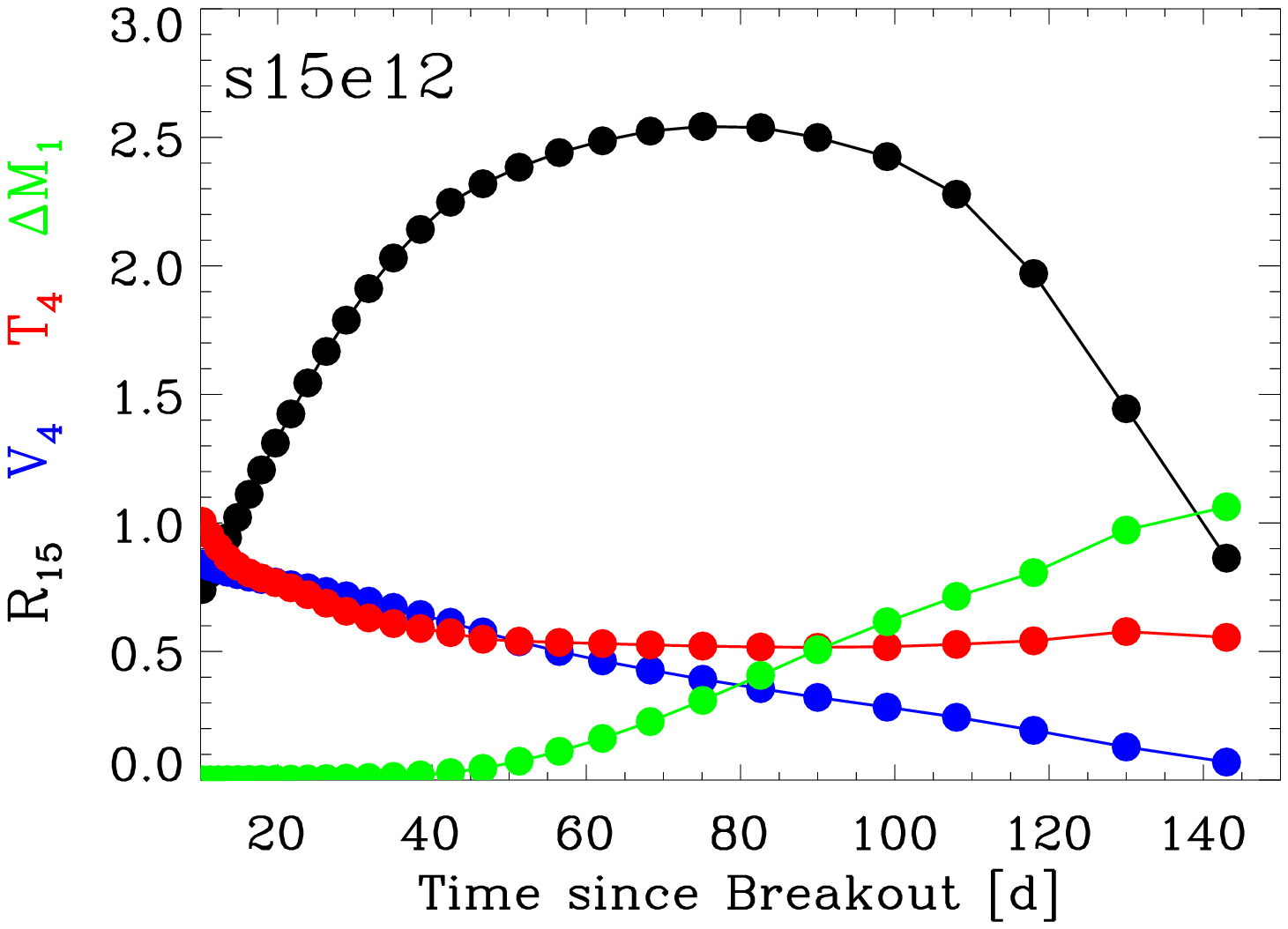,width=8.5cm}
\epsfig{file=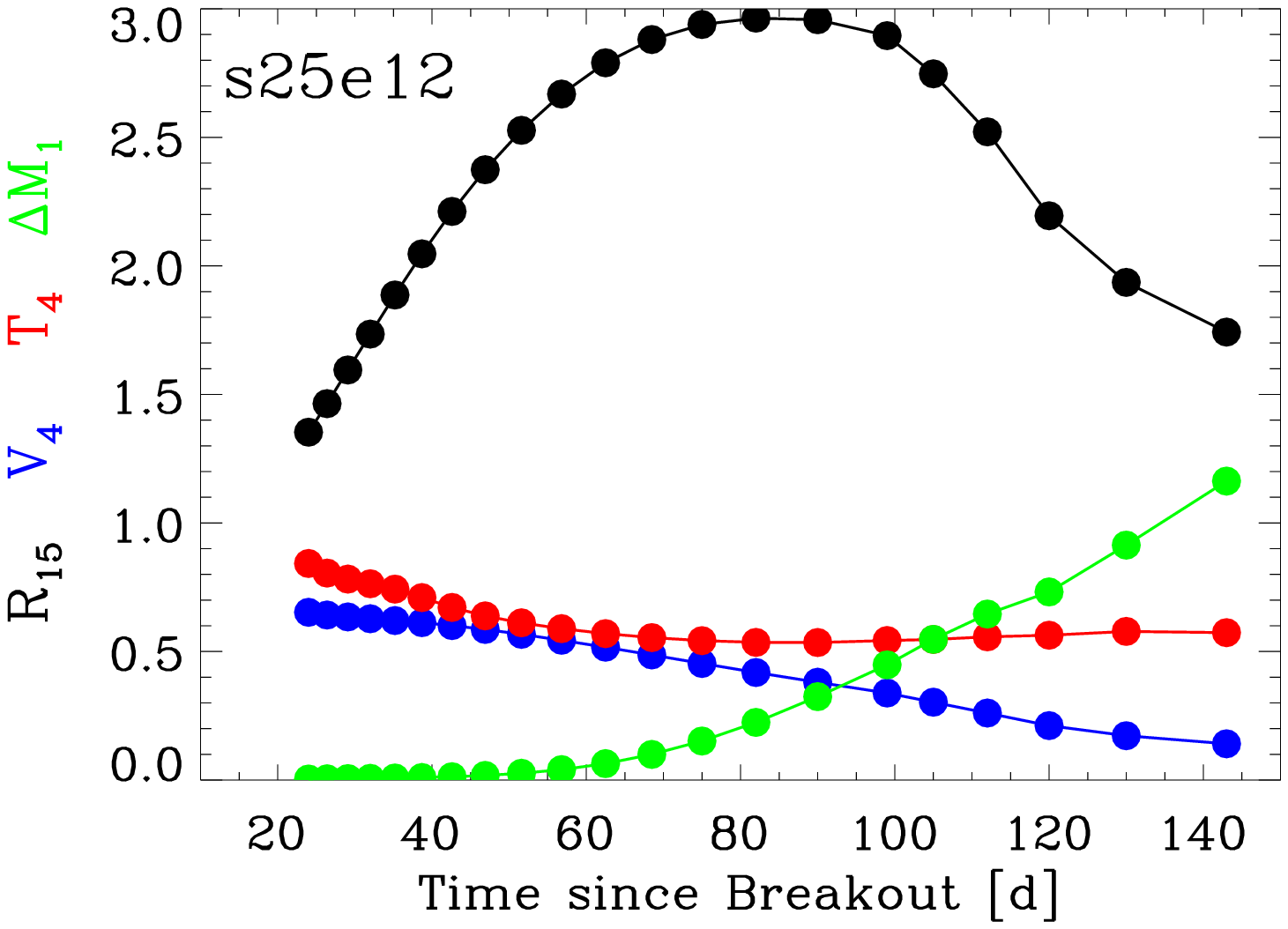,width=8.5cm}
\caption{
Evolution of the photospheric radius $R_{15}$ (black; $R_{15}=R_{\rm phot}/10^{15}$\,cm),
photospheric  velocity $V_4$ (blue; $V_4=V_{\rm phot}/10^4$\,\kms), photospheric temperature
$T_4$ (red; $T_4=T_{\rm phot}/10^4$\,K), and the mass above the photosphere $\Delta M_1$
(green; $\Delta M_1=\Delta M_{\rm phot}/10\,$\msun) until 150\,d after explosion,
and shown for simulations based on model s15e12 (left) and s25e12 (right).
The dots correspond to the actual times at which the radiation-transfer simulations were performed.
The evolution at nebular times is excluded since no photosphere exists at such epochs.
\label{fig_phot}
}
\end{figure*}

One issue of concern is the starting conditions. In our approach we fix the temperature to that of the hydrodynamical
simulation and solve the relativistic transfer equation and statistical equilibrium equations neglecting time dependent effects.
For the models presented here this is likely to be a reasonable approximation. Steady-state {\sc cmfgen} models showed that time-dependent effects for SN 1999em did not become noticeable in spectra until day 16 \citep{DH06_SN1999em}.  On the other hand our starting times would be unreasonable for SN 1987A, where time dependent effects are already important at day 4 \citep{UC05_time_dep, DH10}.  The reasons for starting these exploratory simulations late were two fold --- the need to run many additional models at earlier time steps was considered unnecessarily wasteful, and the assumption of a homologous expansion at earlier times is increasingly problematic.

\begin{figure}
\epsfig{file=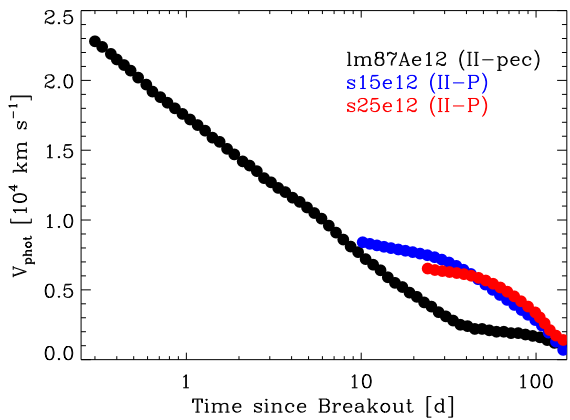,width=8.5cm}
\caption{Evolution of the photospheric velocity obtained with our simulations based on
SNe II-P models s15e12 and s25e12 (blue and red lines; this work) and the SN II-pec model
lm87Ae12 of SN1987A \citep[black line;][]{DH10}. \label{fig_vphot_comp}
}
\end{figure}

The post-explosion time corresponding to our ``adjusted'' inputs has been modified by 5-8\% (because of
the enforcement of homology) so the ejecta age is 8.51\,d (22.01\,d)  for our initial simulation based on model
s15e12 (s25e12). To step in time, we adopt a fixed logarithmic increment, $\Delta t/t \sim 0.1$
and continue the simulations just past 1000 days (we also run simulations
using a fixed 2-d time increment to cover the rapid fading at
the end of the plateau phase). For the simulation based on model s15e12 (s25e12),
the sequence covers until 1040\,d (1037\,d) in 52 (43) steps and took about 3 months to complete
for each. As in \citet{DH10} the gas temperature is held fixed, and set  to that provided by the hydrodynamical input model,
for the first model in the sequence. For subsequent models, the temperature is solved for everywhere. Hence, only
results from the second time step and onwards are shown or discussed here (corresponding to ages 9.3 and 24.0\,d for simulations
based on models s15e12 and s25e12).
We impose a floor value to the gas temperature to avoid problems with the evaluation of exponentials
involving high ionisation stages.
Since the ejecta cool as time proceeds, it eventually becomes unnecessary to include
high-ionization stages. As we discard them, starting from the highest ones,
we can lower the floor temperature and not face the floating-point-errors
we would otherwise get in the evaluation of exponentials.
In practice, we use a floor temperature of 2000--4000\,K during the plateau phase,
lowered to 1000-2000\,K during the nebular phase. A lower floor temperature, or none at all, would be more suitable.
Indeed, our tests suggest that this is in part the reason for the overestimated flux in the UV and $UB$ bands.

The spatial grid used for the radiative-transfer solution is set to have (approximately) a constant increment
in optical-depth (computed using the Rosseland-mean opacity), with typically 5 points per optical-depth decade,
and a maximum of 100 points. This is much less than the original 750-900 data points of the hydrodynamical-input
grid. However, our grid is much finer for the regions where radiation decouples from matter, the transition from
Eddington factors of 1/3 to 1 occurs over a few tens of points, while in the hydro
input this transition takes place over merely a few points. Furthermore, as the ejecta recombine, our grid adjusts
to track the recombination fronts (across which the optical depth varies steeply with radius). In SNe II, three
main fronts form, associated with the recombination to He\,{\sc ii}, He\,{\sc i}, and H\,{\sc i}. However, these fronts
never steepen as fiercely as in our simulations of SN~1987A \citep{DH10} or as in steady-state SNe II-P simulations
(e.g., \citealt{dessart_etal_08}). Convergence is not inhibited by this feature and is indeed quite steady
in the simulations we describe here.

For the treatment of radioactive decay, we assume only one decay chain corresponding to
$^{56}$Ni $\rightarrow$ $^{56}$Co $\rightarrow$ $^{56}$Fe.
In a forthcoming paper, we will present a Monte Carlo
code that solves for the transport of $\gamma$-ray photons from the site of emission out through the ejecta
and potentially towards escape.
We find that in SNe II, the deep-seated location of nucleosynthesised $^{56}$Ni
and the large overlying mass prevent the escape of $\gamma$-ray photons up to $\gtrsim$500\,d.
Non-local energy deposition starts however to operate at a few hundred days.
As the ejecta become thin to optical photons near the end of the photospheric phase,
high-energy (i.e., non-thermal) electrons, Compton-scattered by $\gamma$-rays,
will affect the excitation and ionisation level of the gas.

A proper and accurate treatment of radioactive decay requires first the determination
of the energy deposition versus depth \citep{AS88,swartz_etal_95}, and second the heating/excitation/ionisation
effects of the high-energy electrons produced \citep{axelrod_80,L91_HeI,swartz_91,KF92,swartz_etal_93,KF98a,KF98b}.
Although aware of these important issues, we assume in this first study of SNe II-P
that all decay energy is deposited locally as heat. We anticipate this will overestimate the SN visual brightness at nebular
times, and underestimate the excitation/ionisation level of the gas. In our simulations, this neglect seems to affect primarily
hydrogen, causing Balmer lines and the associated continuum to vanish abruptly beyond $\sim$130\,d after explosion,
contrary to observations (see \S~\ref{sect_comp_obs}).

\begin{figure*}
 \epsfig{file=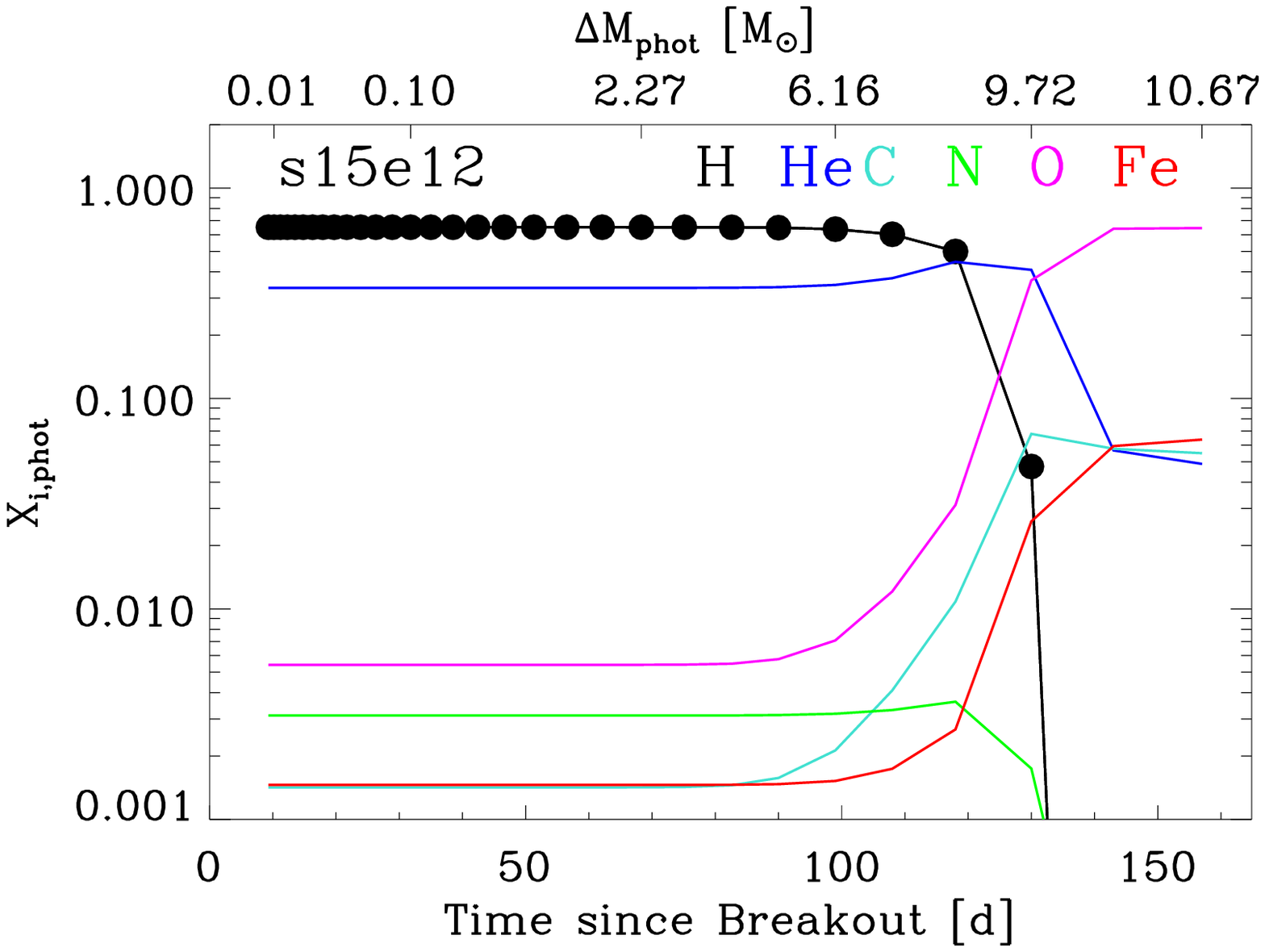,width=8.5cm}
 \epsfig{file=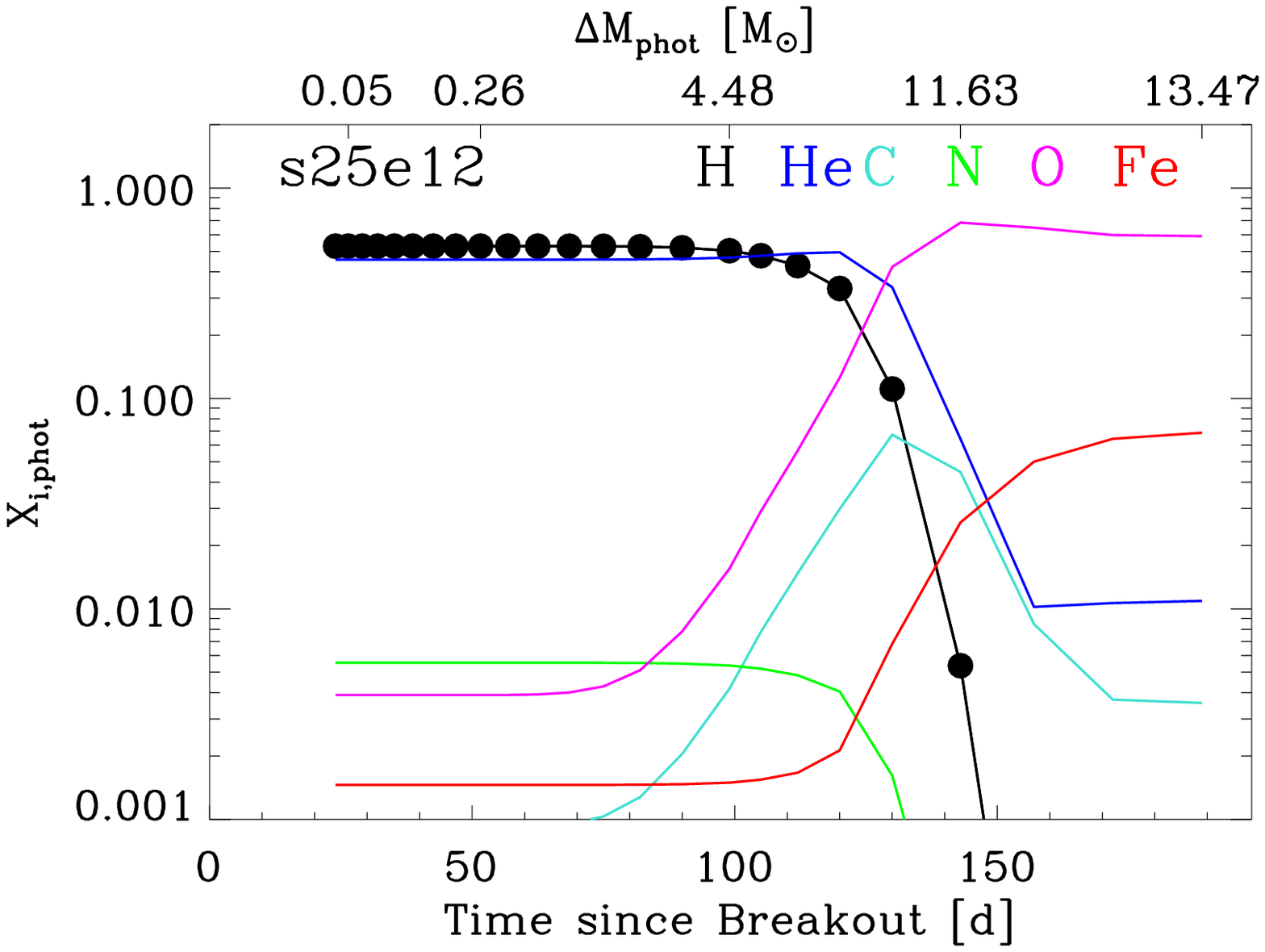,width=8.5cm}
\caption{Evolution of the composition at the photosphere for a few important species
in our simulations based on models s15e12 (left) and s25e12 (right).
The dots give the epoch for each simulation in  the time sequence, and $\Delta M_{\rm phot}$ in Fig.~\ref{fig_comp_phot}
is the mass contained between the photosphere and the outer-model boundary.
Notice the coincidence of the end of the plateau phase (shown further below in Fig.~\ref{fig_lbol}) with the recession
of the photosphere into the helium-rich (hydrogen-poor) layers of the ejecta.
}
\label{fig_comp_phot}
\end{figure*}

\section{Model Atoms, atomic data, and departures from LTE}
\label{model_atoms}

The model atoms used in our simulations are similar to those adopted in \citet{DH10}. The choice is a compromise
between physical accuracy, completeness, and computational burden. Our simulations require about 6GB of RAM
and one time step takes about 2-3 days on a single CPU.
In {\sc cmfgen}, we can in principle treat all species up to $Z=28$ (Nickel), as well as Barium ($Z$=56).
However, to limit the computational burden, our early {\sc cmfgen} simulations based on models s15e12/s25e12
neglected some species and/or ions.
In Table~\ref{tab_atom}, we document these details, including for each ion species the number of super
and full levels, the number of transitions, and the uppermost level included.\footnote{Level designations
are generally according to the National Institute of Standards and Technology (NIST), except we use
``z'' to refer to states where several high ``l'' states have been merged (say $l=4$ to $n-1$), and ``w'' states refer
to a level where all ``l'' states for a given multiplicity have been merged.}
 To capture the higher level of detail of the ejecta composition for model s15e12iso and compute
more accurate spectra for comparison with observations, we additionally
include in our treatment Ne\two, Al\two, Al\three, Ar\three, Sc\one, Sc\two, Sc\three, Cr\two, Cr\three, Fe\one,
Ni\two, Ni\three, Ni\four, Ba\one, and Ba\two\ (these additional ions appear with a superscript $a$).
As we discuss below, our simulations based on s15e12iso show the importance of treating Na\,{\sc i}, Sc\,{\sc ii},
Cr\,{\sc ii}, and Fe\,{\sc i}, as well as considering the corresponding elemental abundance stratification with depth.

\begin{table}
\begin{center}
\caption[]{Summary of the complete model atom used in our radiative-transfer calculations
for model s15e12iso (our simulations for models s15e12 and s25e12 were done
without the species identified with superscript $a$). The source of the atomic datasets is given in \citet{DH10}.
N$_{\rm f}$ refers to the number of full levels, N$_{\rm s}$ to the number of super levels, and N$_{\rm trans}$
to the corresponding number of bound-bound transitions. The last column refers to the upper level for each ion
treated. The total number of full levels treated is 8273, which corresponds to 173965 bound-bound transitions.}
\label{tab_atom}
\begin{tabular}{l@{\hspace{3mm}}r@{\hspace{3mm}}r@{\hspace{3mm}}r@{\hspace{3mm}}l}
\hline
 Species        &  N$_{\rm f}$  &  N$_{\rm s}$ & N$_{\rm trans}$ & Upper Level \\
\hline
H\,{\sc i}      & 30  &  20   &     435    & $n\le$ 30\\
He\,{\sc i}     & 51  &  40   &     374    & $n\le$ 11\\
He\,{\sc ii}    & 30  &  13   &     435    & $n\le$ 30\\
C\,{\sc i}      & 26  &  14   &     120    & $n\le$ 2s2p$^3$ $^3$P$^o_{0}$\\
C\,{\sc ii}     & 26  &  14   &      87    & $n\le$ 2s$^2$4d $^2$D$_{5/2}$\\
C\,{\sc iii}    & 112 &  62   &     891    & $n\le$ 2s8f $^1$F$^o$\\
C\,{\sc iv}     & 64  &  59   &    1446    & $n\le$ 30\\
N\,{\sc i}      & 104 &  44   &     855    & $n\le$ 5f $^2$F$^o$\\
N\,{\sc ii}     & 41  &  23   &     144    & $n\le$ 2p$^3$d $^1$P$_{1}$\\
O\,{\sc i}      & 51  &  19   &     214    & $n\le$ 2s$^2$2p$^3$($^4$S)4f $^3$F$_{3}$\\
O\,{\sc ii}     & 111 &  30   &    1157    & $n\le$ 2s$^2$2p$^2$($^3$P)4d $^2$D$_{5/2}$\\
O\,{\sc iii}    & 86  &  50   &     646    & $n\le$ 2p4f $^1$D\\
O\,{\sc iv}     & 72  &  53   &     835    & $n\le$ 2p$^2$($^3$P)3p $^2$P\\
O\,{\sc v}      & 78  &  41   &     523    & $n\le$ 2s5f $^1$F$^o_{3}$\\
Ne\,{\sc ii}$^a$ &     242   & 42  &  5794  & $n\le$ 2s$^2$2p$^4$ $^1$D 4d$\,^2$S            \\
Na\,{\sc i}     & 71  &  22   &    1614    & $n\le$ 30w $^2$W\\
Mg\,{\sc ii}    & 65  &  22   &    1452    & $n\le$ 30w $^2$W\\
Al\,{\sc ii}$^a$        &  44  & 26 &  171 &  $n\le$ 3s5d $^1$D$_{2}$                    \\
Al\,{\sc iiii}$^a$      & 45  &  17 &  362 &  $n\le$ 10z $^2$Z    \\
Si\,{\sc ii}    & 59  &  31   &     354    & $n\le$ 3s$^2$($^1$S)7g $^2$G$_{7/2}$\\
Si\,{\sc iii}   & 61  &  33   &     310    & $n\le$ 3s5g $^1$G$_{4}$\\
Si\,{\sc iv}    & 48  &  37   &     405    & $n\le$ 10f $^2$F$^o$\\
S\,{\sc ii}     &324  &  56   &    8208    & $n\le$ 3s3p$^3$($^5$S$^o$)4p $^6$P\\
S\,{\sc iii}    & 98  &  48   &     837    & $n\le$ 3s3p$^2$($^2$D)3d $^3$P\\
S\,{\sc iv}     & 67  &  27   &     396    & $n\le$ 3s3p($^3$P$^o$)4p $^2$D$_{5/2}$\\
Ar\,{\sc iiii}$^a$    & 346 & 32 &  6898 &  $n\le$ 3s$^2$3p$^3$($^2$D$^o$)8s $^1$D$^o$            \\
Ca\,{\sc ii}    & 77  &  21   &    1736    & $n\le$ 3p$^6$30w $^2$W\\
Sc\,{\sc i}$^a$      &  72  &   26 & 734  & $n\le$ 3d4s($^3$D)5s $^2$D$_{5/2}$          \\
Sc\,{\sc ii}$^a$      &  85   &  38 & 979  & $n\le$ 3p$^6$3d4f $^1$P$_{1}$              \\
Sc\,{\sc iii}$^a$     &  45   &  33  & 235  & $n\le$ 7h $^2$H$_{11/2}$                  \\
Ti\,{\sc ii}    & 152 &  37   &    3134    & $n\le$ 3d$^2$($^3$F)5p $^4$D$_{7/2}$\\
Ti\,{\sc iii}   & 206 &  33   &    4735    & $n\le$ 3d6f $^3$H$^o_{6}$\\
Cr\,{\sc ii}$^a$    &   196 &   28 & 3629  & $n\le$ 3d$^4$($^3$G)4p x$^4$G$_{11/2}$          \\
Cr\,{\sc iii}$^a$    &   145  &  30  & 2359 &  $n\le$ 3d$^3$($^2$D2)4p $^3$D$^o_{3}$             \\
Fe\.{\sc i}$^a$       &  136 &   44 & 1900 &  $n\le$ 3d$^6$($^5$D)4s4p x$^5$F$^o_{3}$           \\
Fe\,{\sc ii}    & 115 &  50   &    1437    & $n\le$ 3d$^6$($^1$G1)4s d$^2$G$_{7/2}$\\
Fe\,{\sc iii}   & 477 &  61   &    6496    & $n\le$ 3d$^5$($^4$F)5s $^5$F$_{1}$\\
Fe\,{\sc iv}    & 294 &  51   &    8068    & $n\le$ 3d$^4$($^5$D)4d $^4$G$_{5/2}$\\
Fe\,{\sc v}     & 191 &  47   &    3977    & $n\le$ 3d$^3$($^4$F)4d $^5$F$_{3}$\\
Fe\,{\sc vi}    & 433 &  44   &   14103    & $n\le$ 3p5($^2$P$^o$)3d$^4$($^1$S) $^2$P$_{3/2}$\\
Fe\,{\sc vii}   & 153 &  29   &    1753    & $n\le$ 3p5($^2$P$^o$)3d$^3$($^2$D1) $^1$P$_{1}$\\
Co\,{\sc ii}    & 144 &  34   &    2088    & $n\le$ 3d$^6$($^5$D)4s4p $^7$D$^{o}_{1}$\\
Co\,{\sc iii}   & 361 &  37   &   10937    & $n\le$ 3d$^6$($^5$D)5p $^4$P$_{3/2}$\\
Co\,{\sc iv}    & 314 &  37   &    8684    & $n\le$ 3d$^5$($^2$P)4p $^3$P$_{1}$\\
Co\,{\sc v}     & 387 &  32   &   13605    & $n\le$ 3d$^4$($^3$F)4d $^2$H$_{9/2}$\\
Co\,{\sc vi}    & 323 &  23   &    9608    & $n\le$ 3d$^3$($^2$D)4d $\,^1$S$_{0}$\\
Co\,{\sc vii}   & 319 &  31   &    9096    & $n\le$ 3p5($^2$P$^o$)3d$^4$($^3$F) $^2$D$_{3/2}$\\
Ni\,{\sc ii}$^a$ &   93      &   19  & 842  & $n\le$ 3d$^7$($^4F$)4s4p $^6$D$^o_{1/2}$          \\
Ni\,{\sc iii}$^a$ &      67    &   15  &  379  & $n\le$ 3d$^7$($^4$F)4p $^3$D$^o_{1}$              \\
Ni\,{\sc iv}$^a$ &      200  &   36 & 4085 & $n\le$ 3d$^6$($^3$D)4p $^2$D$_{5/2}$            \\
Ni\,{\sc v}   & 183 &  46   &    3065    & $n\le$ 3d$^5$($^2$D3)4p $\,^3$F$^o_{3}$\\
Ni\,{\sc vi}  & 314 &  37   &    9569    & $n\le$ 3d$^4$($^5$D)4d $^4$F$_{9/2}$\\
Ni\,{\sc vii} & 308 &  37   &    9225    & $n\le$ 3d$^3$($^2$D)4d $\,^3$P$_{2}$\\
Ba\,{\sc i}$^a$       &      31     &  17 & 30      &  $n\le$ 5d6p $^1$P$_{1}$                  \\
Ba\,{\sc ii}$^a$      &      100  &  38 & 2514  &  $n\le$ 3s30w $^2$W                      \\
\hline
\end{tabular}
\end{center}
\end{table}

The model atoms described in Table~\ref{tab_atom} are used to start the time sequences. As time proceeds, the ejecta
recombine and cool so that the highest ionisation species can progressively be discarded.
During the nebular phase, our model atom typically contains species that are at most three-times ionised.
Concomitantly, the low density and weaker radiation field at late times favours the appearance of
forbidden lines (associated with Ca\,{\sc ii} and O\,{\sc i} in particular).
We therefore adapt the super-level assignment to capture this change in conditions as ions get
less and less excited. In practice, past 100-200 days, we typically split the lower 10 super levels for all species
(splitting the lower 20 super levels does not alter the results visibly and supports this choice).
The set of atomic processes, most of the atomic data, and their use, have been summarised in \citet{DH10} and
discussed in more detail in \citet{HM98_lb}; they are thus not repeated here. For Ba\,{\sc ii}, not included in earlier models,
we used the atomic data discussed by \cite{MLB92_SN1987A} (kindly supplied by K.~Butler), while
for Sc\,{\sc ii} and Sc\,{\sc iii} we used oscillator strengths from \cite{Kur09_ATD}\footnote{Data is available online at
http://kurucz.harvard.edu} and crude approximations for the photoionisation cross-sections.
Energy levels, when available, were taken from NIST.\footnote{Data is available online at
http://physics.nist.gov/asd3}

\subsection{Departure Coefficients}
\label{fig_dep_coef}

In Fig.~\ref{fig_dep_coef} we show the departure coefficients as a function of the Rosseland-mean optical depth (\taur)
for H\one, O\one, and Fe\two\ at two different epochs. These departure coefficients are defined as the ratio of the non-LTE population
to the LTE population, where the LTE population has been computed using the electron temperature, and the actual values of
the ion and electron densities.  As expected, LTE is recovered at large optical depth (\taur$ > 10$). Below \taur=10,
significant departures from LTE begin to occur, and these are well developed around the photospheric layers. At small optical depths,
the departure from LTE can be orders of magnitude. In general the behavior of
the departure coefficients is complex, and highly dependent on the level under consideration.

\begin{figure}
\epsfig{file=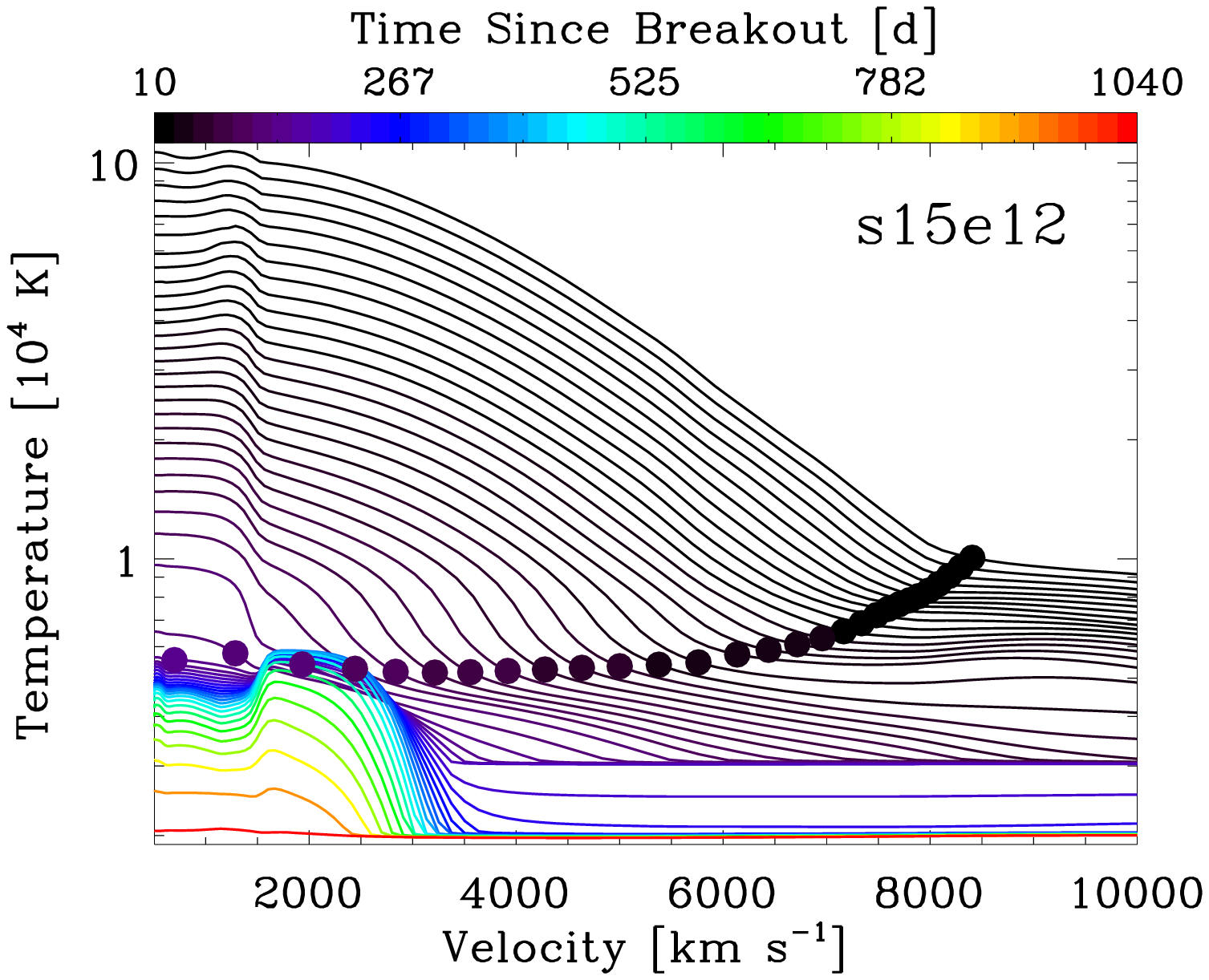,width=8.5cm}
\epsfig{file=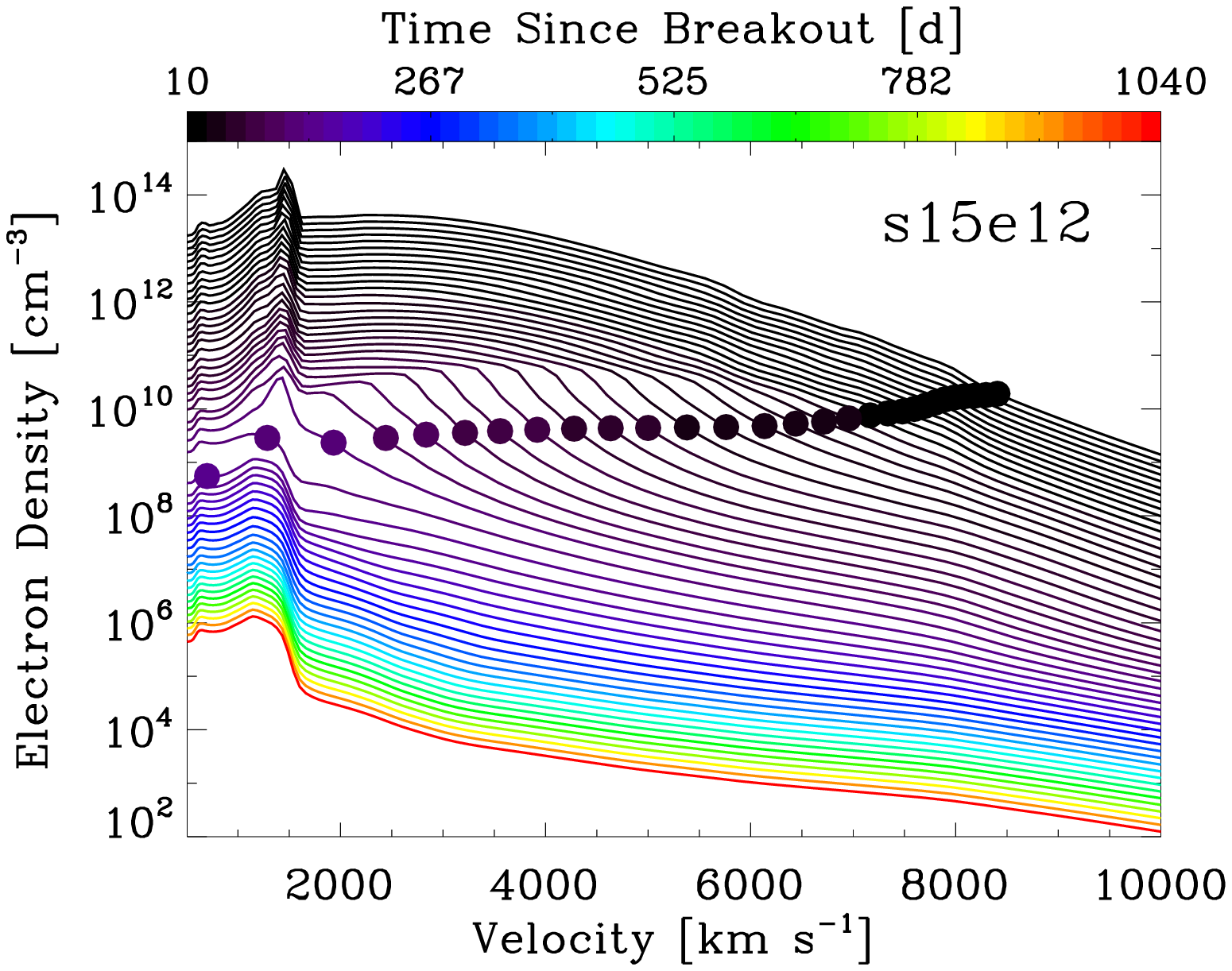,width=8.5cm}
\caption{{\it Top:} Evolution of the ejecta gas temperature from 10 to 1040\,d after explosion in
the simulations based on model s15e12. Dots give the position of the photosphere, which by definition
exists as long as the base electron-scattering optical depth is greater than 2/3.
A color coding is used to differentiate the epochs.
{\it Bottom:} Same as top, but now showing the electron density. \label{fig_t_ed}
}
\end{figure}

The very significant departures of the populations from their LTE values is not surprising. In general electron scattering, and line scattering
dominate the opacity, and these do not strongly couple the radiation field with the electron temperature.
Further, the electron density is low at the photosphere ($\sim10^{10}$\,cm$^{-3}$), and hence collisions are generally
ineffective at causing a significant coupling of the majority of the level
populations and ionisation state of the gas, and cannot overcome their
stronger coupling with the radiation field.

Above the photosphere the electron temperature is strongly decoupled from the radiation field temperature,
and is instead determined by an equilibrium between many different processes, including adiabatic cooling,
collisional cooling and heating, and heating by photoionization. Indeed, at the outer boundary of s15e12iso  at, e.g., $\sim$50\,d,
the temperature has reached the floor value of 3000\,K, while the photospheric temperature is somewhat lower than 6000\,K.
If we assume that the ground-state population is determined by the radiation temperature,
we would expect a departure coefficient of order $\exp(157,850/T_{\rm rad} - 157,850/T_{\rm e})$ or $\sim 3.0 \times 10^{-12}$
assuming $T_e=3000$\,K, $T_{\rm rad}=6000$\,K and a fixed electron density; hence very small departure coefficient
for the lower levels are to be expected. Of course the situation is more complex than assumed above  --- the radiation field
is a diluted black-body, and in the outer regions the H ionization is not in equilibrium because the recombination time-scale
is longer than the expansion time scale.

For H{\sc i}, the departure coefficients are close to unity for high principal
quantum number, as expected. For Fe\,{\sc ii}, all the levels shown have very
small departure coefficients at low \taur. This is not surprising --- although
we have shown 115 levels (with a smaller number of super levels),  the
principal quantum number of the valence electron for levels illustrated is less than 5.
For Fe\,{\sc ii} the lower metastable levels are coupled by collisions with the ground state, and hence tend to be
in LTE with respect to the ground state (i.e., they have the same departure coefficient as the ground state).
This can be seen in Fig.~\ref{fig_dep_coef}  --- many levels have similar departure coefficients  around \taur=1.
A similar statement applies to metastable levels of other elements.

In Monte-Carlo simulations of SN spectra it is common practice to use approximate formula to determine
the ionization structure, and to use an LTE assumption to determine populations \citep[e.g.,][]{KTN06_SN_MC,KW09}.
As the photon scattering is treated explicitly, the calculations are neither LTE, or rigorously non-LTE.
As apparent from the above, and indeed from our need to explicitly allow for the time-dependence when solving for
the hydrogen ionization structure, non-LTE populations need to be allowed for when modeling Type II SN.
Experience indicates that as we include more processes, more lines, and more levels, the departures from LTE will
probably become smaller, but the departures will still be highly significant.

\section{Ejecta evolution}
\label{sect_gas}

   Over the 10-1000\,d covered by each simulation, the ejectum expands by two orders
   of magnitude and changes from being rather dense and optically-thick to being very tenuous and optically-thin.
   Over the same time period, the ejectum cools from a few 10$^5$\,K at depth, to temperatures  everywhere less than
   3000\,K. 
    These dramatic changes cause an equivalently dramatic change in the radiated light, which we describe in
   \S~\ref{sect_LC} and \S~\ref{sect_spec}.

   During the photospheric phase, the electron-scattering radial optical depth is, by definition, greater than a few.
   Thus during this phase there always exists a ``photosphere'', which we formally define as the location where that optical depth is 2/3.
   The behaviour of the photosphere determines to a large extent the SN light curve, and
   we illustrate the evolution of its properties in Fig.~\ref{fig_phot}.
   Starting $\gtrsim$10\,d past explosion, our simulations miss the phase of rapid evolution of photospheric properties
   which immediately follows shock breakout \citep{dessart_etal_08,gezari_etal_08,tominaga_etal_09,rabinak_waxman_10}.
   Indeed, at the start of our simulations, the photosphere has already expanded by about a factor of ten, its temperature
   dropped to $\lesssim$10000\,K, and its velocity has been reduced to $\lesssim$8500\,\kms.
   Subsequently, for each simulation, and over the first $\sim$80\,d, we find a slow decrease of the photospheric temperature
   and velocity, but a systematic increase in radius.

   Importantly, the photosphere resides within the outer 0.5\,\msun\ of the ejecta for about $\sim$50\,d after the explosion
   (Fig.~\ref{fig_comp_phot}).
   Over that initial expansion, the ejecta has sufficient thermal energy to remain ionised while expanding, so
   that the photosphere remains close to the progenitor surface layers. As recombination sets in during the plateau phase,
   the photosphere eventually recedes in mass, but this recession is initially compensated by expansion so that the change in photospheric
   radius is small ---  in the present models the photospheric radius only begins to shrink after $\sim$ 80\,d.

   \begin{table}
\caption{Summary of photospheric-phase properties for the non-LTE time-dependent
simulations based on model s15e12.
For each epoch, we give the total ejecta electron-scattering
optical depth, the photospheric radius (in $10^{15}$\,cm), the photospheric
velocity (in \kms), the photospheric temperature (in K), the amount of mass
above the photosphere ($\Delta M_{\rm phot}$, given in \msun),
as well as the emergent bolometric luminosity (in $10^8$\,L$_{\odot}$).
Results for the time-sequence based on s15e12iso are the same as for those based on
s15e12, and hence are not shown here. \label{tab_prop_s15}
}
\begin{tabular}{l@{\hspace{1.6mm}}c@{\hspace{1.6mm}}c@{\hspace{1.6mm}}c@{\hspace{1.6mm}}c@{\hspace{1.6mm}}c@{\hspace{1.6mm}}c@{\hspace{1.6mm}}}
\hline
\multicolumn{7}{c}{s15e12} \\
\hline
 Age       & $\tau_{\rm base,es}$ &    $R_{\rm phot}$      &   $V_{\rm phot}$   &   $T_{\rm phot}$  &   $\Delta M_{\rm phot}$   & $L_{\rm bol}$ \\
 \hline
 [d]      &                    &     [$10^{15}$\,cm]   &  [\kms]  &   [K]  & [\msun] & [$10^8$\,L$_{\odot}$] \\
\hline
          9.30   &    14034.20  &         0.69 &  8545 & 10661  &        0.008 &  7.94  \\
         10.20   &    11053.89  &         0.74 &  8406 & 10063  &        0.010 &  7.64  \\
         11.20   &     9134.01  &         0.80 &  8294 &  9527  &        0.011 &  7.54  \\
         12.30   &     7567.96  &         0.87 &  8183 &  9057  &        0.014 &  7.50  \\
         13.50   &     6259.40  &         0.94 &  8083 &  8642  &        0.016 &  7.44  \\
         14.80   &     5192.15  &         1.02 &  7988 &  8317  &        0.019 &  7.37  \\
         16.30   &     4252.72  &         1.11 &  7890 &  8054  &        0.023 &  7.24  \\
         17.90   &     3490.37  &         1.21 &  7800 &  7859  &        0.027 &  7.11  \\
         19.70   &     2823.76  &         1.31 &  7701 &  7678  &        0.032 &  6.96  \\
         21.70   &     2287.51  &         1.42 &  7595 &  7477  &        0.038 &  6.78  \\
         23.90   &     1863.47  &         1.54 &  7480 &  7208  &        0.046 &  6.65  \\
         26.30   &     1519.30  &         1.67 &  7337 &  6878  &        0.057 &  6.56  \\
         28.90   &     1233.58  &         1.79 &  7164 &  6564  &        0.075 &  6.49  \\
         31.80   &      993.68  &         1.91 &  6957 &  6305  &        0.101 &  6.42  \\
         35.00   &      796.48  &         2.03 &  6715 &  6087  &        0.141 &  6.35  \\
         38.50   &      630.28  &         2.14 &  6438 &  5893  &        0.200 &  6.30  \\
         42.40   &      495.07  &         2.25 &  6135 &  5735  &        0.290 &  6.29  \\
         46.60   &      388.62  &         2.32 &  5760 &  5478  &        0.455 &  6.19  \\
         51.30   &      304.74  &         2.38 &  5378 &  5403  &        0.733 &  6.19  \\
         56.50   &      240.12  &         2.44 &  4999 &  5355  &        1.118 &  6.15  \\
         62.10   &      190.81  &         2.49 &  4634 &  5306  &        1.619 &  6.03  \\
         68.30   &      149.01  &         2.52 &  4277 &  5255  &        2.274 &  5.85  \\
         75.10   &      113.56  &         2.54 &  3918 &  5211  &        3.097 &  5.61  \\
         82.60   &       83.02  &         2.54 &  3556 &  5174  &        4.071 &  5.32  \\
         90.00   &       58.83  &         2.50 &  3214 &  5162  &        5.069 &  5.02  \\
         99.00   &       37.12  &         2.43 &  2835 &  5185  &        6.159 &  4.52  \\
        108.00   &       21.80  &         2.28 &  2441 &  5276  &        7.144 &  3.98  \\
        118.00   &        9.53  &         1.97 &  1932 &  5415  &        8.069 &  3.31  \\
        130.00   &        2.19  &         1.44 &  1286 &  5775  &        9.723 &  2.18  \\
        143.00   &        0.75  &         0.86 &   698 &  5550  &       10.628 &  1.11  \\
\hline
\end{tabular}
\end{table}

\begin{table}
\caption{
Same as Table~\ref{tab_prop_s15}, but now for the time sequence based on model s25e12. \label{tab_prop_s25}
}
\begin{tabular}{l@{\hspace{1.6mm}}c@{\hspace{1.6mm}}c@{\hspace{1.6mm}}c@{\hspace{1.6mm}}c@{\hspace{1.6mm}}c@{\hspace{1.6mm}}c@{\hspace{1.6mm}}}
\hline
\multicolumn{7}{c}{s25e12} \\
\hline
 Age       & $\tau_{\rm base,es}$ &    $R_{\rm phot}$      &   $V_{\rm phot}$   &   $T_{\rm phot}$  &  $\Delta M_{\rm phot}$   & $L_{\rm bol}$  \\
 \hline
  [d]      &                    &     [$10^{15}$\,cm]   &  [\kms]  &   [K]  & [\msun] &  [$10^8$\,L$_{\odot}$] \\
\hline
        24.0 &      2124.66 &         1.35 &         6524 &         8423 &        0.040  &    12.64 \\
        26.4 &      1709.78 &         1.46 &         6419 &         8050 &        0.048  &    11.87 \\
        29.1 &      1356.11 &         1.60 &         6344 &         7815 &        0.057  &    11.50 \\
        32.0 &      1080.90 &         1.73 &         6274 &         7635 &        0.067  &    11.21 \\
        35.2 &       849.14 &         1.89 &         6203 &         7426 &        0.080  &    10.91 \\
        38.7 &       665.03 &         2.05 &         6121 &         7097 &        0.098  &    10.55 \\
        42.6 &       523.98 &         2.21 &         6009 &         6714 &        0.128  &    10.19 \\
        46.9 &       414.59 &         2.37 &         5858 &         6385 &        0.179  &     9.81 \\
        51.6 &       324.49 &         2.53 &         5668 &         6116 &        0.263  &     9.42 \\
        56.8 &       251.96 &         2.67 &         5437 &         5889 &        0.405  &     9.05 \\
        62.5 &       197.03 &         2.79 &         5166 &         5696 &        0.639  &     8.69 \\
        68.5 &       157.38 &         2.88 &         4866 &         5540 &        0.997  &     8.33 \\
        75.0 &       123.25 &         2.94 &         4535 &         5422 &        1.518  &     7.98 \\
        82.0 &        94.47 &         2.96 &         4183 &         5354 &        2.243  &     7.64 \\
        90.0 &        67.84 &         2.96 &         3803 &         5349 &        3.244  &     7.28 \\
        99.0 &        45.06 &         2.90 &         3385 &         5427 &        4.480  &     6.86 \\
       105.0 &        33.80 &         2.75 &         3028 &         5467 &        5.473  &     6.55 \\
       112.0 &        24.62 &         2.52 &         2605 &         5571 &        6.457  &     5.98 \\
       120.0 &        16.11 &         2.19 &         2117 &         5633 &        7.317  &     5.43 \\
       130.0 &         8.67 &         1.94 &         1723 &         5781 &        9.138  &     4.81 \\
       143.0 &         3.30 &         1.74 &         1410 &         5732 &       11.628  &     3.22 \\
       157.0 &         1.16 &         1.57 &         1154 &         5434 &       12.923  &     1.91 \\
       172.0 &         0.84 &         1.03 &          693 &         5270 &       13.365  &     1.51 \\
       189.0 &         0.66 &         0.39 &          238 &         5181 &       13.473  &     1.30 \\
\hline
\end{tabular}
\end{table}

   The photospheric composition does not change for the first  $\sim$70\,d in these two SNe II-P
   models and is identical to that of the progenitor surface. The same finding by \citet{eastman_etal_94}
   suggests that this is a generic property of SNe II-P.
   It implies that the spectral evolution of SNe II-P during the first half
   of the plateau phase does not stem from composition changes, but instead from changes in ionisation.
   The same applies to the early evolution of SN~1987A \citep{DH10}.

  The near uniform photospheric composition with time during the photospheric phase, as well as the modest
  composition contrast between the hydrogen envelopes of RSG models (see, e.g., \citealt{WHW02} or \citealt{WH07}),
  highlights the difficulty of constraining the progenitor identity with photospheric-phase spectra.
  It also partially explains the uniformity of SNe II-P spectra, which tend to  show lines from the same ions,
  with the same strength and shapes, and generally only differing in width \citep{DH06_SN1999em,dessart_etal_08}.
   This suggests that differences between SNe II-P are difficult to extract after the onset of recombination in the
  photospheric phase. They may instead be constrained from early-time observations, when the photospheric
  regions are partially ionised \citep{DH06_SN1999em,baron_etal_07}, or from nebular-phase observations
  during which the progenitor core, whose properties differ greatly with progenitor main-sequence mass,
  is revealed \citep{WHW02,DLW10b}.

  The location of the SN II-P photosphere in the outer $\lesssim$1\,\msun\ of the ejecta suggests that a mass buffer of $\sim$10\,\msun\
   of material separates the radiating layer from the birth place of the explosion.
  As long as the photosphere remains in those outer ejecta shells, which seems to be during the first half of the plateau phase,
  radioactive decay cannot influence the emergent radiation of SNe II-P.
  Altogether, these properties give support to the notion that distance determinations to SNe II-P using spectroscopy
  are best determined from early- rather than late-time observations \citep{baron_etal_07,dessart_etal_08,DH09_review}.

  A comparison with simulations of SN~1987A \citep{DH10} illustrates further the ingredients controlling the behaviour of
  the photosphere. At early post-breakout times, the rapid recombination of the highly-ionised fast-expanding outer
  ejecta leads to a steep decrease of the photospheric velocity (Fig.~\ref{fig_vphot_comp}).
  In SNe II-pec, this fast decrease is caused in part by the
  relatively small progenitor-star radius, and it slows down only when the photosphere reaches the inner ejecta heated
  by decay energy (at $\sim$30\,d in SN1987A). In SNe II-P, a similar, albeit more modest, decrease in photospheric velocity is seen at early times
  \citep{gezari_etal_08}, but it slows down as recombination sets in. A rapid decrease kicks in again when the
  photosphere reaches the base of the hydrogen envelope (at $\gtrsim$100\,d in the two present simulations).

   Compared to our analyses of SNe II-P 1999em and 2006bp \citep{DH06_SN1999em,dessart_etal_08},
   characterised by explosion energies close to but likely larger than the 1.2\,B value of our models \citep{utrobin_07,DLW10b},
   the photospheric
   quantities we obtain with these two sets of simulations are in good agreement for the temperature and the velocity.
   However, our models reach larger maximum radii of 2.5--3$\times$10$^{15}$\,cm, compared to $\sim$1.6$\times$10$^{15}$\,cm
   \citep{dessart_etal_08}.  We speculate on the origin of this discrepancy in \S~\ref{sect_comp_obs}.

\begin{figure}
\epsfig{file=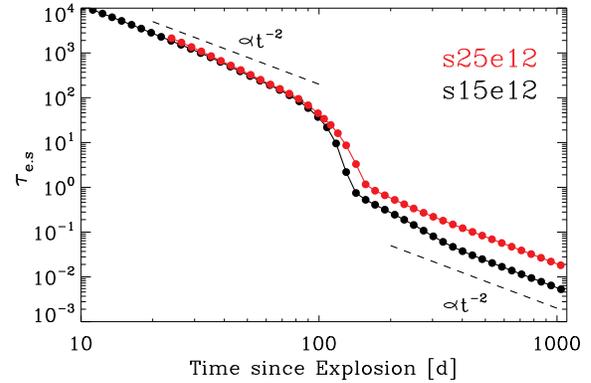,width=8.5cm}
\caption{Evolution of the electron-scattering optical depth at the base of the ejecta for the simulations
based on model s15e12 (black) and s25e12 (red). Dots represent the epochs at which our computations
are carried out. We draw a line of $1/t^2$ slope, which gives the rate of change of the optical depth for
a homologously expanding gas of non-changing ionisation. The faster decline that we obtain in both simulations
is caused by ionisation changes (i.e., recombination, and primarily of hydrogen),
while the step at $\sim$100\,d is caused by a composition change as the photosphere enters the hydrogen-deficient
helium-rich core. \label{fig_tau}
}
\end{figure}

   As the homologously expanding SN evolves through the plateau phase, the ejecta electron-scattering optical depth
   decreases (Fig~\ref{fig_tau}).
   It does so initially at constant ionisation, so its rate of decline follows a $1/t^2$ slope, but as recombination sets in after a few weeks,
   this rate becomes steeper. At about 100\,d after explosion, the optical depth suddenly  declines much faster in both simulations, as the
   photosphere enters the hydrogen-deficient helium-rich regions of the ejecta, which possess a higher mean-atomic weight and
   hence a smaller mass-absorption coefficient.

   The photosphere vanishes when the total ejecta electron-scattering optical depth drops below 2/3
   (last entry in Tables~\ref{tab_prop_s15} and \ref{tab_prop_s25}) and the SN enters the nebular phase.
   The ejecta are optically thin initially in the optical continuum, although not yet in the lines.
   All ejecta mass shells are visible, with emission biased towards the denser/hotter inner layers.
   During the nebular phase,
   the maximum electron density decreases to values on the order of 10$^5$--10$^6$\,cm$^{-3}$, and
   forbidden lines such as [O{\sc i}]\,6303--6363\AA\  and [Ca{\sc ii}]\,7291--7323\,\AA\ become strong.
   The nebular ejecta does not change its ionisation
   much with time, and the rate of change of the now very low electron-scattering optical depth follows approximately
   a $1/t^2$ slope.
   For completeness, we show the full evolution of the ejecta temperature and electron density over the whole
   time sequence for the simulation based on model s15e12 in Fig.~\ref{fig_t_ed}.

\begin{figure*}
\epsfig{file=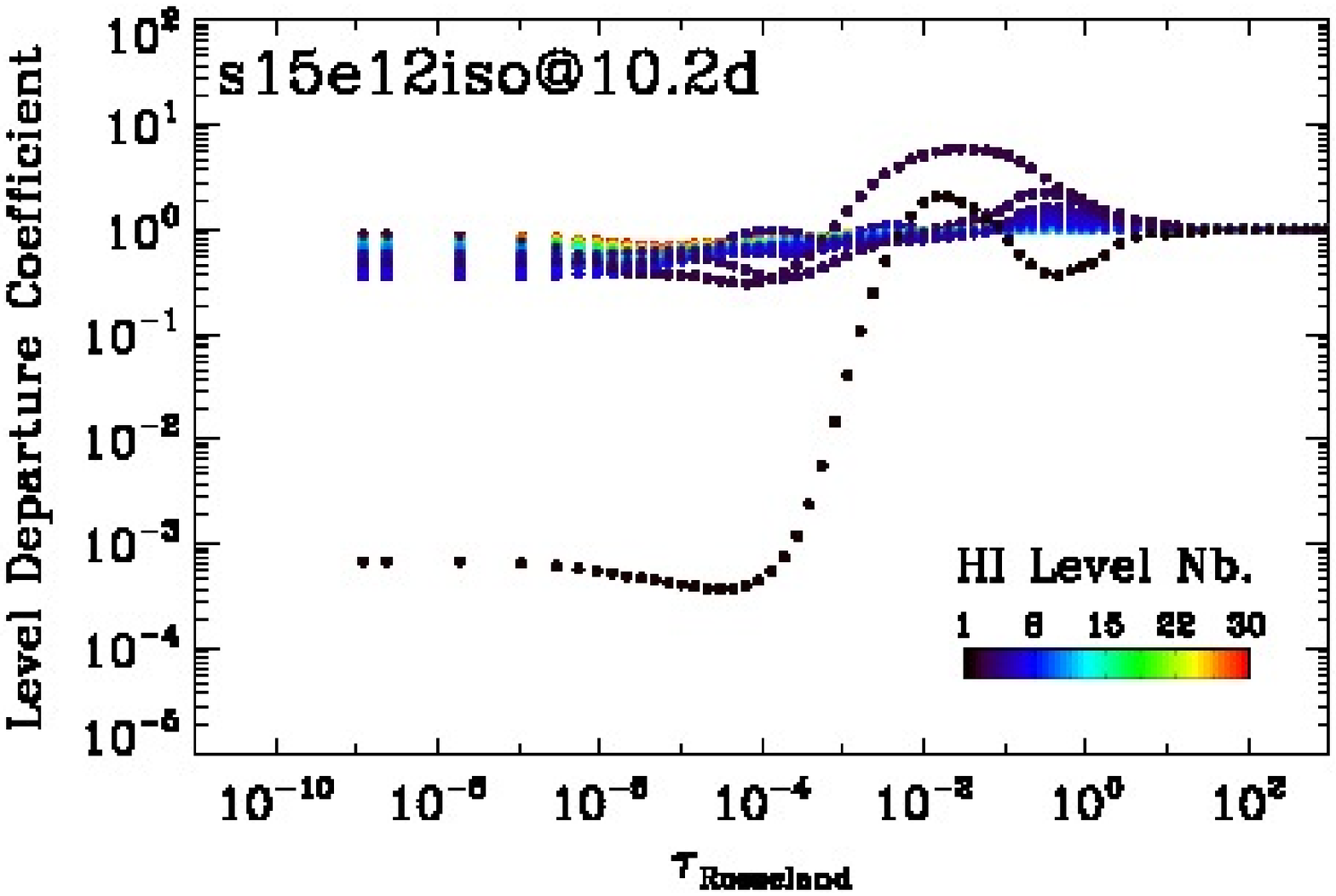,width=8.5cm}
\epsfig{file=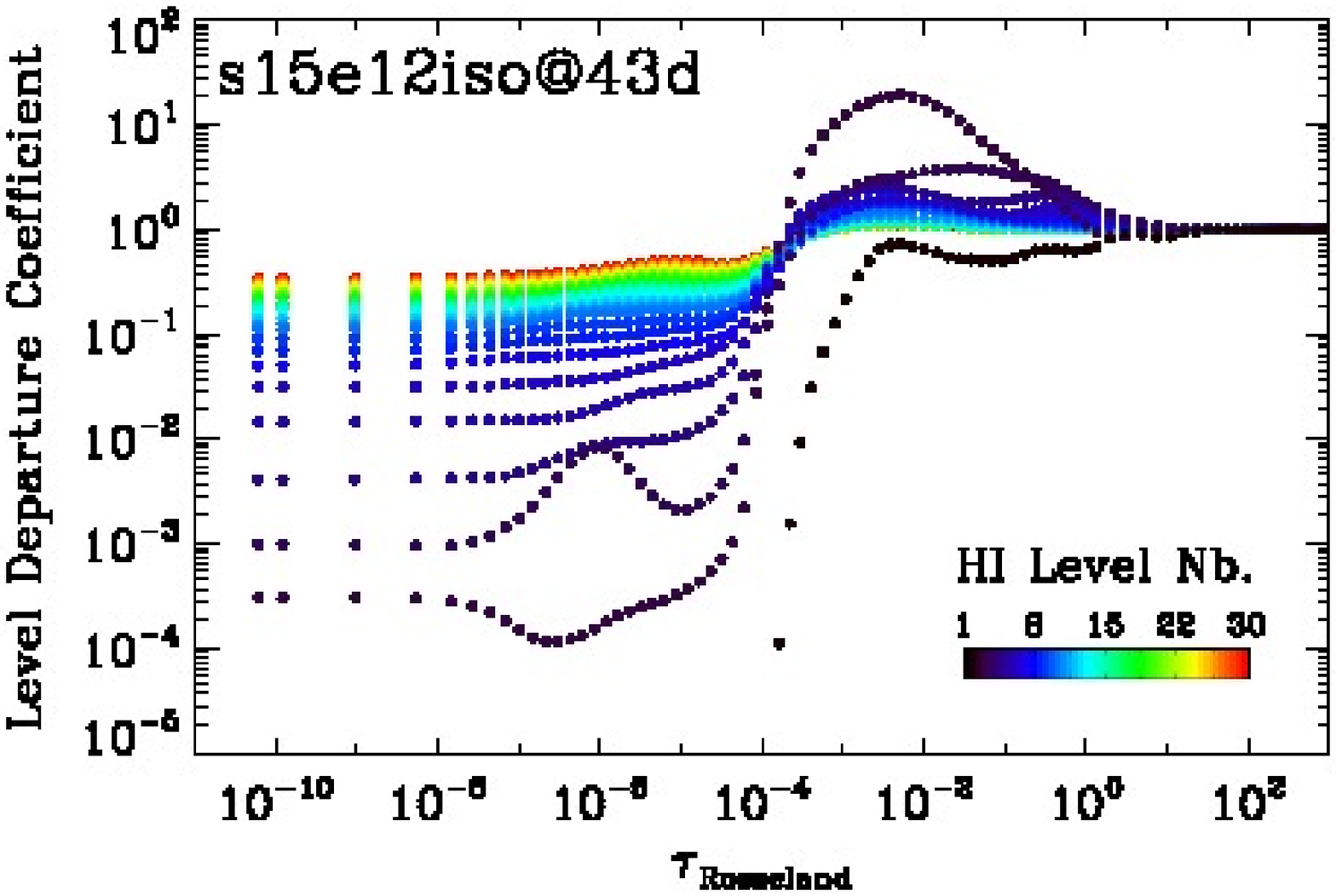,width=8.5cm}
\epsfig{file=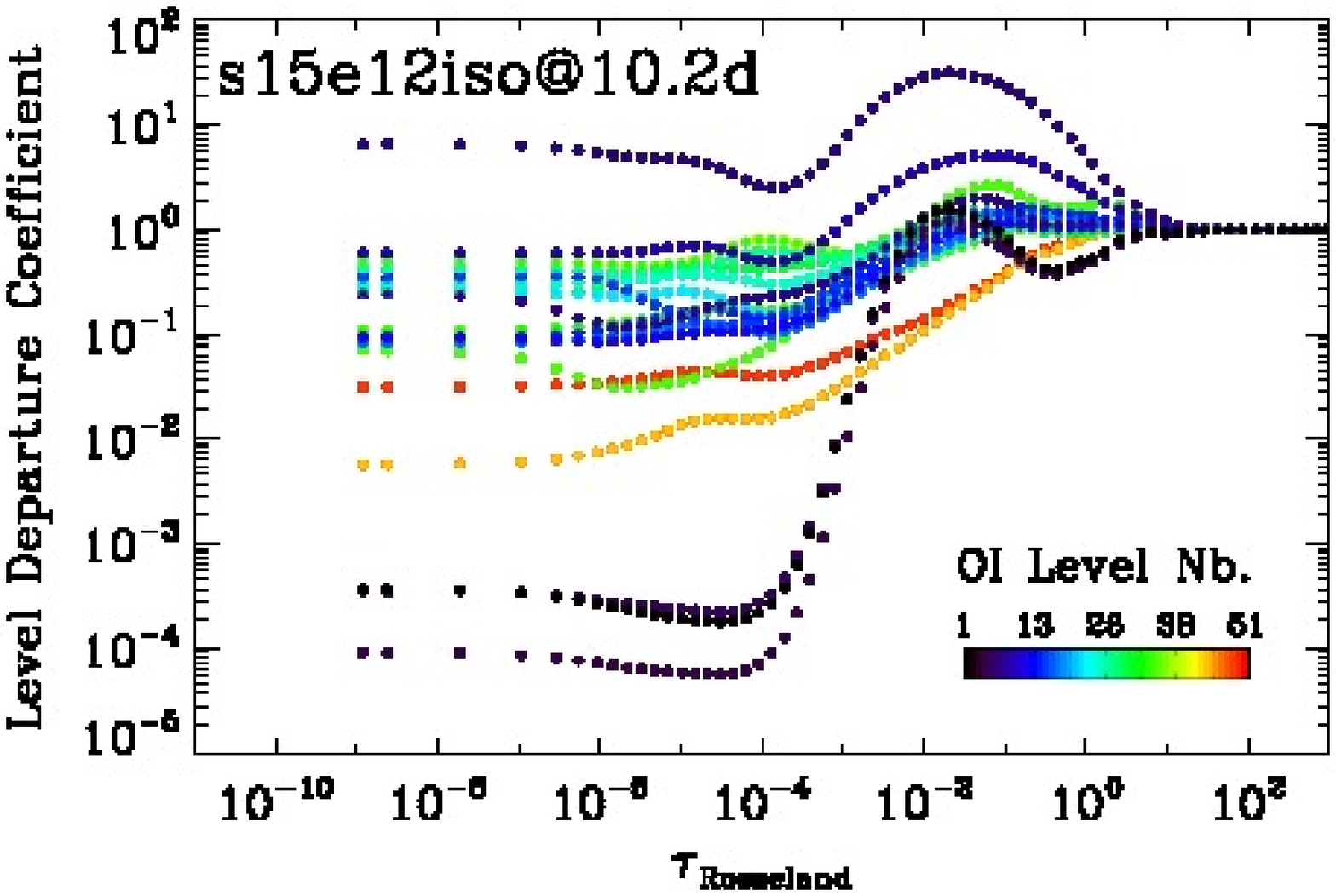,width=8.5cm}
\epsfig{file=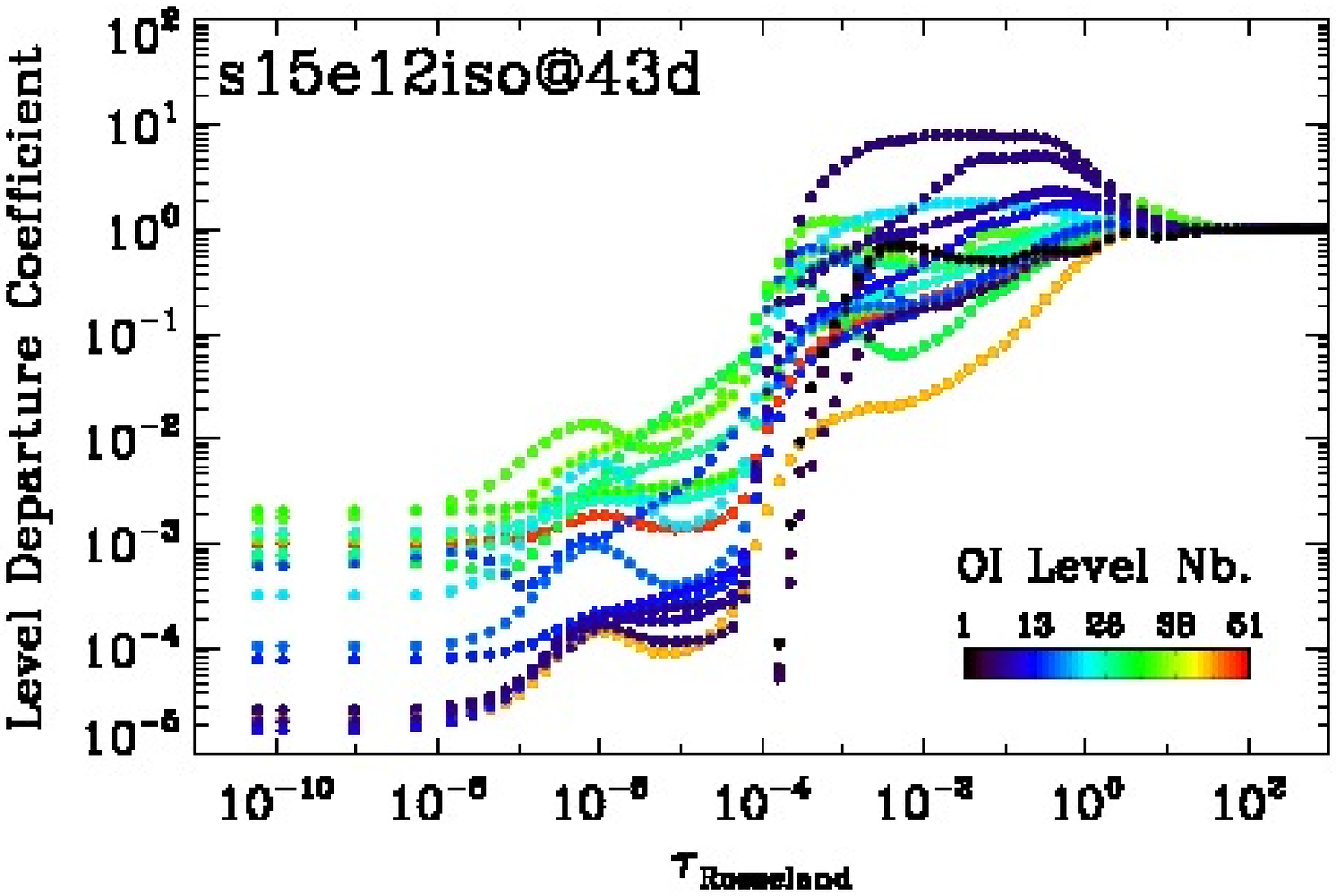,width=8.5cm}
\epsfig{file=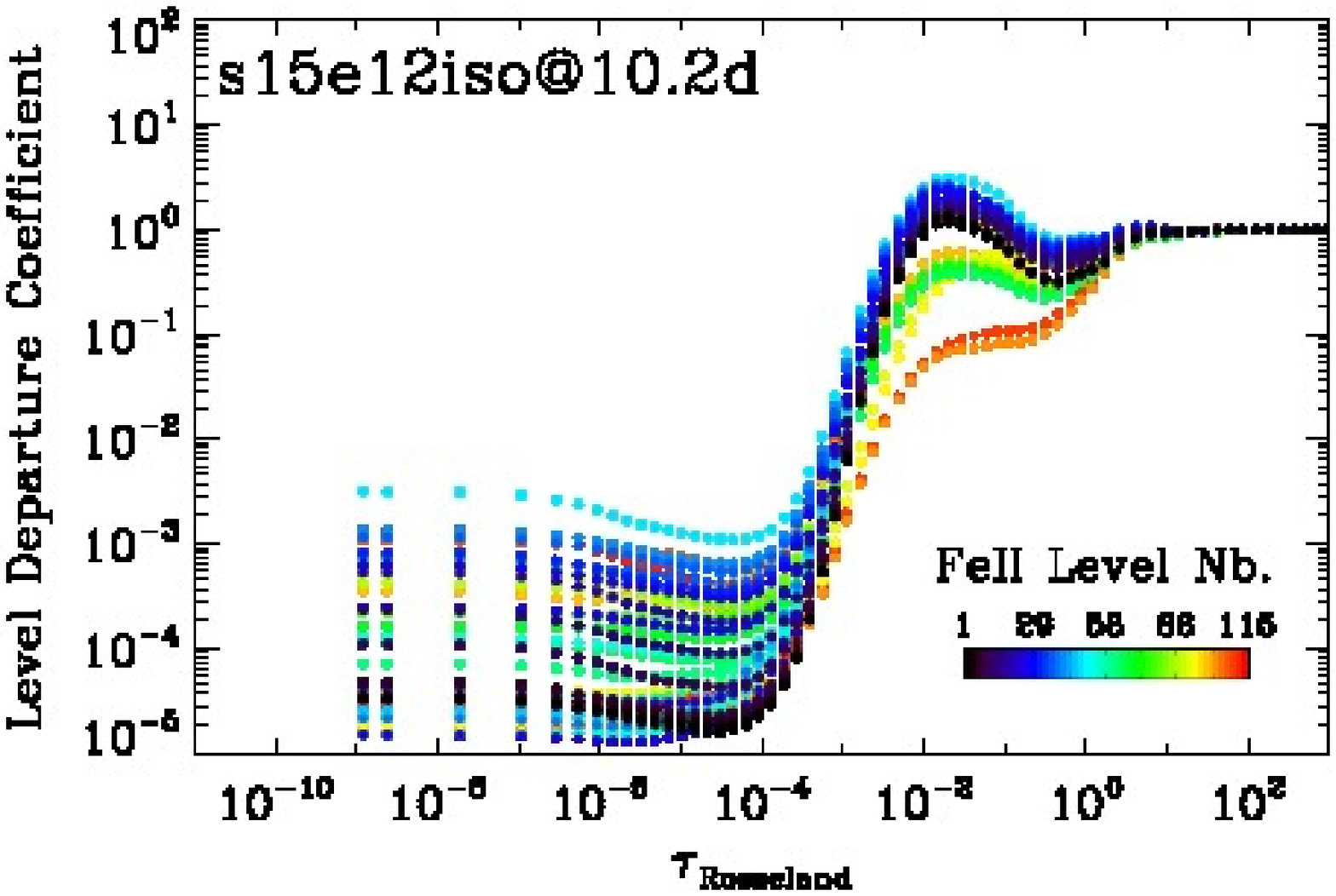,width=8.5cm}
\epsfig{file=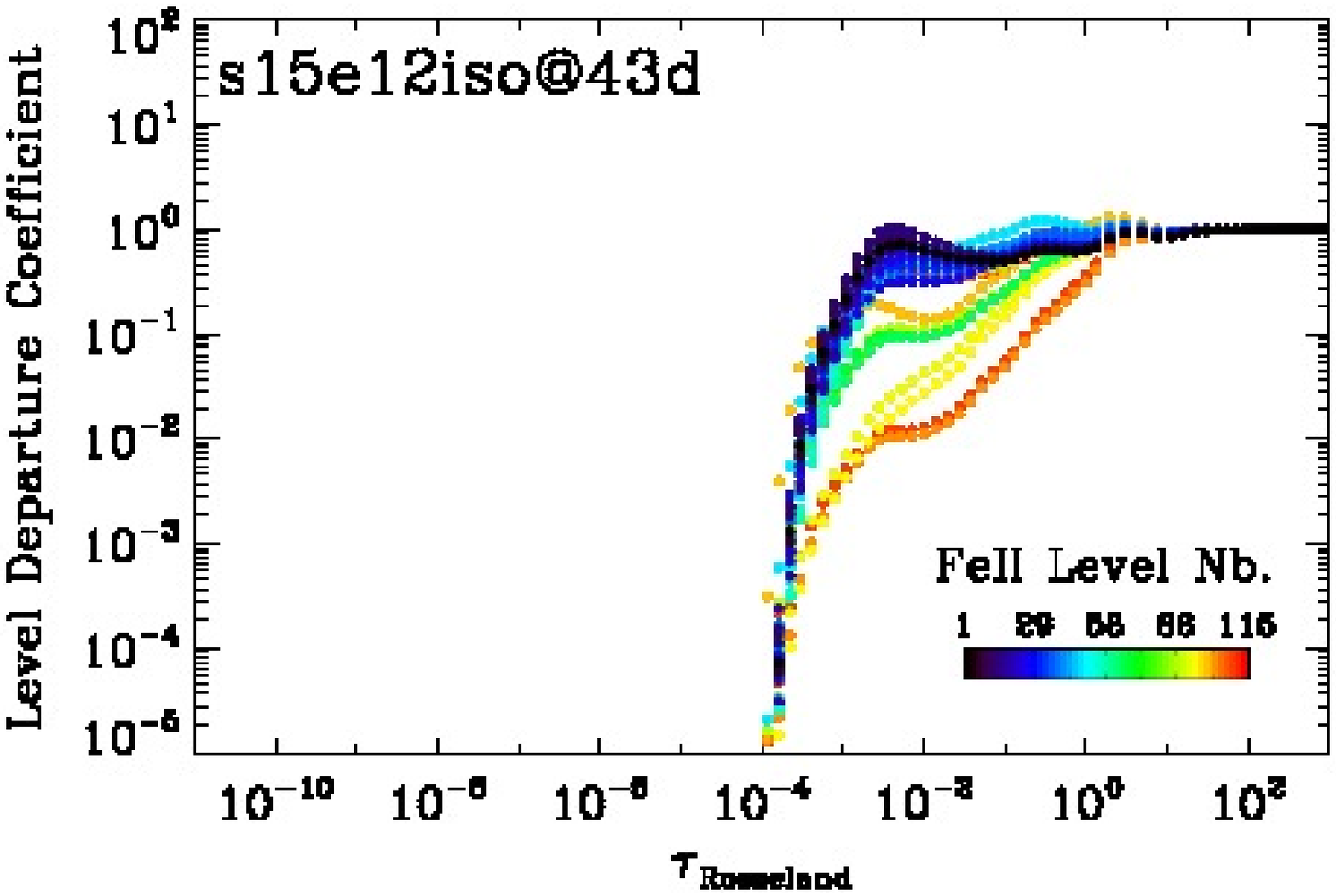,width=8.5cm}
\caption{{\it Left column:} Variation of the departure coefficients of H{\sc i} (top), O{\sc i} (middle), and
Fe{\sc II} levels versus Rosseland optical depth for models s15e12iso at 10.2\,d after shock breakout.
{\it Right column:} Same as left, but for the model s15e12iso at 43\,d after shock breakout.
Non-LTE level populations computed in our simulations match closely their LTE counterpart
only in regions where the Rosseland mean optical depth is greater than a few. Non-LTE prevails
everywhere else.\label{fig_dep_coef}}
\end{figure*}

\section{Synthetic bolometric luminosity and light curves}
\label{sect_LC}

\begin{figure}
\epsfig{file=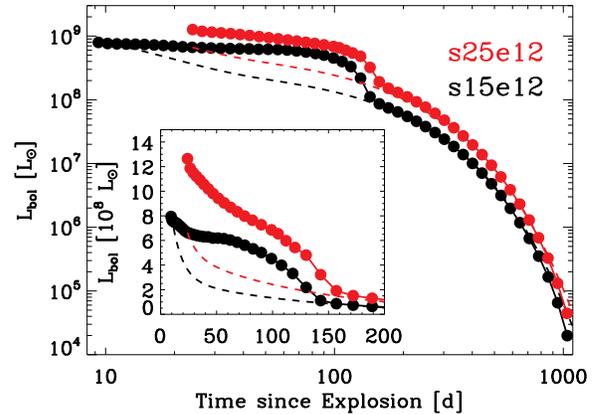,width=8.5cm}
\caption{{\it Main panel:} Log-log plot of the evolution of the bolometric luminosity for our simulations based on
model s15e12 (black; model contains 0.0866\,\msun\ of $^{56}$Ni initially)
and s25e12 (red; model contains 0.163\,\msun\ of $^{56}$Ni initially). Dots indicate the time of each computation.
We also over-plot as a dashed line the mean $\gamma$-ray energy released per time interval, computed independently
for each model using our Monte Carlo code (see Hillier \& Dessart in prep.), and assumed in this set of calculations
to be deposited locally as a heat source (we do not treat non-thermal ionisation/excitation).
Note that due to computation over different time intervals, the solid/dashed curves diverge somewhat at very late times,
when time increments are large and on the order of 100 days.
{\it Insert:} Same as main panel, but now zooming-in on the plateau phase and the onset of the nebular phase.
We now use a linear scale for both axes.
\label{fig_lbol}
}
\end{figure}

   In Fig.~\ref{fig_lbol}, we show the bolometric luminosity for each time sequence
   from 10-20 days until about 1000 days after the explosion.
   As expected for the explosion of RSG progenitors, the initial photospheric phase is characterised
   by a quasi-plateau luminosity, with a length on the order of 100-150\,d. The longer plateau duration in
   the simulation based on model s25e12 is caused in part by the larger production of $^{56}$Ni during the explosion
   (i.e., 0.163\,\msun) compared to that in model s15e12 (0.0866\,\msun). The bolometric luminosity decreases
   at the end of the plateau phase and meets the (dashed) curve representing the energy-production rate through radioactive decay
   at $\sim$140\,d (160\,d) for the simulation based on model s15e12 (s25e12).\footnote{The decay-energy production rate
   shown in Fig.~\ref{fig_lbol} corresponds to the average energy produced per time interval, and thus differs from the instantaneous
   value at a given post-explosion time.}

\begin{figure}
\epsfig{file=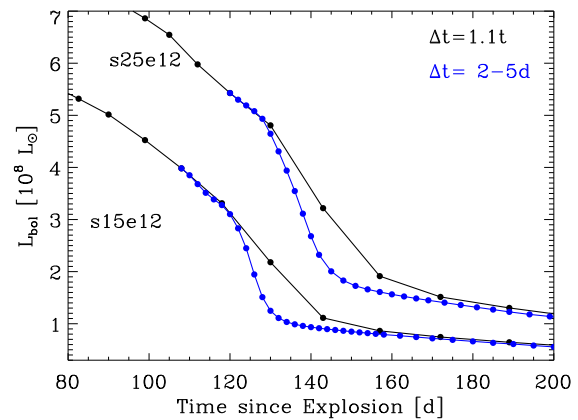,width=8.5cm}
\caption{Same as Fig.~\ref{fig_lbol} for models s15e12 and s25e12, but now comparing results from {\sc cmfgen} when
we adopt a time increment of 10\% of the current time (black; this is our standard choice
already shown in Fig.~\ref{fig_lbol}, which yields large time steps at late times), or a constant
time increment of 2 to 5 days (blue curves). For each curve, the dots give the times at which
the {\sc cmfgen} calculations were carried out.
Using a fixed and finer time increment yields a different result  at the end of the plateau phase, which is better resolved and
found to last about 10\,d in both models.
\label{fig_lbol_fine}
}
\end{figure}

   In our models the fading from the plateau is not abrupt,  but instead takes about a month, and corresponds to a luminosity
   decrease by a factor of a few.
   We find that this prolonged fall-off is in part an artefact of the large time step of $\sim$12\,d
   (i.e., 10\% of 120 days) in our simulations
   at such times.  Recomputing this phase with {\sc cmfgen} using a finer and constant time increment of 2\,d
   yields a faster fading of $\sim$10\,d but still gives the same fading magnitude (Fig.~\ref{fig_lbol_fine}).
   Our neglect of light-travel-time effects is secondary here, since it would only occasion a smearing of the bolometric luminosity
   over a timescale of $\lesssim$1\,d.
   Comparison between the evolution of the bolometric luminosity and the composition at the photosphere (Figs.~\ref{fig_comp_sn}
   and \ref{fig_lbol}) indicates that the plateau or high-brightness phase of the SN corresponds to the epoch during
   which the photosphere resides within the hydrogen-rich region of the SN ejecta. As soon as
   the photosphere reaches the outer edge of the helium core, the plateau brightness ebbs.
   The subsequent light-curve evolution is qualitatively similar in both models, suggesting little
   sensitivity to the differing helium-core masses (4.27\,\msun\ in model s15e12 and 8.19\,\msun\ in model s25e12).

   Bolometric light curves for four SN II-P (SN 1999em, SN 2003gd, SN 2004et, SN 2005cs) are presented by \cite{smartt_09}.
   All four show a drop of around 1 magnitude close to 120d from maximum, and for one of them (SN 2005cs) the bolometric
   luminosity  changes by almost a factor of 2 on a time scale of $\sim$10 days, while for SN 1999em the change is more
   gradual. As the end of the plateau phase is determined by the transition of the photosphere from the hydrogen rich
   core to the helium core, the steepness of the change will depend on any mixing that has occurred between these layers,
   as well any departure from spherical symmetry.

   The bolometric luminosity for the simulation based on model s25e12 is greater at all times than for the simulation
   based on model s15e12.  The larger plateau brightness stems
   from the larger progenitor radius in model s25e12, while the larger nebular flux is caused by the larger amount
   of $^{56}$Ni produced by the explosion in model s25e12. The magnitude of the fading at the end of the plateau is
   thus a function of both the progenitor size and the amount of $^{56}$Ni produced.
   As the progenitor radius and mass of nucleosynthesised $^{56}$Ni are unlikely to be fundamentally
   related (because of the influence of abundance variations, rotation, binary evolution, and the complexities
   of the explosion mechanism), there should be no clear correlation between plateau and nebular brightness,
   and thus the observed post-plateau fading should show a large scatter.
   The approximate treatment of radioactive decay in this
   study cannot alter this result since no $\gamma$-rays are predicted to escape up to $\gtrsim$500\,d.
   At later times, our neglect of $\gamma$-ray photon escape will lead
   to an overestimate of the bolometric luminosity shown here, by up to a factor of two at 1000\,d.

\begin{figure}
\epsfig{file=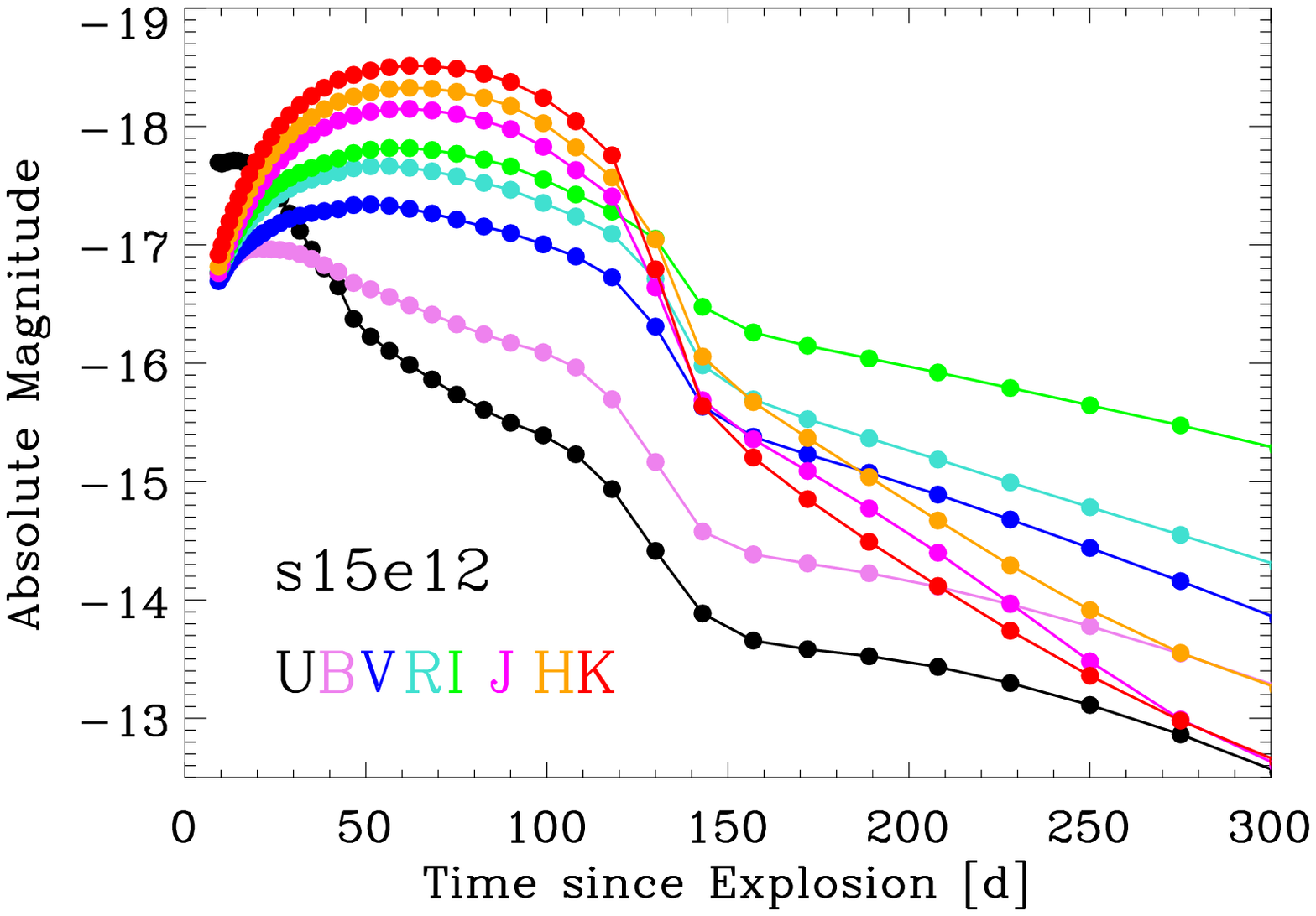,width=8.5cm}
\epsfig{file=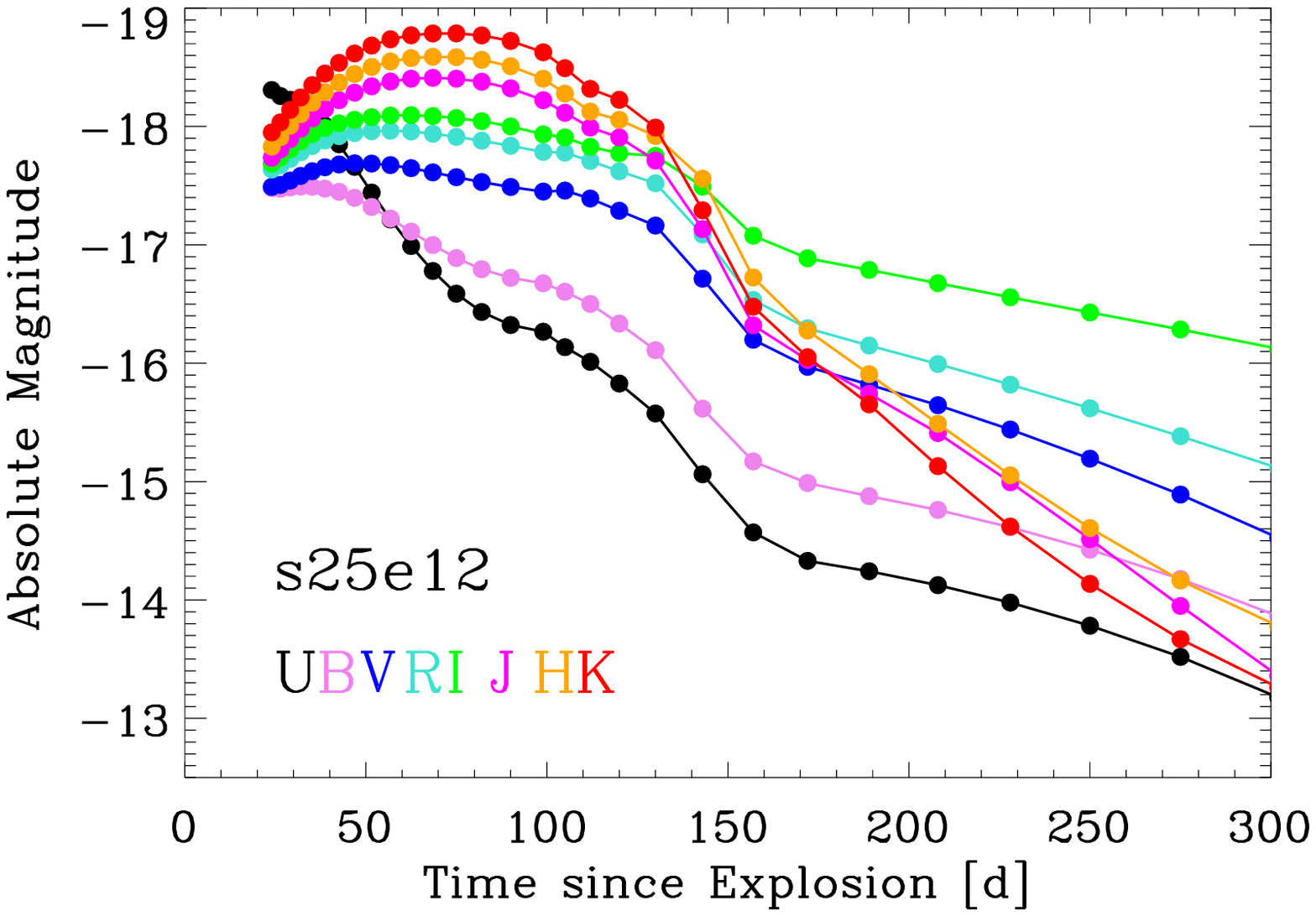,width=8.5cm}
\caption{{\it Top:} Synthetic optical and near-IR light curves for the simulation based on model s15e12 and
obtained by integrating our non-LTE time-dependent spectra at each epoch over the filter-transmission functions
of \citet{landolt_92} and \citet{cohen_etal_03}.
{\it Bottom:} Same as top, but now for the simulation based on model s25e12.
Dots indicate the time of each computation.
\label{fig_lc}
}
\end{figure}

   The bolometric luminosity is not directly observable because not all spectral bands are accessible. Moreover,
filter bandpasses are separated by gaps, sometimes sizeable, that miss regions of non-zero flux (and potentially strong lines).
Regions of intense atmospheric
absorption prevent the observation of certain spectral regions altogether, as in some parts of the near-infrared (near-IR).
Hence, in Fig.~\ref{fig_lc}, we show multi-band synthetic photometry for our two simulations based on models s15e12 and s25e12.
These are obtained by integrating our non-LTE time-dependent synthetic spectra at each epoch with the filter
transmission functions of \citet{landolt_92} for the optical bands, and with those of \citet{cohen_etal_03} for the near-IR bands.

   While the bolometric luminosity shows a steady and slow decline at most epochs, the multi-band light-curves show
   a contrasting behaviour affected by ejecta cooling.
   In particular, the photospheric cooling after shock breakout leads to a continuous decrease of the flux at short wavelengths,
   as the peak of the distribution sweeps through the far-UV, to the UV, and approaches the optical bands \citep{gezari_etal_08}.
   This translates here into a ``bell" shape for the $B-$band light curve, but a plateau followed by a steady fading in the $U-$band.
   In contrast, the $VRIJHK$ magnitudes all rise at all times prior to about 50 days after explosion, before turning over and declining.
   Both simulations based on models s15e12 and s25e12 show a similar behaviour, the latter being visually brighter at all photospheric
   and nebular epochs. As discussed previously, this contrast stems from the larger progenitor radius, and the larger amount of $^{56}$Ni produced in model s25e12.

   Besides differences in progenitor properties, we also find that magnitudes/colors are affected by the completeness of the
   model atom employed for the calculations. Model s15e12iso is identical in ejecta structure
   to that of model s15e12 (same radius/velocity/density/temperature distribution), but differs in composition due to the treatment
   of under-abundant species, ignored in model s15e12. We find stronger line blanketing in simulations based on model s15e12iso
   at the recombination epoch, in particular associated with the treatment
   of Sc\,{\sc ii}, Cr\,{\sc ii}, Fe\,{\sc i}, and Ni\,{\sc ii}, which leads to a reduction of the already faint $UB$-band
   magnitudes by $\lesssim$0.5\,mag.
   As already realised for the modelling of SN 1987A \citep{DH10}, including all species, and in particular all under-abundant
   intermediate-mass-elements/metals up to Fe/Co/Ni, and as many ionisation stages as possible is critical
   for obtaining converged results. In SNe II modelling, however,
   it is not critical to include a huge number of levels, as we find that increasing the number of levels
   by a factor of a few from the choice given in Table~\ref{tab_atom} does not alter our results significantly.

\section{Synthetic spectra}
\label{sect_spec}

\subsection{Global evolution}

\begin{figure*}
\epsfig{file=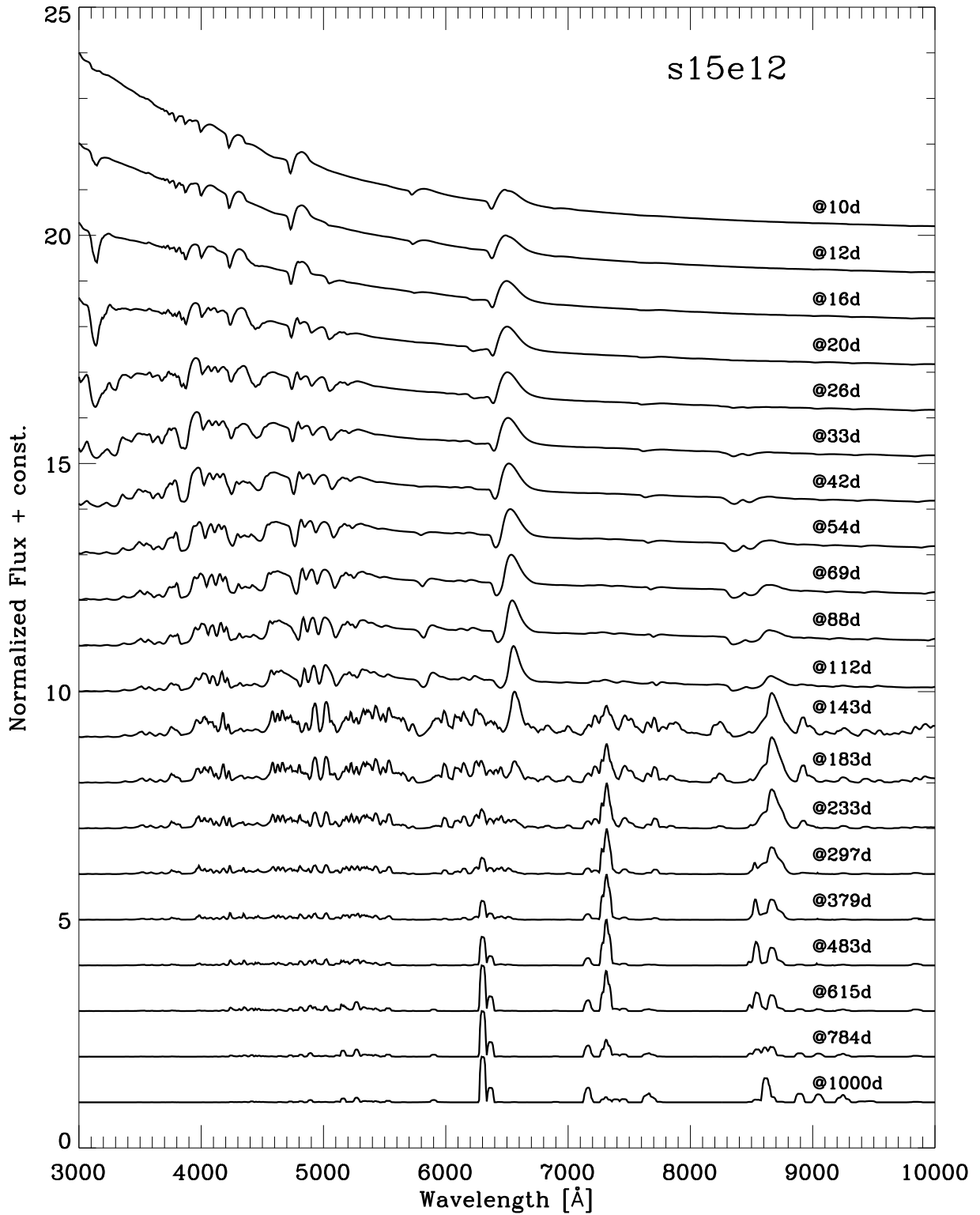}
\vspace{-2.75cm}
\caption{Montage of synthetic optical spectra covering epochs between
10 and 1000\,d after explosion for our simulation based on model s15e12.
The selected times, indicated by the label on the right, are such that the time
interval between two consecutive spectra is constant in the log.
For better visibility, each spectrum is first normalised to the peak flux in the region 6000-8000\AA, which
introduces a variable scaling between epochs, and then shifted vertically for better visibility.
During the photospheric phase, lines features are associated with H{\sc i} Balmer lines, resonance lines like the triplet of
Ca{\sc ii} at $\sim$8500\,\AA\ and forests of metals lines (e.g., Fe{\sc ii} and Ti{\sc ii})
causing severe line blanketing at wavelengths smaller than $\sim$5000\,\AA.
During the nebular phase, forbidden lines like [O{\sc i}]\,6303--6363\,\AA\  and [Ca{\sc ii}]\,7291--7323\,\AA\
progressively strengthen relative to recombination lines and the fading continuum.
\label{fig_montage_spec}
}
\end{figure*}

   We now turn to a presentation of the synthetic spectra for our two simulations. Because the general spectroscopic
   evolution from the photospheric phase to the nebular phase is similar for both sequences, we show in Fig.~\ref{fig_montage_spec}
   a montage for one simulation only, that based on model s15e12.
   Our continuous modelling is permitted by the treatment of the {\it entire} ejecta at all times.
   The inner-grid boundary is at the inner edge of the innermost ejecta shell; any material between that shell
   and the neutron star fell back and is thus not part of the ejecta by definition.
   In the full solution of the radiative transfer with {\sc cmfgen}, we adopt two inner-boundary types.
   The inner boundary is treated as diffusive (with an imposed inner-boundary luminosity
   of \lsun/10$^6$) as long as the base optical depth is greater than $\sim$10. Afterwards, we
   switch to a zero-flux nebular-like inner-boundary.
   For the computation of spectra, we assume a hollow core.
   With such a full-ejecta approach, we can continuously model the ejecta as they evolve
   from optically thick to optically thin.

\begin{figure*}
\epsfig{file=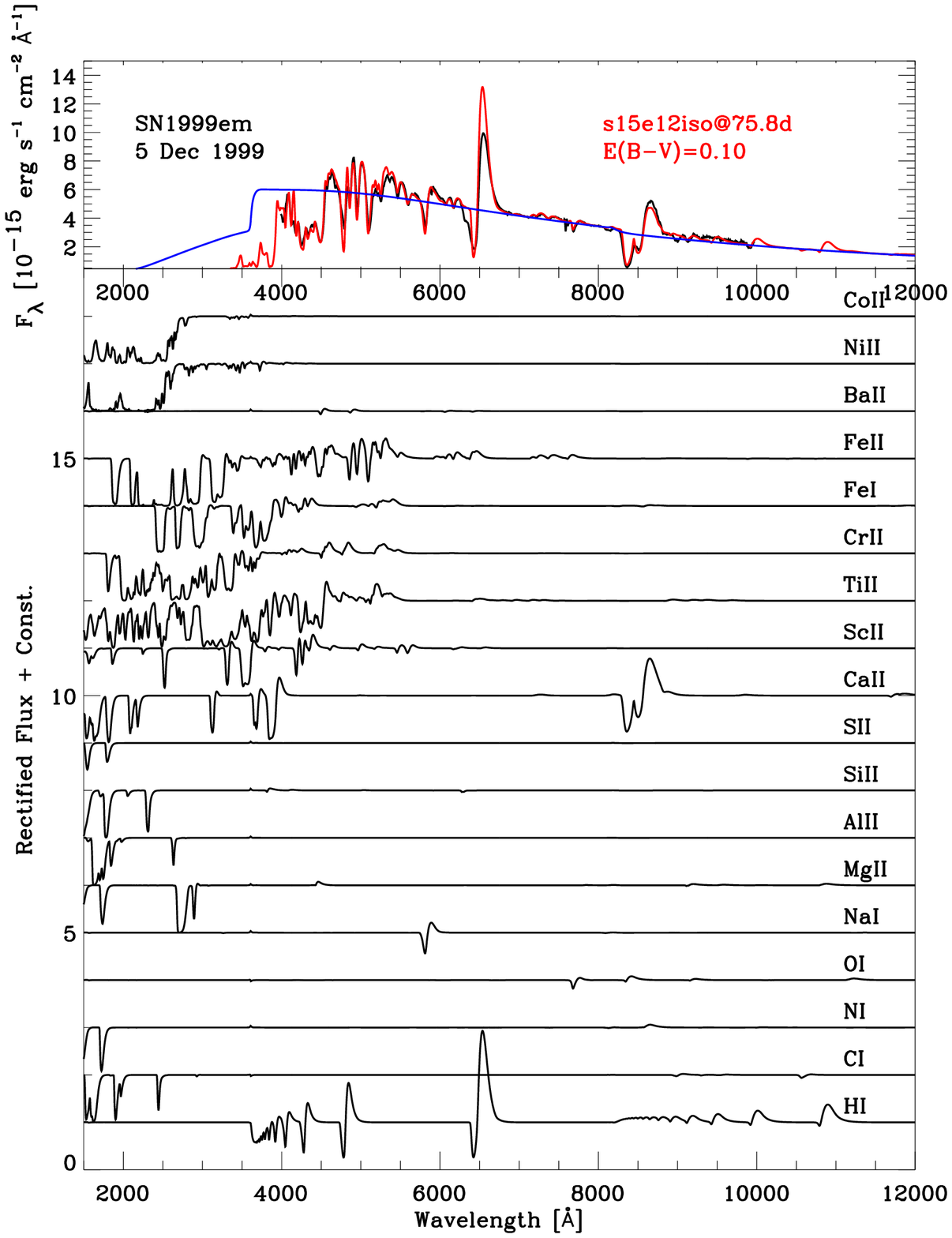}
\vspace{-2.5cm}
\caption{Illustration showing bound-bound contributions of selected species (bottom; fluxes are normalised
to the continuum and shifted vertically) to the full synthetic spectrum (red curve, top panel) obtained
in the simulation based on model s15e12iso at a post-explosion time of 75.8\,d.
The full and continuum-only (blue curve) synthetic spectra have been reddened ($E(B-V)=0.1$) and distance
scaled ($D=11.5$\,Mpc; an additionnal scaling by a factor of 2.3 is added to compensate for the higher brightness
of the model), and in the top panel are compared for illustration to the Keck observations
of SN1999em on 5 Dec. 1999 (black; \citealt{leonard_etal_02a})
\label{fig_ladder_plot}
}
\end{figure*}

   Figure~\ref{fig_montage_spec} shows the evolution of a spectral-energy
   distribution that initially resembles that of a thermal emitter, with a decreasing continuum flux at 100-150\,d after explosion
   when the ejecta become optically thin. As time progresses in the nebular phase, the P-Cygni profile morphology of lines
   transitions to that of pure emission, with a box-like structure typical of optically-thin line formation from a hollow region
   expanding at near-constant speed \citep{castor_70}. Simultaneously, the continuum flux vanishes.

   During the SN evolution, line features evolve considerably, reflecting the change in composition, ionisation/temperature
   and density. At early times, the photospheric layers are hot and ionised and H\,{\sc i} Balmer lines and He\,{\sc i}\,5878\AA\
   are present. As recombination sets in, line blanketing caused by once-ionised intermediate-mass elements and metal species
   appears at all wavelengths smaller than $\sim$5000\,\AA\ (Fig.~\ref{fig_ladder_plot}). This entire photospheric-phase evolution
   takes place at essentially constant composition, so that these spectral changes are governed by ionisation changes.
   As the SN progresses into the nebular phase, lines forming through
   recombination (following photoionisation) weaken because of the fading radiation field, and in particular the quasi absence of
   ionising photons, either in the Lyman or in the Balmer continuum (in reality, this is exaggerated by our neglect of non-thermal
   ionisation by high-energy electrons).
   In contrast, forbidden lines, which form at low density by radiative de-excitation of collisionally-excited levels,
   strengthen on top of an ever-fading continuum. The two most important forbidden lines that appear in essentially all
   nebular-phase SNe II spectra are the doublets [O\,{\sc i}]\,6303--6363\,\AA\  and [Ca\,{\sc ii}]\,7291--7323\,\AA.

 \begin{figure*}
\epsfig{file=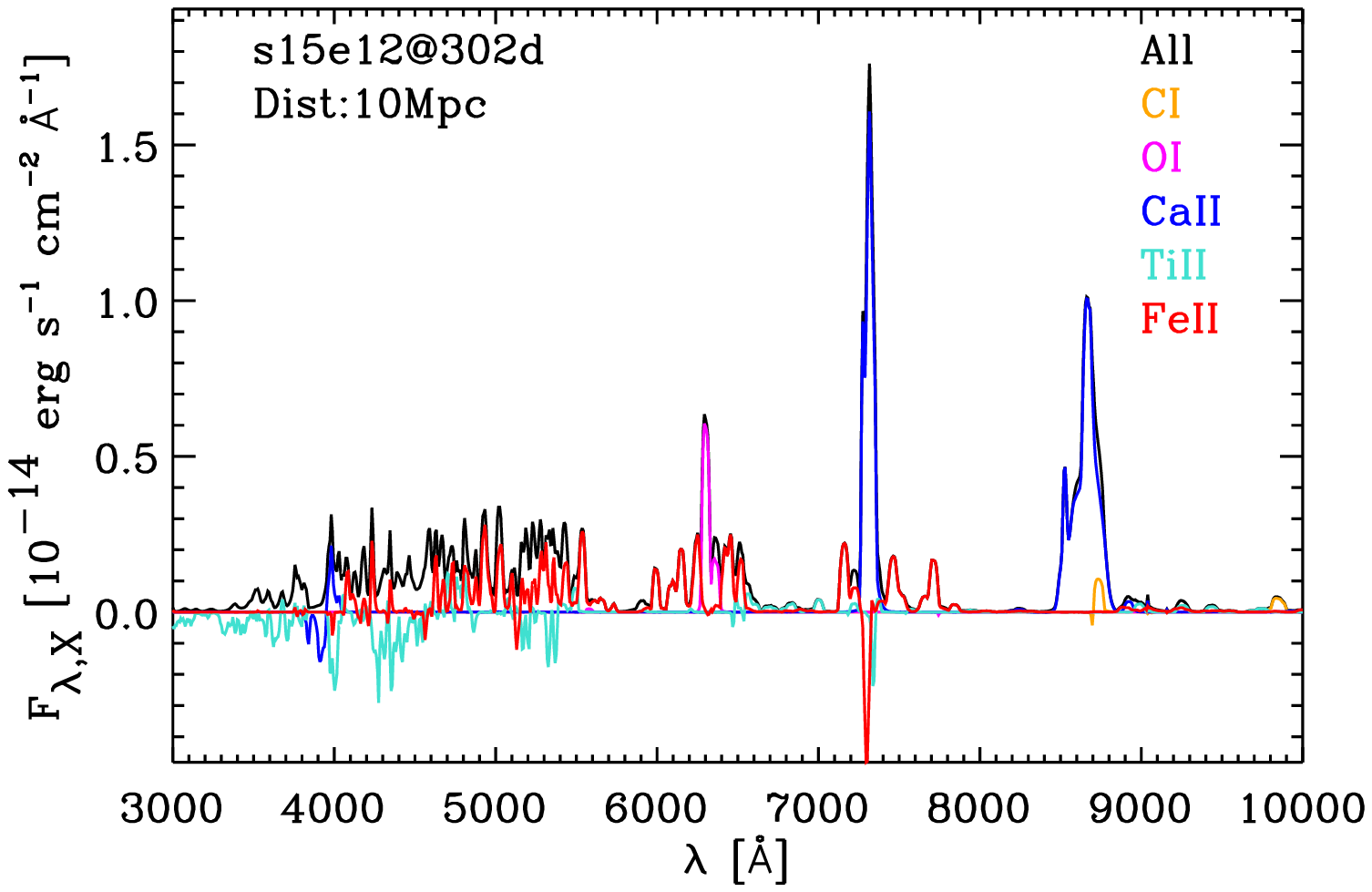,width=8.5cm}
\epsfig{file=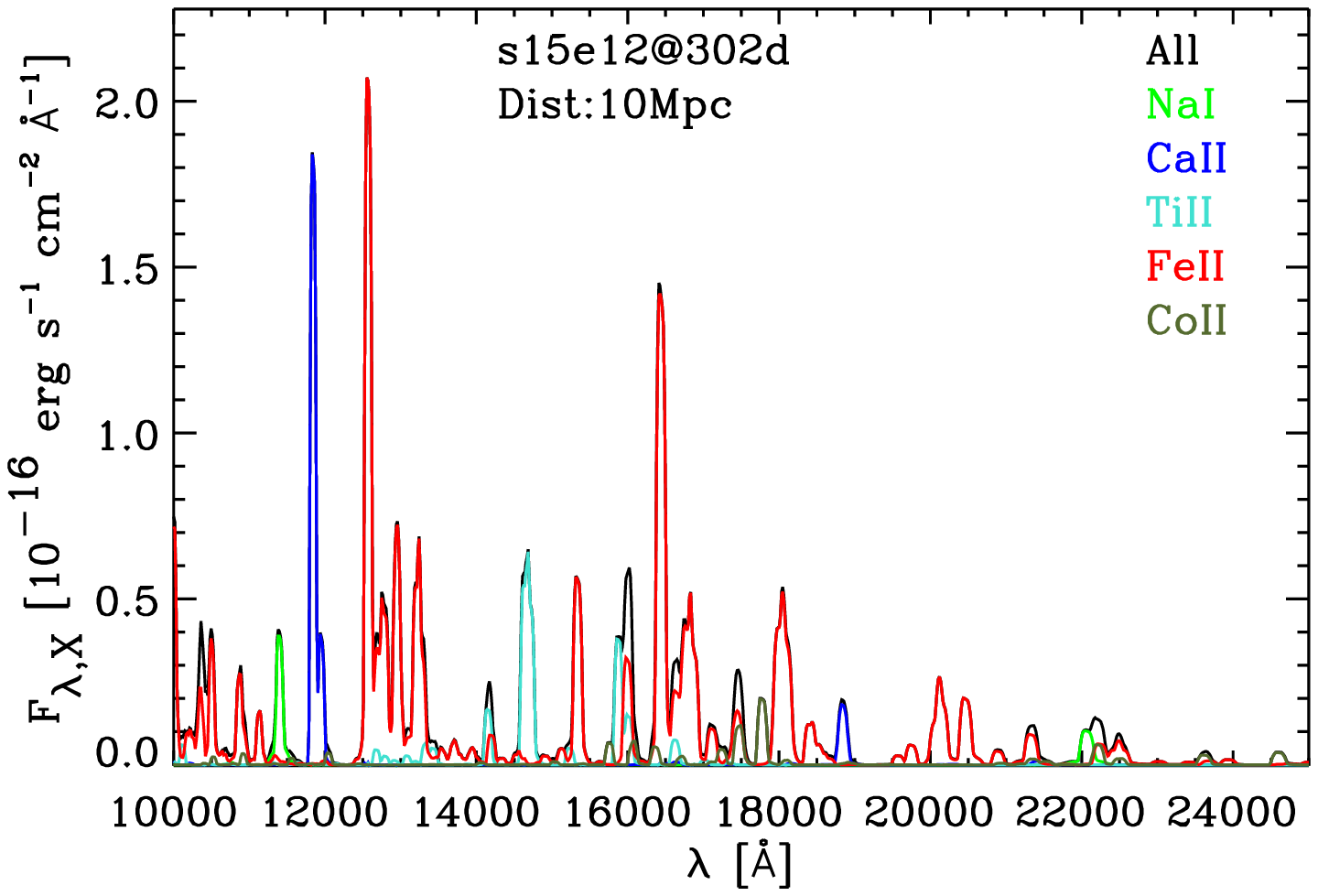,width=8.5cm}
\epsfig{file=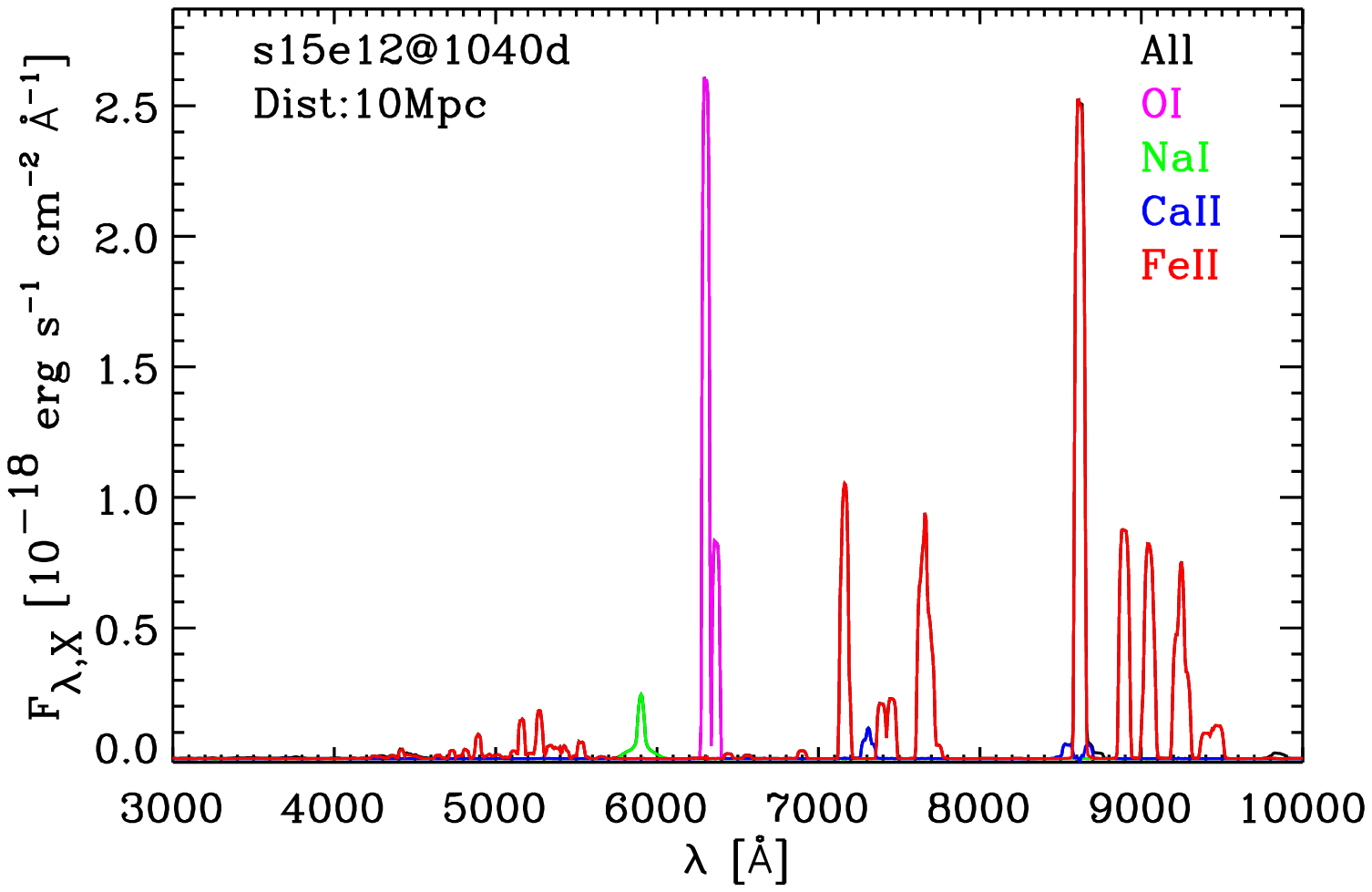,width=8.5cm}
\epsfig{file=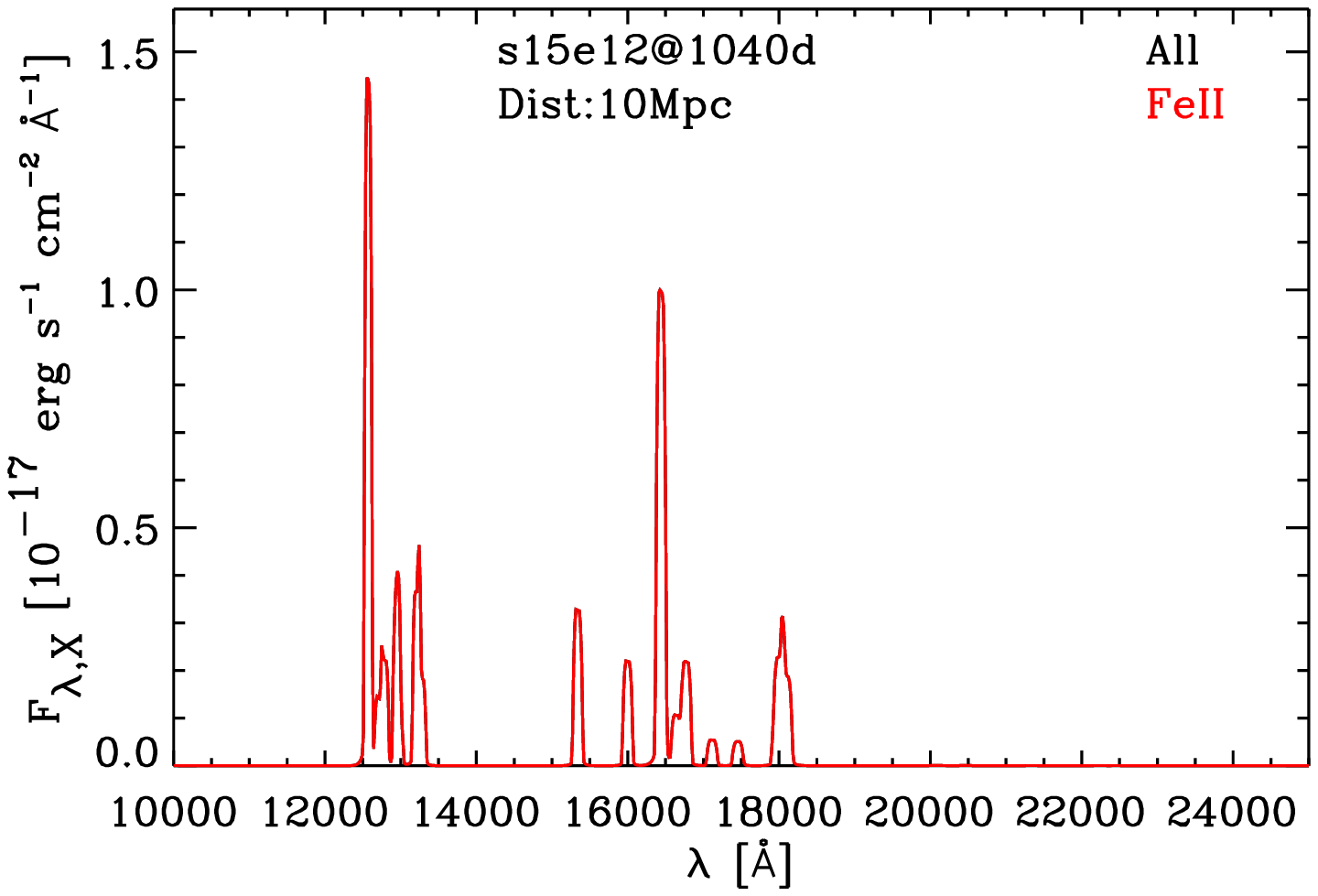,width=8.5cm}
\caption{Illustration of optical (left column) and near-IR (right column) synthetic spectra
obtained for the simulation based on model s15e12 at post-explosion times of 302\,d (top row) and 1040\,d (bottom row).
The full synthetic spectrum is shown as a black curve, while overplotted coloured lines refer to the flux obtained
when including bound-bound transitions exclusively of the species labelled on the right of each panel.
We only display the results for those species that matter to the emergent flux, hence the selection varies
between panels (i.e., for different times and spectral ranges). At late times, only forbidden lines, forming
in the inner SN-ejecta regions corresponding to the former progenitor helium core, contribute to the SN light.
The continuum flux is a few orders of magnitude smaller than the full synthetic flux due to lines.
\label{fig_spec_neb}
}
\end{figure*}

\subsection{Nebular phase}

 Due to their faintness, and unimportance for cosmological studies, there have been relatively few nebular studies of SN.
The most extensive nebular studies are for SN1987A.   For example, \cite{XRM91_compt} studied the Comptonization of
gamma rays by cold electrons, \cite{XM91_87A_energetic} studied how non-thermal electrons are produced and downgraded,
\cite{XMO92_H_SN1987A} studied the formation of the H\,{\sc i} lines, \cite{LM95_SN87A_HeI} studied the formation of
the He\,{\sc i} emission lines, while \cite{LM95_SN87A_HeI,LM93_CaII} studied the formation of the Ca\,{\sc ii}
emission lines. Additional extensive studies of the nebular phase in SN1987A are by \cite{KF98a, KF98b} who discussed
in detail the temperature and ionisation structure, and the formation of the most important emission lines.
A major result of the above studies is the need for mixing of hydrogen down to 700\,\kms, with oxygen and helium present
at even lower velocities. \cite{KF98b} analyzed the O\,{\sc i} lines in SN 1987A and found that the bulk of the oxygen
was located between 400 and 2000\,\kms, a result in broad agreement with the earlier studies of \cite{LM92_SN1987A_OI}.

During the nebular phase the continuum is optically thin and line formation is simplified. Line profiles during the
nebular phase provide information about the distribution of material as a function of velocity, and its geometrical
distribution. For example, an isolated emission line arising from a spherical hollow shell will exhibit a flat top,
with its minimum width determined by the velocity at the inner edge of the shell \citep[e.g.,][]{McC_SN1987A_rev}.
As the Sobolev escape probability is independent of angle in a Hubble flow \citep[see expressions of][]{castor_70},
this result holds independent of the line's optical depth (neglecting continuum absorption).

   During the nebular phase, one probes the entire optically-thin ejecta, although emission
   is biased towards the denser/slower/hotter inner regions where unstable isotopes
   are naturally more abundant. This inner region contains few tens of percent of
   the total ejecta mass, and $\lesssim$10\,\% of the total ejecta kinetic energy in our models (Fig.~\ref{fig_comp_sn}).
   The faintness of all SNe at such epochs makes them difficult to observe at late times.
   However, because the nebular phase reveals the core of SNe II-P progenitors, it is
   a key source of information on the explosion, as well as the core properties of the progenitor
   star. Despite this importance, nebular-phase spectra of SNe II-P are scarce and thus not much has been
   extracted yet from this important phase \citep{DLW10b}. Dedicated campaigns to secure nebular
   spectra of SNe II-P are needed to improve upon this matter.

   We show in Fig.~\ref{fig_spec_neb} the optical and near-IR synthetic spectra obtained for the simulation
   based on the model s15e12 at 302 and 1040\,d after explosion. While the black curve gives the full spectrum,
   coloured curves give the bound-bound contributions from individual species. At such late times, not only
   $^{56}$Ni, but also $^{56}$Co, have decayed and the metal mixture is dominated by  $^{56}$Fe.
   At 302\,d, the synthetic spectrum is dominated by a forest of Ti\,{\sc ii} and Fe\,{\sc ii} lines in the region
   blue-ward of 5500\,\AA. Further to the red, fewer Fe\,{\sc ii} lines are present (though in the simulation based on model s15e12iso,
   lines of Fe\,{\sc i} are predicted), and three main features
   dominate the spectral landscape, the forbidden doublet lines of [O\,{\sc i}]\,6303--6363\,\AA\
   and [Ca\,{\sc ii}]\,7291--7323\,\AA, and the Ca\,{\sc ii} triplet at 8498, 8542, 8662\,\AA.
   As time proceeds, the decreasing ejecta electron/ion density favours the forbidden and ``resonance''
   transitions and so the features associated with [O\,{\sc i}], [Ca\,{\sc ii}], and Ca\,{\sc ii} strengthen while the
   recombination transitions associated with Fe{\sc ii} weaken in the region 4000--5500\,\AA.
   At very late times, only [O\,{\sc i}] and [Fe\,{\sc ii}] persist.

  In our simulations we do not predict  H\,{\sc i} Balmer-line emission at nebular times. The Balmer lines fade abruptly
when the ejecta become optically thin, immediately at the end of the plateau phase, a result of the sudden disappearance
of Balmer-continuum photons and the de-excitation of H\,{\sc i}.
  The treatment of non-thermal ionisation/excitation would counteract the fading of the H{\sc i} Balmer recombination
lines  at the end of the plateau phase,
  since it is precisely at that time that high-energy electrons can impact level populations and the ionisation state
  of the gas in SNe II \citep{KF92,KF98a,KF98b}. We suspect this neglect is less critical for forbidden lines which form through
  transitions between levels that have a low excitation potential, and hence can be excited by thermal electrons.

The rapid disappearance of the H\,{\sc i} lines at the end of the plateau phase indicates that non-thermal processes must
be crucial for the hydrogen ionisation, even at $\sim150$ days. This is in apparent contradiction with the conclusions of
\cite{KF98b} who indicate that in SN1987A, H ionisation is determined by photoionisation from the $n$=2 level up to 500\,d.
A likely explanation is that non-thermal excitation from the ground state, together with the large optical of Ly$\alpha$,
helps to maintain (set) the $n$=2 population from which ionisation occurs. The viability of this option can be verified
by solving the two-level atom plus continuum problem \citep[see, e.g., equation 39 of][]{KF92}.

  In model s15e12,  and at a representative post-explosion time of 332\,d,
  we find that the doublet lines [O\,{\sc i}]\,6303--6363\,\AA\ and [Ca\,{\sc ii}]\,7291--7323\,\AA\
  form primarily in the inner denser ejecta. This is illustrated by the similar and relatively modest
  broadening, of Doppler origin, of the corresponding individual line components.
  And indeed, the full-width-at-half-maximum of $\sim$2500\,\kms\ corresponds quite closely to the velocity of $\sim$1500\,\kms\
  of the outer helium-core (see Fig.~\ref{fig_comp_sn}, but also Fig.~\ref{fig_comp_kepler_cmfgen}).

\begin{figure}
\epsfig{file=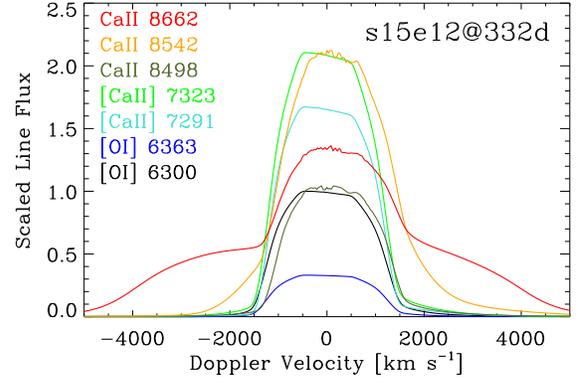,width=8.5cm}
\caption{Illustration of the synthetic line flux for the simulation based on model s15e12 at 332\,d after explosion,
and associated with individual transitions giving rise to the forbidden lines of [O\,{\sc i}] and [Ca\,{\sc ii}],
as well as the ``resonance'' transitions of Ca\,{\sc ii} appearing around 8500\,\AA.
All fluxes are normalised to the peak flux of the O{\sc i}\,6303 line.
A colour coding is used to differentiate individual line transitions.
Notice the similar and relatively narrow width of most of these lines, reflecting the expansion
speed of the helium-core in which they form, while resonance transitions of Ca{\sc ii} form not only
in that same inner region, but also in the outer hydrogen-rich ejecta. Individual profiles were computed
ignoring interactions with overlapping lines.
 \label{fig_line_formation}
}
\end{figure}

  In contrast, the strong permitted transitions of Ca\,{\sc ii} at 8498, 8542, 8662\,\AA\ show a hybrid formation, with contributions
  from both the inner and outer ejecta. This gives a very peculiar shape to the Ca\,{\sc ii}\,8662\,\AA\ line,
  with a relatively narrow emission part associated with the ejecta region corresponding to the progenitor helium-core,
  and a broad emission part associated with the hydrogen-rich region of the SN ejecta (Fig.~\ref{fig_line_formation}).
  The second, broad, component suggests that emission comes from the entire volume of the ejecta, all the way
  to the outermost mass shells (see Fig.~\ref{fig_comp_sn}).
  Excluding Ca\,{\sc ii}\,8662\,\AA\ line (discussed below), simulated emission line profiles  show that the bulk of their emission
  at nebular times comes from the same region of space, which corresponds to the progenitor helium core.

\begin{figure}
\epsfig{file=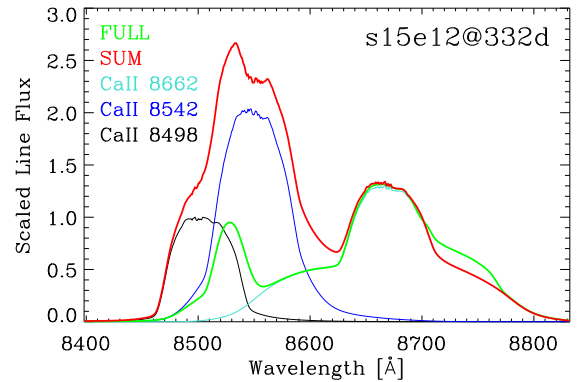,width=8.5cm}
\caption{A comparison of the theoretical spectrum in the neighbourhood of the Ca{\sc ii} triplet (``FULL''; green), and that obtained
by summing line profiles for individual members of the triplet (``SUM''; red).
Individual components are drawn as thin coloured lines.
The ``summed'' spectrum is very different from the true spectrum,
particularly for the 8498 and 8542\,\AA\ lines which contain too much flux in the ``summed'' spectrum.  As explained in the text,
photons in 8498 and 8542\,\AA\ are absorbed and re-emitted in the  8662\,\AA\ Iine.
\label{fig_Ca2_triplet}
}
\end{figure}

The Grotrian diagram for Ca\,{\sc ii} is rather simple and it  is not initially obvious why Ca\,{\sc ii} 8662\,\AA\ should
show such a different line profile from the other lines. A hint comes from comparing the Ca\,{\sc ii} triplet spectrum to
that obtained by summing the individual profiles. As shown in Fig.~\ref{fig_Ca2_triplet}, there is a lot more flux in the
two bluest lines than in the full spectrum. This suggests that the photons emitted in Ca\,{\sc ii} at 8498 and 8542 are
scattered in the  Ca\,{\sc ii} 8662\,\AA\ elsewhere in the envelope. This is reasonable since the triplet transitions
are optically thick, and the velocity shift between neighbouring lines is only $\sim 1500$\,\kms\ and $\sim 4200\,$\kms.
Examination of the Ca\,{\sc ii} triplet profile on day 198 for SN1987A also shows a complex profile, with the 8662\AA\
line being much broader than the 8498/8542\,\AA\ blend \citep{PHH90_87A, LM93_CaII}.

\begin{figure}
\epsfig{file=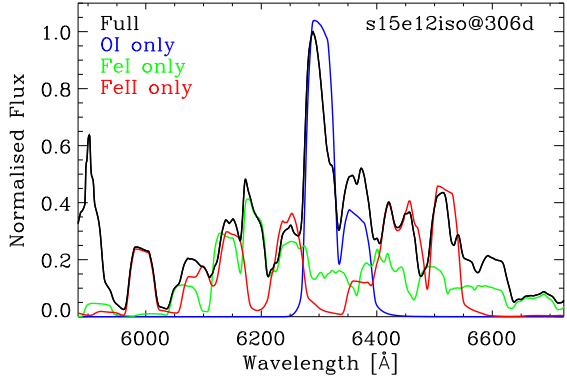,width=8.5cm}
\caption{Full synthetic spectrum (black) for model s15e12iso at 306\,d after explosion and showing
the [O{\sc i}]\,6303--6363\AA\ region. To illustrate the presence of line overlap, we overplot the synthetic
spectrum obtained by accounting for bound-bound transitions of  O{\sc i} (blue), Fe{\sc i} (green), or
Fe{\sc ii} (red) exclusively.
Line overlap and associated line-optical depth effects compromise any eye-ball estimate of the flux contribution
associated with O{\sc i}.
\label{fig_zoom_oi}
}
\end{figure}

Our theoretical spectra also reveal a  wealth of weak emission lines which blend with many of the stronger features.
This was previously highlighted by \cite{KF98b} for O\,{\sc i}. They noted that the ratio of O\.{\sc i} $\lambda$6300
to $\lambda$6363 was significantly less than the theoretical value of 3., which they attribute to a blend of Fe\,{\sc i}
6361\,\AA. In our calculations, we find
that this [O{\sc i}] region overlaps with numerous Fe{\sc i} and Fe{\sc ii} lines (Fig.~\ref{fig_zoom_oi}), and thus may complicate
the determination of the oxygen-mass ejected \citep{DLW10b}.

\subsection{Comparison between simulations based on the s15 and s25 models}

\begin{figure*}
\epsfig{file=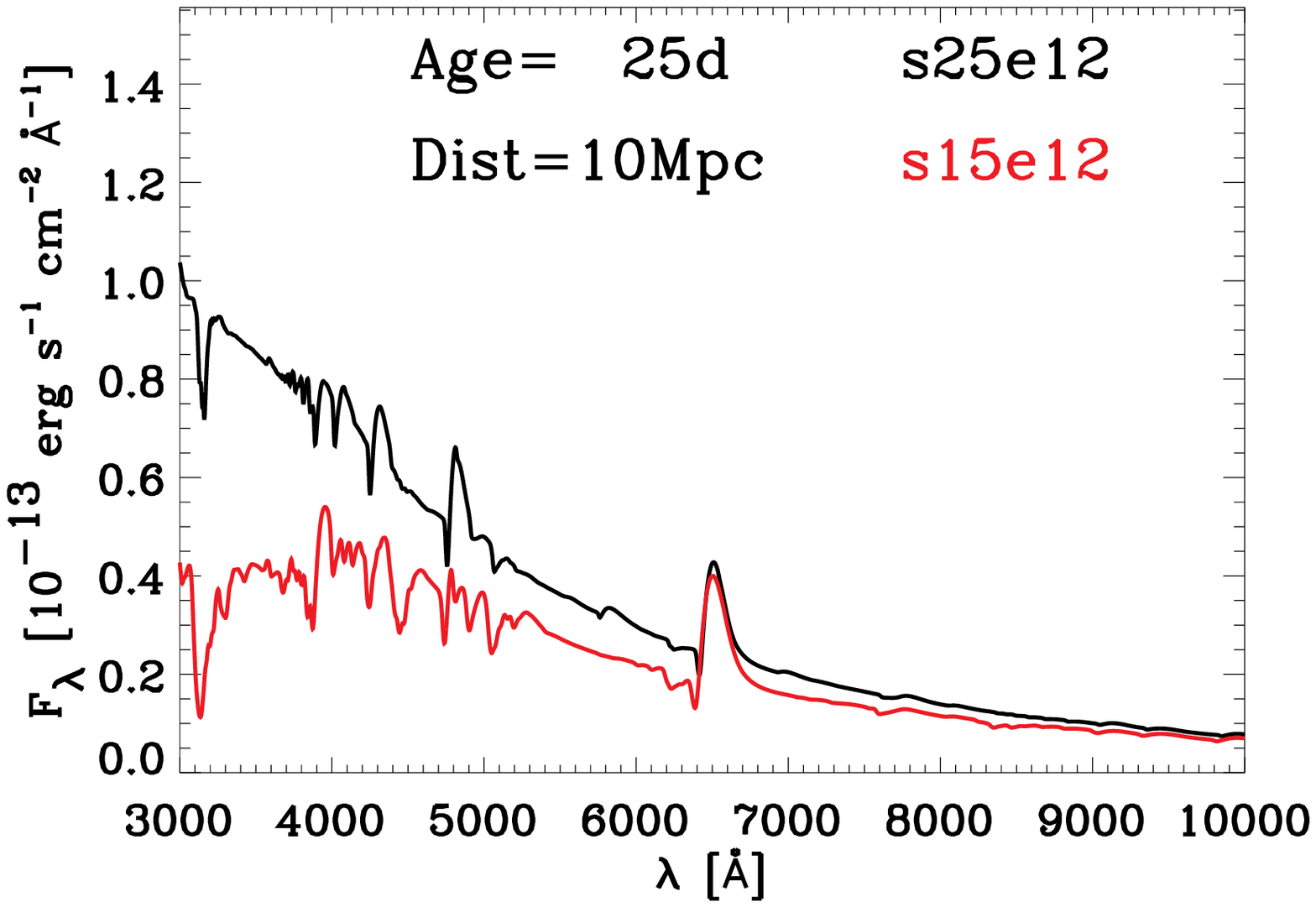,width=8.5cm}
\epsfig{file=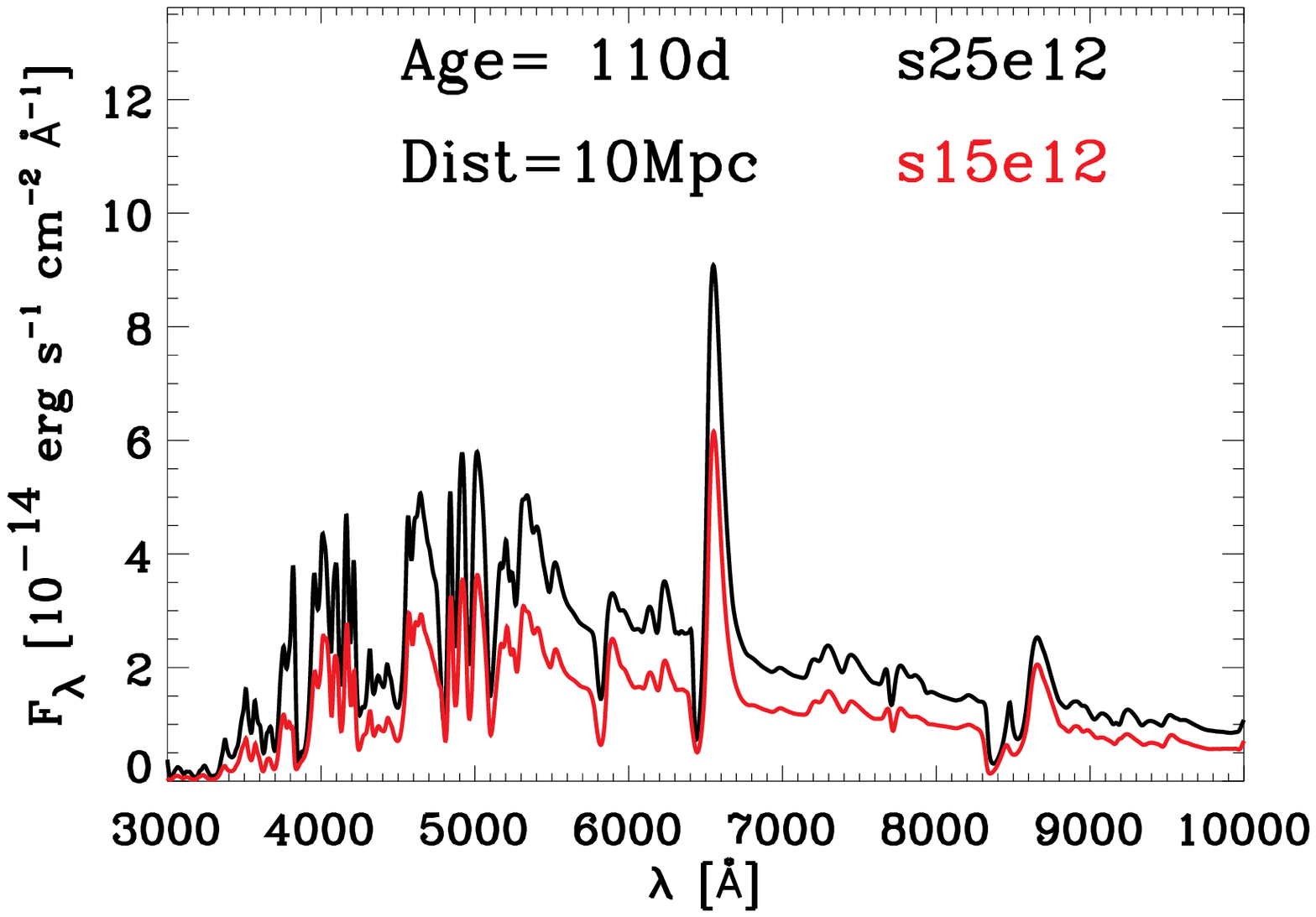,width=8.5cm}
\epsfig{file=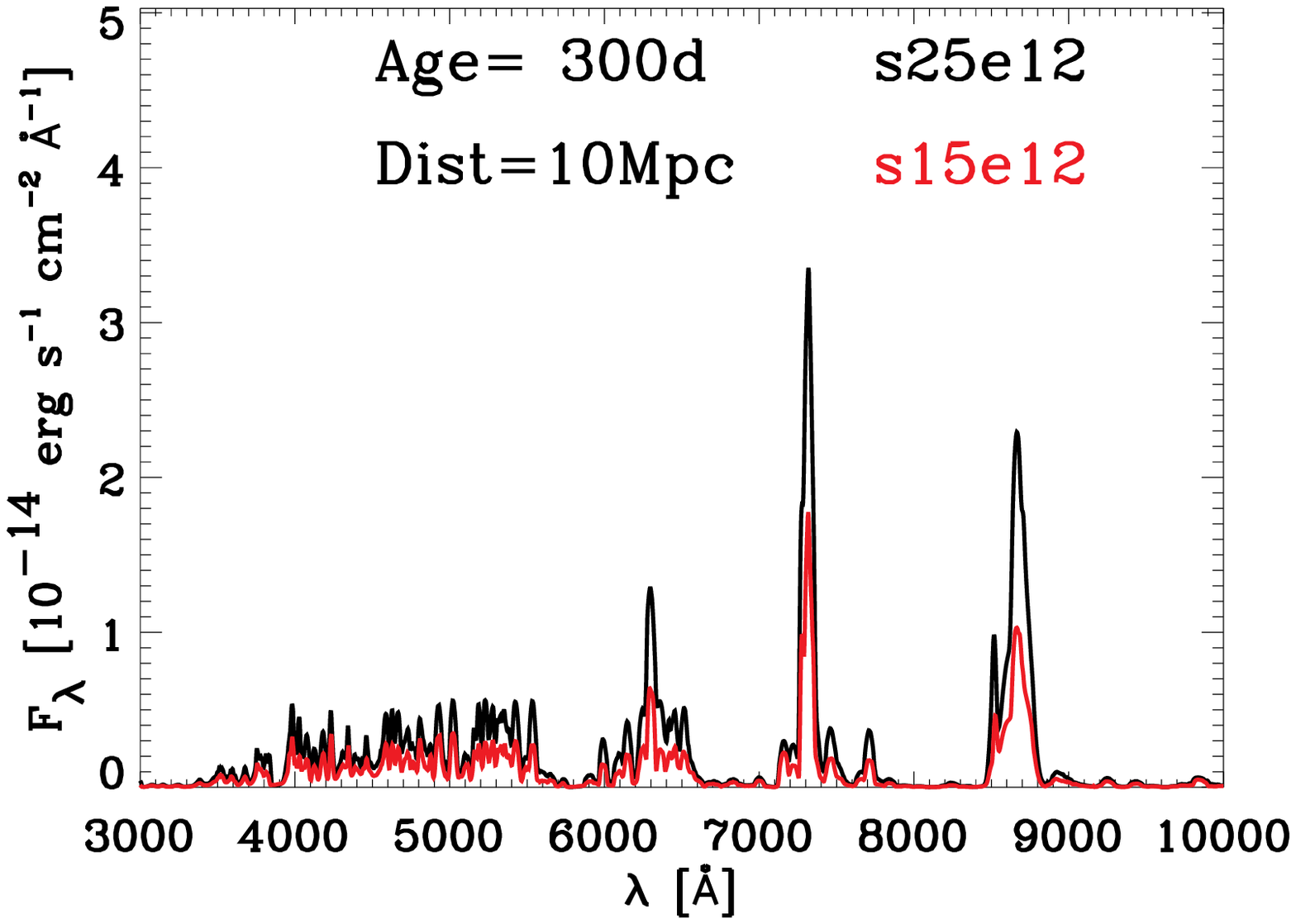,width=8.5cm}
\epsfig{file=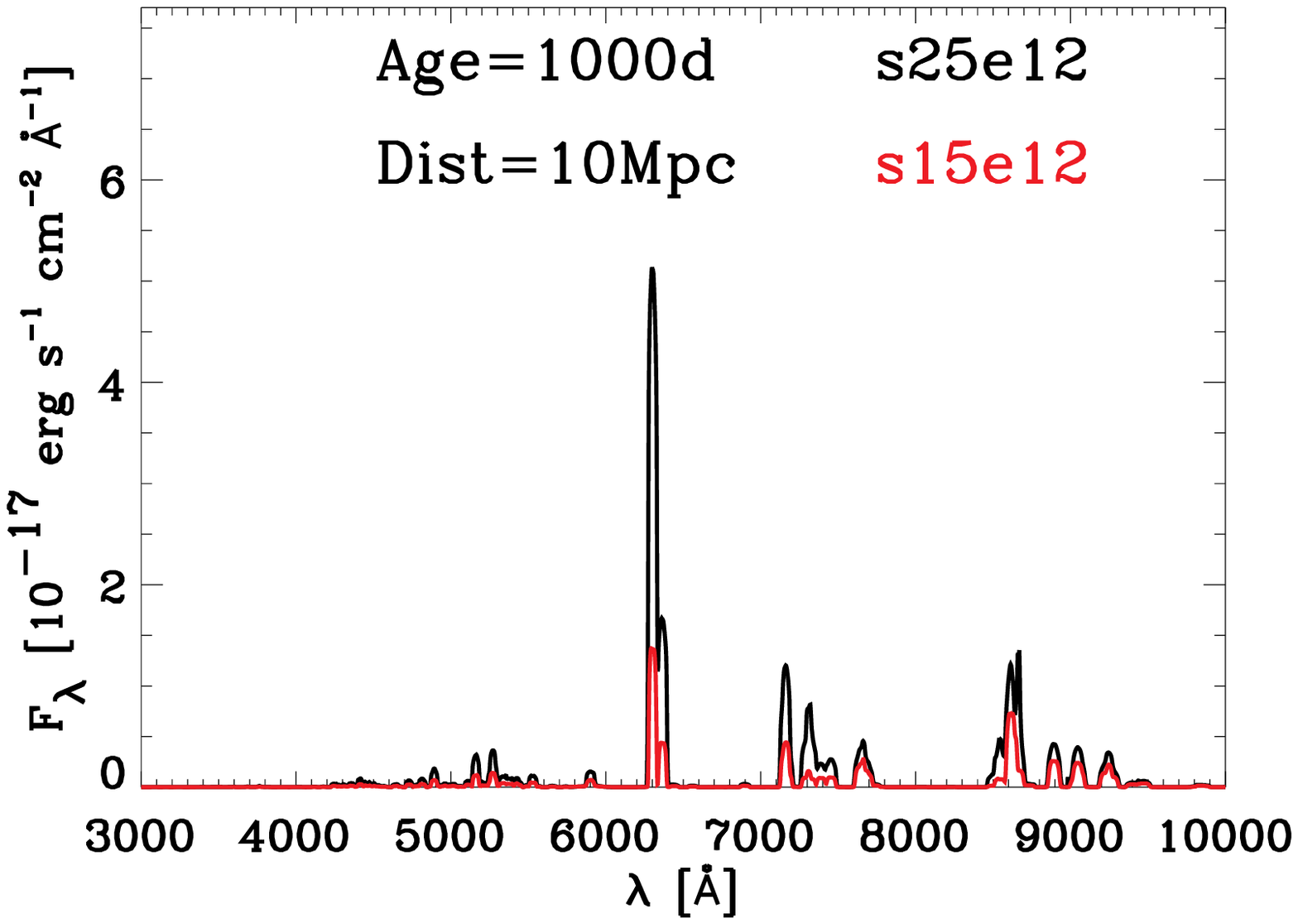,width=8.5cm}
\caption{Comparison between optical synthetic spectra for the simulations based on model s15e12 (red) and
s25e12 (black) and shown at post-explosion times of 25\,d (top left), 110\,d (top right),
300\,d (bottom left), and 1000\,d (bottom right).
Only a distance scaling is performed on these synthetic spectra, adopting a value of 10\,Mpc.
\label{fig_comp_spec1}
}
\end{figure*}

     We have so far focused on the spectroscopic properties obtained for one time sequence.
We now compare the spectroscopic differences between the simulations based on models
s15e12 and s25e12, which both have the same explosion energy, were evolved under
the same conditions (no rotation, solar-metallicity environment), but started from stars with main-sequence
masses of 15 and 25\,\msun. In \S~\ref{sect_setup}, we discussed the key differences between
the two pre-SN stars, which are the larger progenitor radius in model s25e12,
the more pronounced He/N enhancement in the envelope of model s25e12, and the larger helium-core
mass in model s25e12. Indeed, model s25e12 ejects 6.29\,\msun\
of helium-core embedded material, while s15e12 ejects only  2.44\,\msun.
Here, we discuss the radiative signatures resulting from such differences in
progenitor and explosion properties.
We show comparative spectra for the two simulations at the same post-explosion time
during the photospheric and nebular phases in Fig.~\ref{fig_comp_spec1}.

\begin{figure*}
\epsfig{file=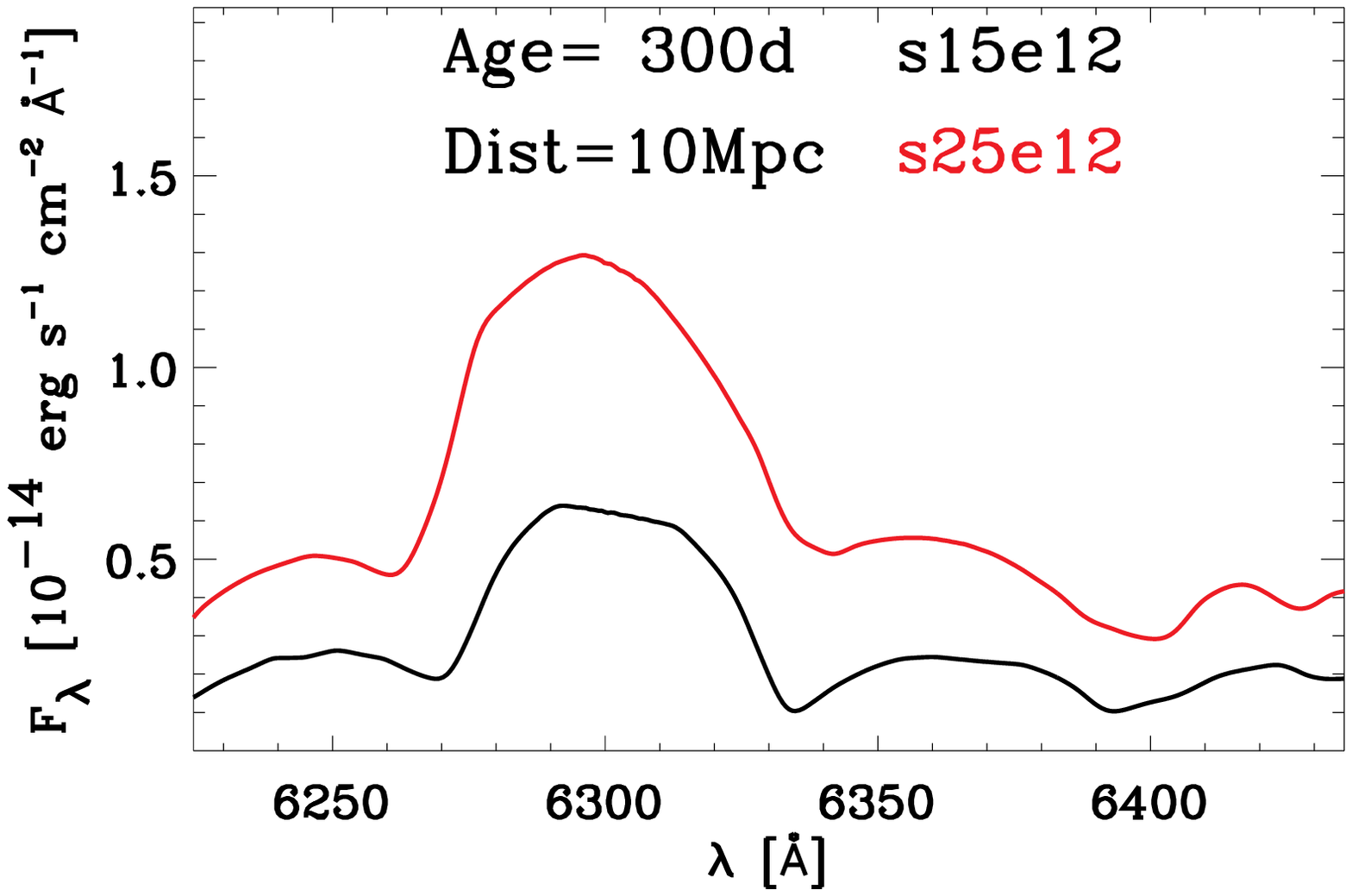,width=8.5cm}
\epsfig{file=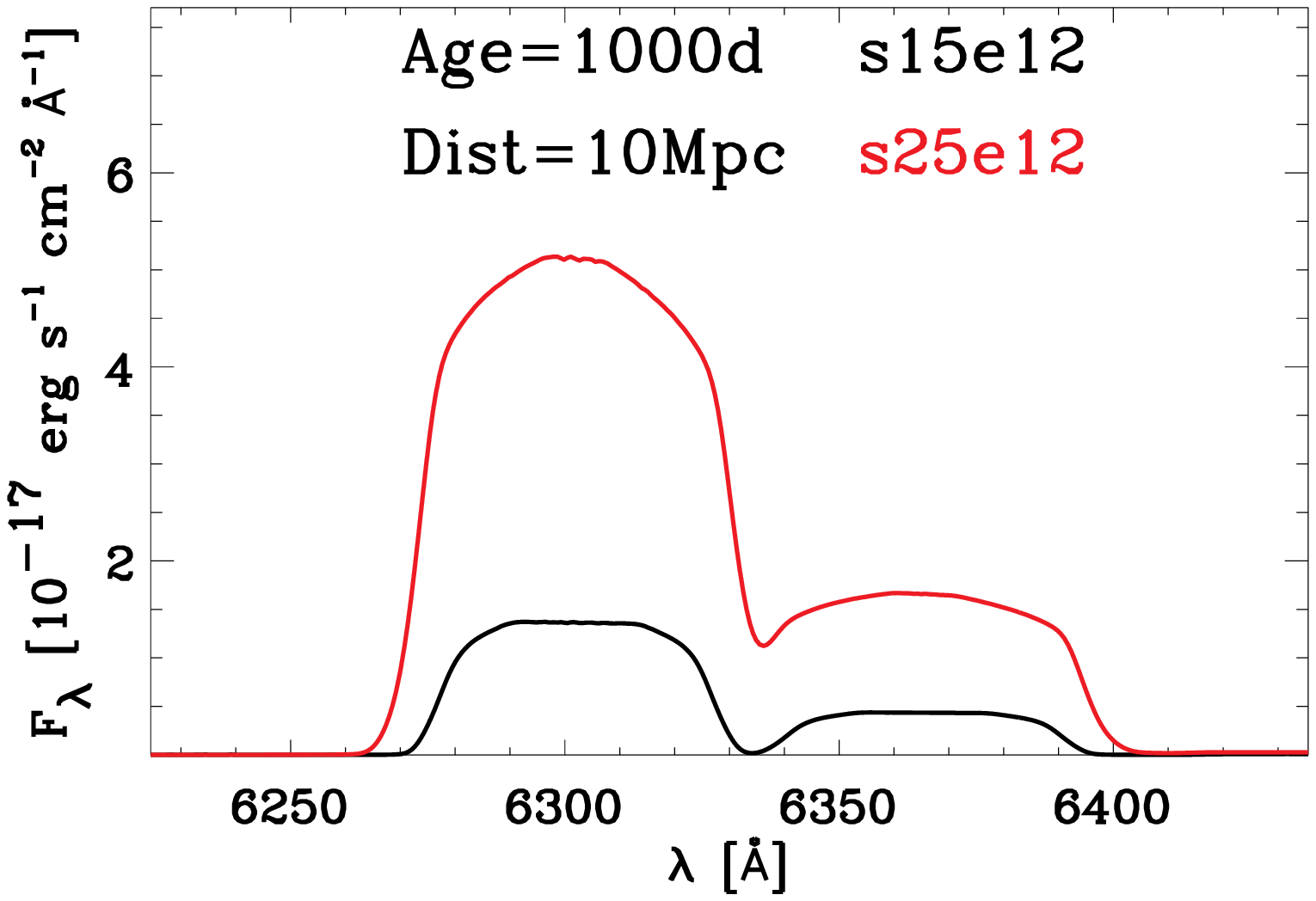,width=8.5cm}
\caption{Same as Fig~\ref{fig_comp_spec1}, but now showing the [O\,{\sc i}]\,6303-6363\AA\ doublet region
at 300\,d (left) and 1000\,d (right). \label{fig_comp_spec2}}
\end{figure*}

As discussed above, the larger progenitor radius for a given
explosion energy gives bluer colours and a brighter bolometric display (larger flux)
during the plateau phase, while the large $^{56}$Ni yields in model s25e12
give a higher nebular luminosity (at a given post-explosion time).
During the recombination epoch, when the simulations show similar colors,
contemporaneous spectra obtained for the two simulations show a remarkable similarity.
The difference in composition, imprinted in H/He/C/N/O is largely masked and
photospheric-phase spectroscopy is unable to lift this apparent degeneracy.
The few C\,{\sc i}, N\,{\sc i} and O\,{\sc i} lines appearing in the red part of the optical
are too weak to allow a reliable and meaningful assessment of CNO abundances.
Past about 15\,d after explosion, He{\sc i} lines disappear, and the weak sensitivity
of the H\,{\sc i} Balmer lines to the H/He abundance ratio compromises even this determination.

Quite paradoxically, most inferences on SNe II-P are based on observations during
the photospheric phase, which, as we have illustrated here, boasts very degenerate
properties that prevent a clear assessment of the progenitor properties.
Having recourse to nebular-phase spectra can help lift this degeneracy in a number
of ways. Because the helium-core mass increases with main-sequence mass,
the relative fraction of the ejecta that it represents grows with main-sequence mass.
\citet{DLW10b} demonstrated that for a given explosion energy, the ejection velocity
of core-embedded oxygen-rich material is an increasing function of main-sequence
mass. The width of [O\,{\sc i}]\,6303--6363\,\AA\ can be used as a diagnostic of this
differential expansion speed. Its strength can then be used to constrain the total
amount of core-embedded oxygen ejected. Together with the constraint on the
helium-core mass (which is set by the main-sequence mass), one may have a means
to constrain the mass of the compact remnant formed.

We plan a forthcoming parameter study using non-LTE time-dependent simulations based on
models of SNe II-P with different main-sequence mass, explosion energies, and $^{56}$Ni.
This will reveal a wealth of information, and document what SNe II-P ejecta properties condition
the  [O\,{\sc i}] strength and width. Based on the two simulations
presented here, we find that the doublet line [O\,{\sc i}]\,6303--6363\,\AA\ is both broader and
stronger in the simulation based on the higher-mass progenitor, and this results from differences in the pre-SN stars
rather than in the explosion properties (Fig.~\ref{fig_comp_spec2}; note that the enforcement of
homology alters somewhat the inner-ejecta kinematics).
The $\sim$ 30\% larger core-embedded oxygen ejection speed in model s25e12
is reflected by the greater width of each component of the [O\,{\sc i}] doublet line,
while the larger core-embedded oxygen ejected mass in model s25e12 (3.32 compared to 0.816\,\msun)
is reflected by the higher line flux at all nebular epochs (this is also influenced by the
$^{56}$Ni mass). Clumping and subtleties associated
with radioactive heating will complicate a quantitative assessment here, but this differential reasoning
at least opens a route towards constraining in a systematic fashion the properties
of the inner ejecta of SNe II-P.

\begin{figure*}
\epsfig{file=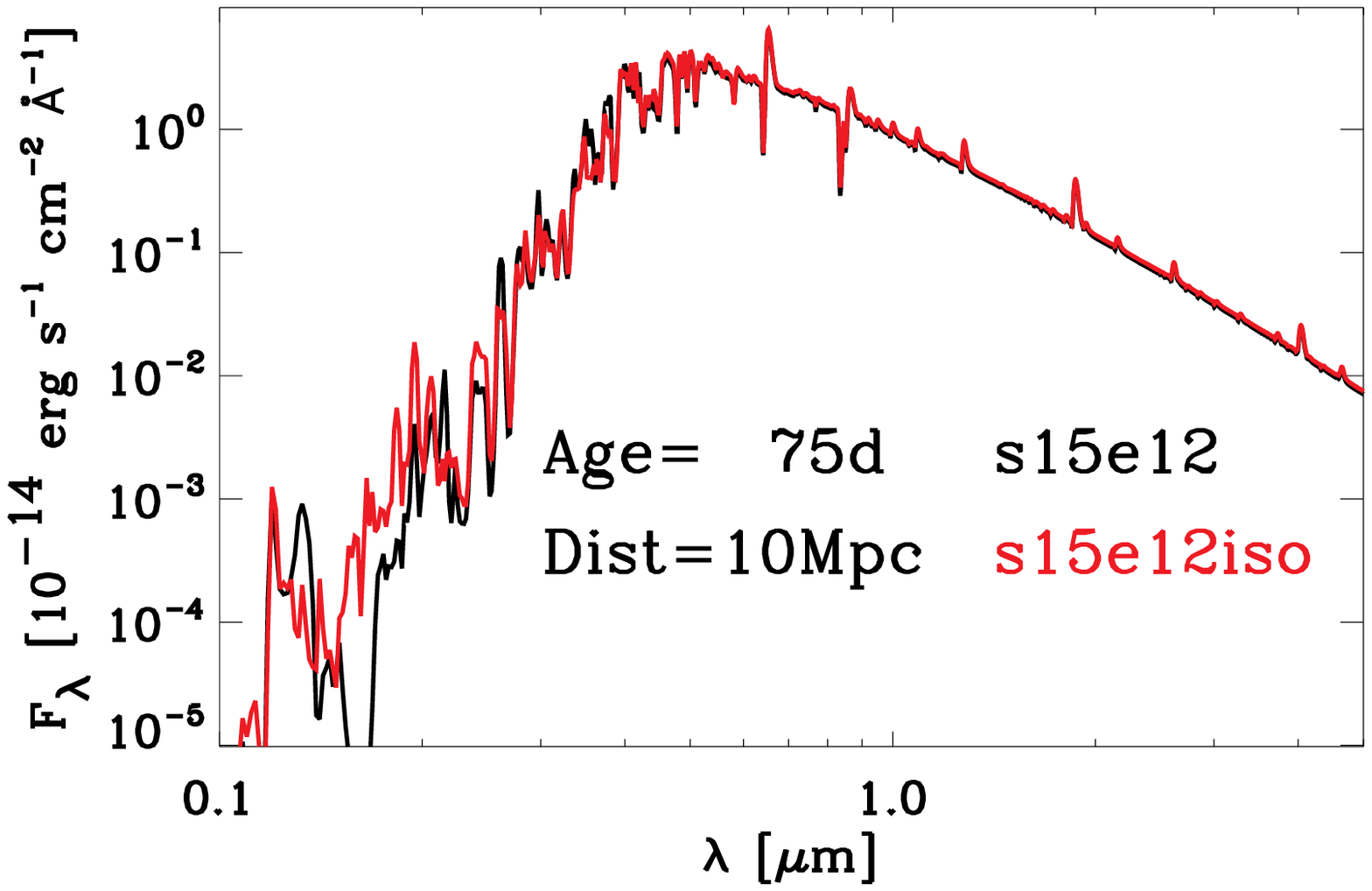,width=8.5cm}
\epsfig{file=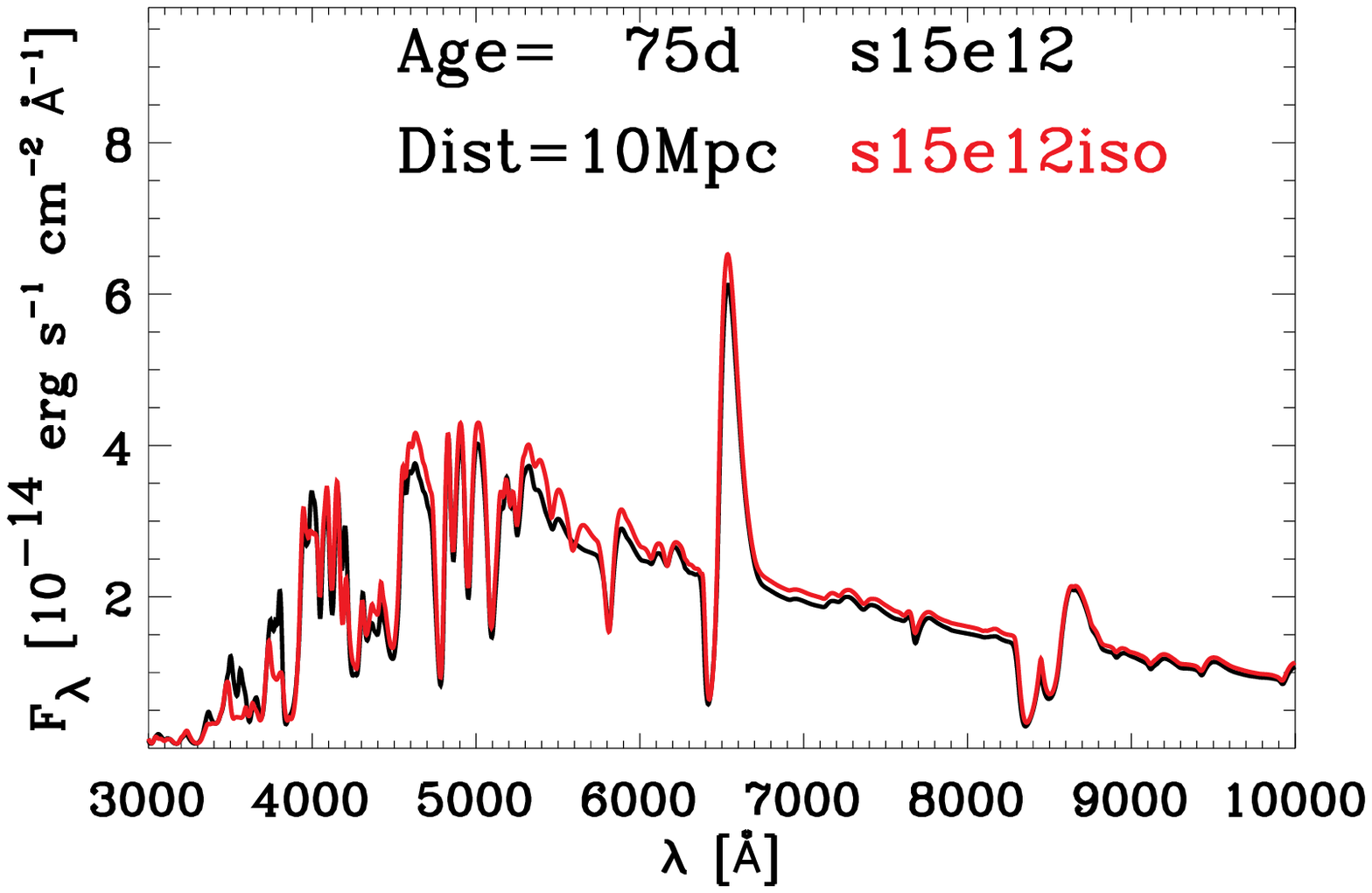,width=8.5cm}
\caption{{\it Left:} Log-log plot showing synthetic spectra at 75\,d (41st time step in the sequence
for each) after explosion for the simulations based on models s15e12 (black) and s15e12iso (red).
The spectral range covers from 1000\,\AA\ to 5\,$\mu$m.
{\it Right:} Same as left, but now showing the optical range only and using a linear scale.
\label{fig_comp_s15}}
\end{figure*}

\subsection{Comparison between simulations based on s15e12 and s15e12iso models}

   As detailed previously in \S~\ref{sect_setup}, models s15e12 and s15e12iso differ in their composition,
and our simulations based on these simulations reflect that distinction, as well as the additional
treatment of ions in the model atom (see Table~\ref{tab_atom}). In our simulation based on model s15e12,
we also neglected some species for which no associated line feature was ever seen, such as
neon or argon. In that case, we applied a scaling on the rest of the elements to normalise the sum of
mass fractions to unity - this scaling modifies by $\sim$10\% the abundances in the helium core.
The difference in composition between the two models contains two components.
First, some species are included in model s15e12iso, like Sc or Ba, which are
missing in model s15e12. Second, model s15e12iso gives the distribution of species versus depth.
It shows, for example, an increase in sodium abundance towards the core, while in model s15e12 sodium
is taken at its solar-metallicity value.

   So, when comparing synthetic spectra from our simulations based on models s15e12 and s15e12iso, both
   composition and model atom differences intervene.
   We show in Fig.~\ref{fig_comp_s15} the differences at the same post-explosion time of 75\,d for each
   simulation. There is $\sim$5\% absolute flux level difference between the two, with significant differences
   appearing in the UV, and essentially none beyond $\lesssim$1\,$\mu$m.
   The enhanced UV flux in the simulations based on model s15e12iso may stem from
   the reduced metal abundances, in favour of the treatment of species like Ar or Ne which have little blocking power,
   while the reduced $UB$-band flux is caused by the treatment of additional blanketing sources (e.g., Fe\,{\sc i} or Ni\,{\sc ii};
   see Fig.~\ref{fig_ladder_plot} for an illustration of individual contributions).
   Lines of Sc\,{\sc ii}, whose abundance is zero in model s15e12, are predicted in the simulation based on model s15e12iso.
   Similarly, the Na\,{\sc i}\,D doublet line is stronger in that simulation, which reflects the increasing sodium mass fraction
   in model s15e12iso with depth.
   During the nebular phase, differences are significant between the two simulations (not shown), but they are primarily associated
   with the treatment of Fe\,{\sc i}.

   Although the spectral evolution computed for simulations based on models s15e12 and s15e12iso are
   essentially the same, the model with a more complete set of metals and model atom yields
   stronger line blanketing (which translates into a weaker flux in the $UB$ bands)
   as well as the prediction of specific lines associated with, for example, Sc{\sc ii} (Fig.~\ref{fig_ladder_plot}).

\section{Comparison with observations}
\label{sect_comp_obs}

\begin{figure*}
\epsfig{file=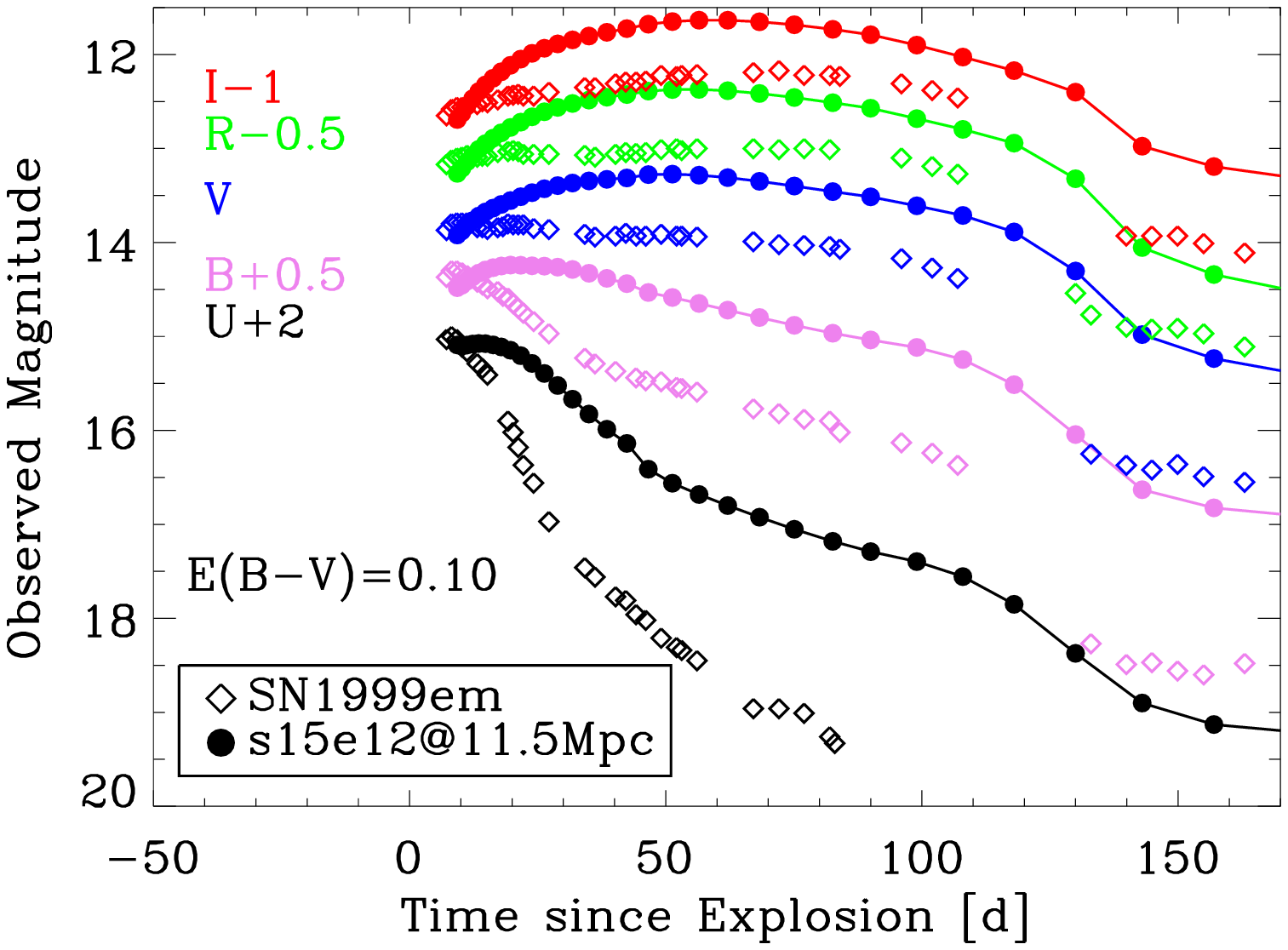,width=8.5cm}
\epsfig{file=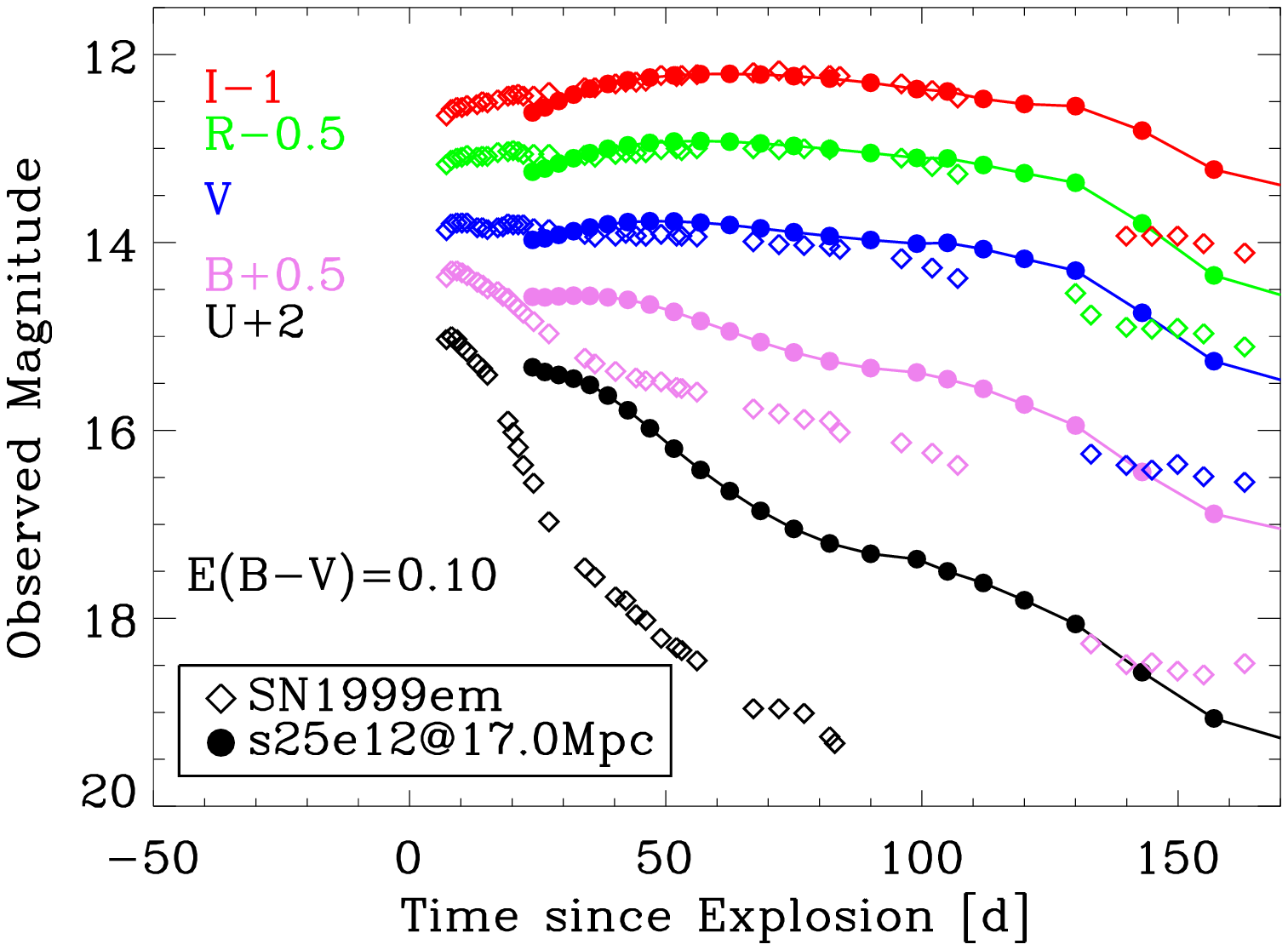,width=8.5cm}
\caption{{\it Left:} Comparison between the observed multi-color light-curve (diamonds) of SN1999em \citep{leonard_etal_02a}
and the corresponding reddened ($E(B-V)=$0.1) and distance scaled (D=11.5\,Mpc; \citealt{DH06_SN1999em})
magnitudes from the s15e12 model (dots). A further vertical shift is applied to avoid overlap, as in \citet{leonard_etal_02a}.
Note the excellent agreement at day 9 for each band (see spectral comparison
at that epoch in Fig.~\ref{fig_spec_99em_1}), but the subsequent divergence between observed and synthetic magnitudes.
This suggests a significant mismatch between the structure of the RSG progenitor model and/or explosion energy
with that corresponding to the progenitor of 99em. This may stem from the mismatch in density distribution,
which is in evidence in Fig.~\ref{fig_spec_99em_1}.
{\it Right:} Same as left, but now for the time sequence based on model s25e12 (we now scale the synthetic magnitudes
adopting a SN distance of 17\,Mpc). \label{fig_comp_obs_LC}
 }
\end{figure*}

\subsection{Bolometric luminosity and light curves}

 Observed $V$-band light curves of a large sample of SNe II-P have been presented in \citet{hamuy_03} or  \citet{poznanski_etal_09}.
These suggest that despite the expected variations in progenitor \citep{levesque_etal_05} and explosion properties
\citep{burrows_etal_07a,burrows_etal_07b,marek_janka_09}, their properties
are quite uniform (those wishing to standardise such SNe for distance determinations will find instead that there is a sizeable scatter).
Quite generally, about 10-20 days after the inferred time of explosion, the visual brightness levels off
or fades slowly (corresponding to a typical luminosity of $\sim$1.2$\times$10$^{42}$\,erg).
When reaching 75-120\,d after explosion, the SN fades by 0.8-1.2\,dex in luminosity
as it transitions on a timescale of a few weeks from the plateau phase
to the nebular phase \citep{bersten_hamuy_09}.

  Our simulations reveal plateau-phase bolometric luminosities (Fig.~\ref{fig_lbol})
  that are a factor of two too large compared to observations.
  Indeed, the absolute $V$-band light curves we obtain are brighter
  than those we derived from detailed modelling of SNe 1999em and 2006bp \citep{DH06_SN1999em,dessart_etal_08}.
  To illustrate this discrepancy, we show a comparison of our reddened and distance-scaled synthetic light curves
  to observations of SN1999em \citep{leonard_etal_02a} in Fig.~\ref{fig_comp_obs_LC}.
  Interestingly, the same discrepancy applies to the simulations of SNe II-P light curves presented in \citet{KW09}.
  In contrast, the simulation of \citet{eastman_etal_94}, which is based on a 15\,\msun\ progenitor star characterised by a smaller
  radius of 480\,\rsun, predicts a representative plateau-phase bolometric-luminosity of $\sim$4$\times$10$^8$\,\lsun\
  and a peak $V$-band magnitude of about $-17$.
  Hence, we surmise that the main problem with the progenitors used in recent SNe II-P simulations (including this work)
  is the progenitor radius, which is systematically too large, and perhaps the helium mass fraction in the hydrogen envelope,
  which is too low.

\begin{figure*}
\epsfig{file=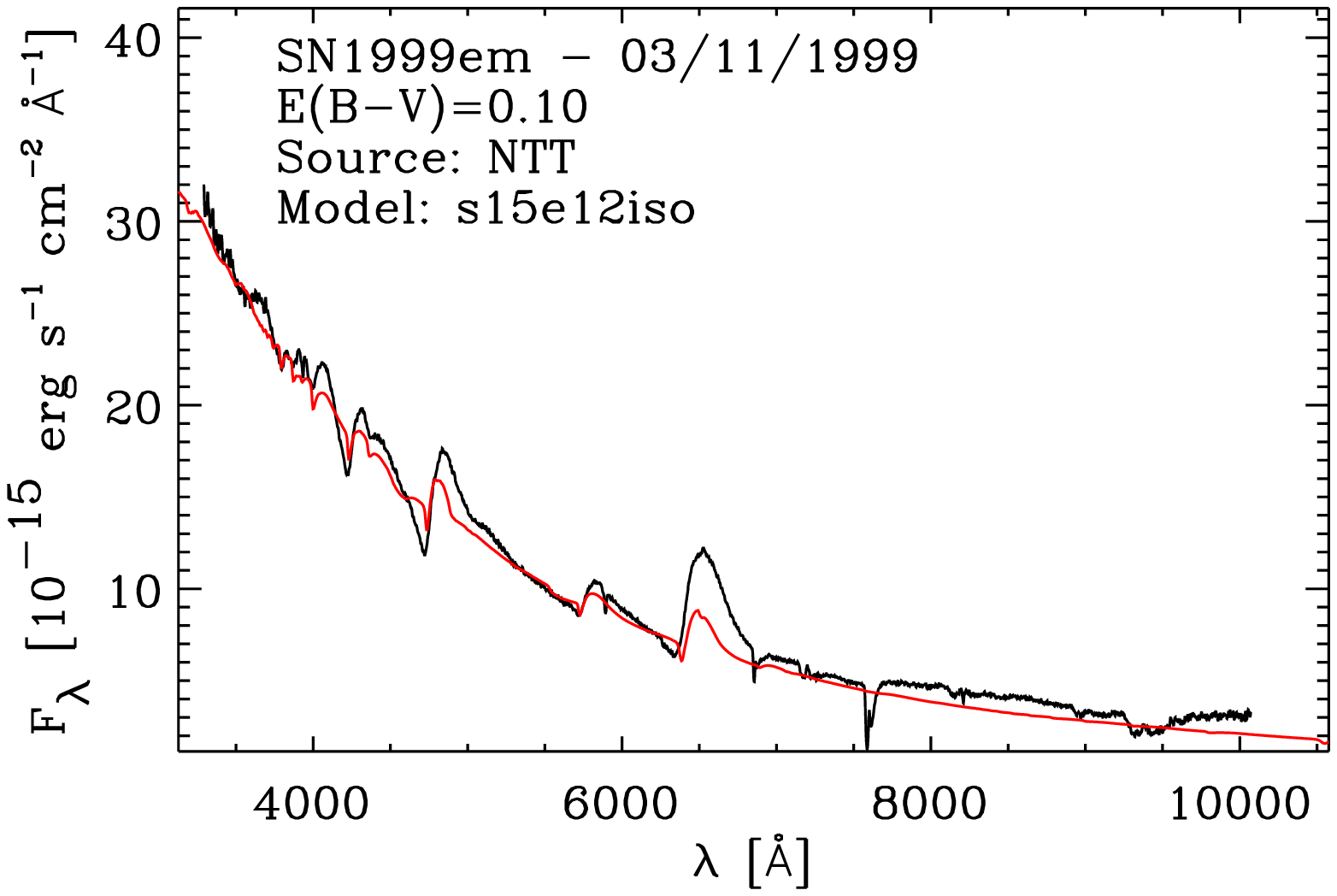,width=8.5cm}
\epsfig{file=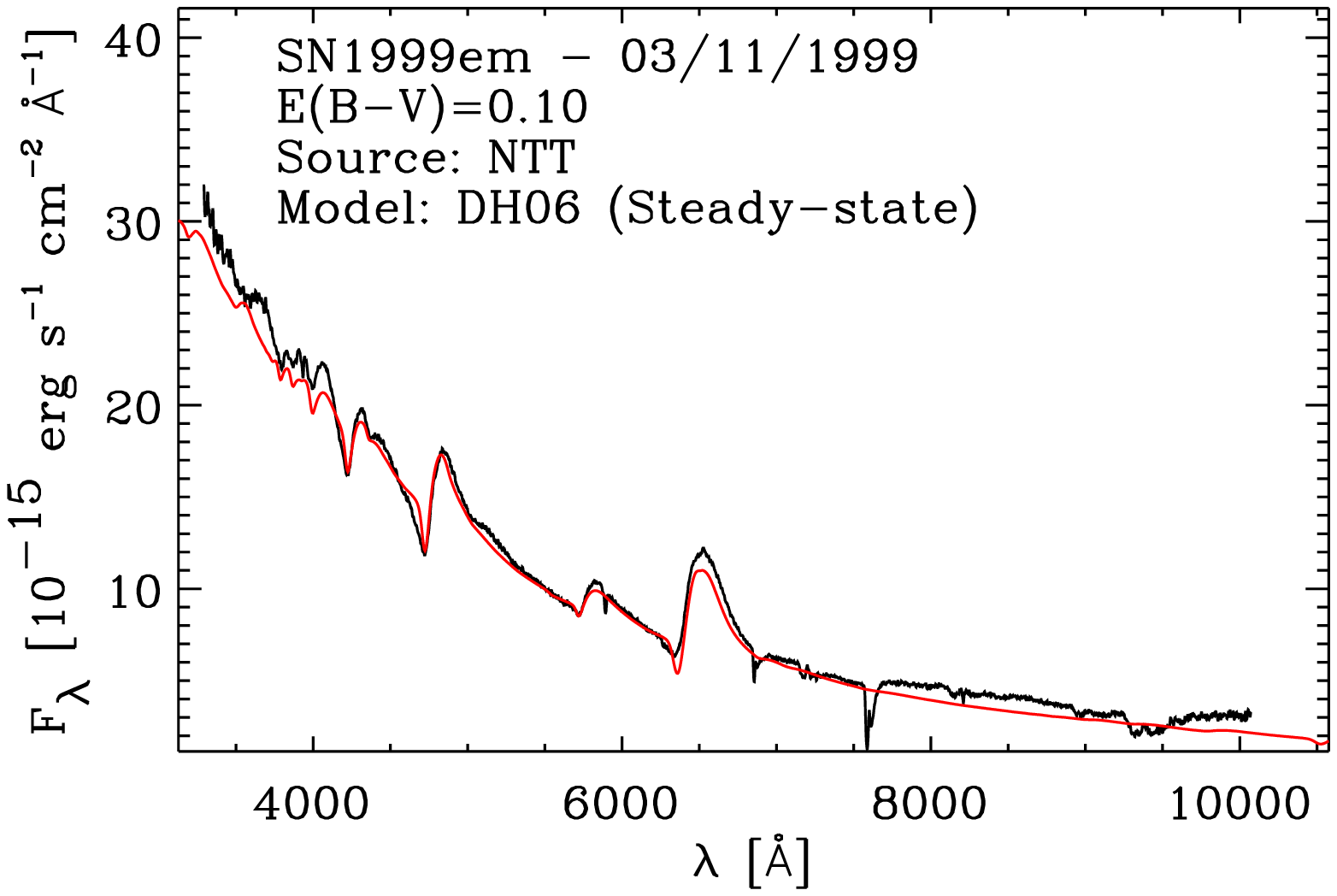,width=8.5cm}
\caption{{\it Left:} Comparison between the observations of SN1999em on the 3rd of November 1999 \citep{hamuy_etal_01} with the
reddened and distance-scaled synthetic spectrum obtained from the s15e12iso model on day 9.3 evolved with the time-dependent solver.
{\it Right:} Same as left, but now using the model of \citet{DH06_SN1999em} for the same date, which is computed with
a steady-state solver (which also accounts for $v/c$ terms appearing in the radiative-transfer equation).
The difference between the two models is caused by the steeper density distribution in the hydrodynamical
input s15e12iso, corresponding to $N_{\rho}\sim25$ in the photospheric region at the time, compared to  $N_{\rho}=$10
used in the steady-state calculation. This affects significantly the flux in recombination lines, as well as the
evolution of the light curve and colours, and suggests that the hydro input is not fully adequate.
\label{fig_spec_99em_1}}
\end{figure*}

  Our simulations have an explosion energy of 1.2\,B and a $^{56}$Ni mass that may not fit specific SNe II-P
  but is representative of the standard and thus should not be the cause of this overall discrepancy.
   Finally, in the simulations of SN1987A performed with the same code \citep{DH10}, we did reproduce the observed
   SN1987A light curve to within 10\% in flux at all times between 0.3 and 21\,d.
   This agreement suggests there is a physical discrepancy with the pre-SN RSG stars we employ at present.
   We speculate that RSG progenitors with a surface radius
   intermediate between the blue-supergiant progenitor of SN1987A ($\sim$50\,\rsun) and those of the RSG
   progenitors of models s15e12/s25e12 (810/1350\,\rsun) would correspondingly reveal light curves
   that are intermediate between those produced from our simulations of SN1987A and those from models
   s15e12/s25e12.

   A faster increase in (or a globally-larger) helium mass fraction in the hydrogen-rich envelope would likely remedy
   the sustained visual brightening obtained with our models during the plateau phase. Indeed, in this context, the progression
   of the recombination wave is faster, leading to a faster decline of the photospheric radius than presently predicted
   (Fig.~\ref{fig_phot}). In our non-LTE steady-state simulations for SNe 2005cs and 2006bp \citep{dessart_etal_08},
   we employed a chemical mixture with depleted H/C/O and enhanced He/N (a similar trend was obtained
   by \citealt{baron_etal_07}), resembling closely the surface composition
   of model s25e12, but not that of model s15e12. As discussed earlier, there is evidence of mixing in RSG stars which
   points in this direction of enhanced He/N \citep{lancon_etal_07}, perhaps as a result of stellar rotation.
   This determination of H/He/C/N/O abundances is best done with early-time observations of SNe II-P, when the photospheric
   layers are fully or partially ionised. This issue is important to resolve, as it has ramifications both for the modelling of SN light
   and for the understanding of mixing processes in massive stars prior to the explosion.

   As in our simulations of SN1987A, there are residual problems with the UV range and $UB$-bands. The flux
   that falls within these regions is typically smaller, or much smaller, that the visual flux. This region is strongly
   affected by line blanketing, which effectively absorbs the relatively small flux appearing blue-ward of $\lesssim$4000\,\AA\
   at the thermalisation depth. This line blanketing is caused by under-abundant metal species, and may also be affected
   by the floor temperature we employ in our simulations. For example, the additional contributions from Sc{\sc ii},
   Cr\,{\sc ii}, Fe\,{\sc i} and Ni\,{\sc ii} in simulations based on model s15e12iso led to a steeper decline and
   $\lesssim$0.5\,mag-fainter $UB$-band magnitudes than obtained with model s15e12. This sensitivity suggests that the
   discrepancy is really connected to completeness of the model atoms, although tests have not revealed any obvious
   candidate species, or a problem with the adopted model atoms. As we neglect many neutral species we will underestimate
   the line blanketing, particularly at late time.  Examination of the
   $U$-band light curve presented by \cite{KW09}
   for their standard II-P model shows less of a discrepancy with the $U-V$ color, further hinting to a possible problem
   with our adopted atomic data.

   The metallicity at the site of the explosion could also play a role  (it may
   not be solar, as assumed here), as could the H/He abundance ratio. The H/He abundance ratio modifies the
   way radiation thermalises so that for
   enhanced helium abundance, the UV flux drops. A higher helium mass fraction in the hydrogen-envelope
   of the RSG progenitors would thus lead to a faster visual-brightness decline and a fainter UV flux, both
   improving the match to standard observations of SNe II-P.

\subsection{Photospheric- and nebular-phase spectra}

   As described above, the absolute brightness and spectral evolution predicted with the simulations
   performed in this paper differ quantitatively from that of standard SNe II-P. However, modulo an absolute
   scaling to get the fluxes in correspondence, and allowing for an adjustment in time, we can compare
   our synthetic spectra to individual observations. Here, we use the spectroscopic observations of SN1999em, for
   which we also have tailored simulations computed with a steady-state approach \citep{DH06_SN1999em}.

   Our simulation at 9.3\,d and based on model s15e12iso matches the SN 1999em observations on 3 Nov. 1999
   \citep{hamuy_etal_01} both in relative and in absolute flux in the optical (Fig.~\ref{fig_spec_99em_1}).
   Similarly, the equivalent steady-state model computed for that date (right panel; \citealt{DH06_SN1999em})
   shows the same color and reproduces the overall flux very well.
   However, the equivalent steady-state model reproduces line profile strengths much better.
   Time dependence might play a role here, but more likely, the mass-density profile in the outer ejecta of model s15e12iso
   is the culprit.

   At very early post-explosion times,
   a density exponent of 15-50 is required to reproduce the observations of SNe 2005cs and 2006bp
   \citep{dessart_etal_08}, but at 10 days after explosion, this density exponent was found to be on the order of 10
   in the photospheric region for these SNe, as well as for SN1999em \citep{DH06_SN1999em}.
   In our simulations (s15e12/s15e12iso), the photosphere resides from day 10 to day 25
   in the outer ejecta layers where the density exponent is approximately 25 (!). This then seems the most probable reason
   for the underestimated line strength (see discussion of the effect of the density exponent on Balmer-line profile
   morphology in \citealt{DH05_qs_SN}).  The progenitor star may have a different structure.
   For example, the density exponent for the corresponding regions (velocity in the range 7000--9000\,\kms)
   in our 87A model ejecta, produced from the explosion of a 50\,\rsun\ BSG star, is 8 \citep{DH10}.
   Furthermore, simulations of SNe II-P ejecta tend to have low resolution in the outermost mass shells,
   which questions the accuracy of these layers and the corresponding early-time SN photospheric properties.

\begin{figure}
\epsfig{file=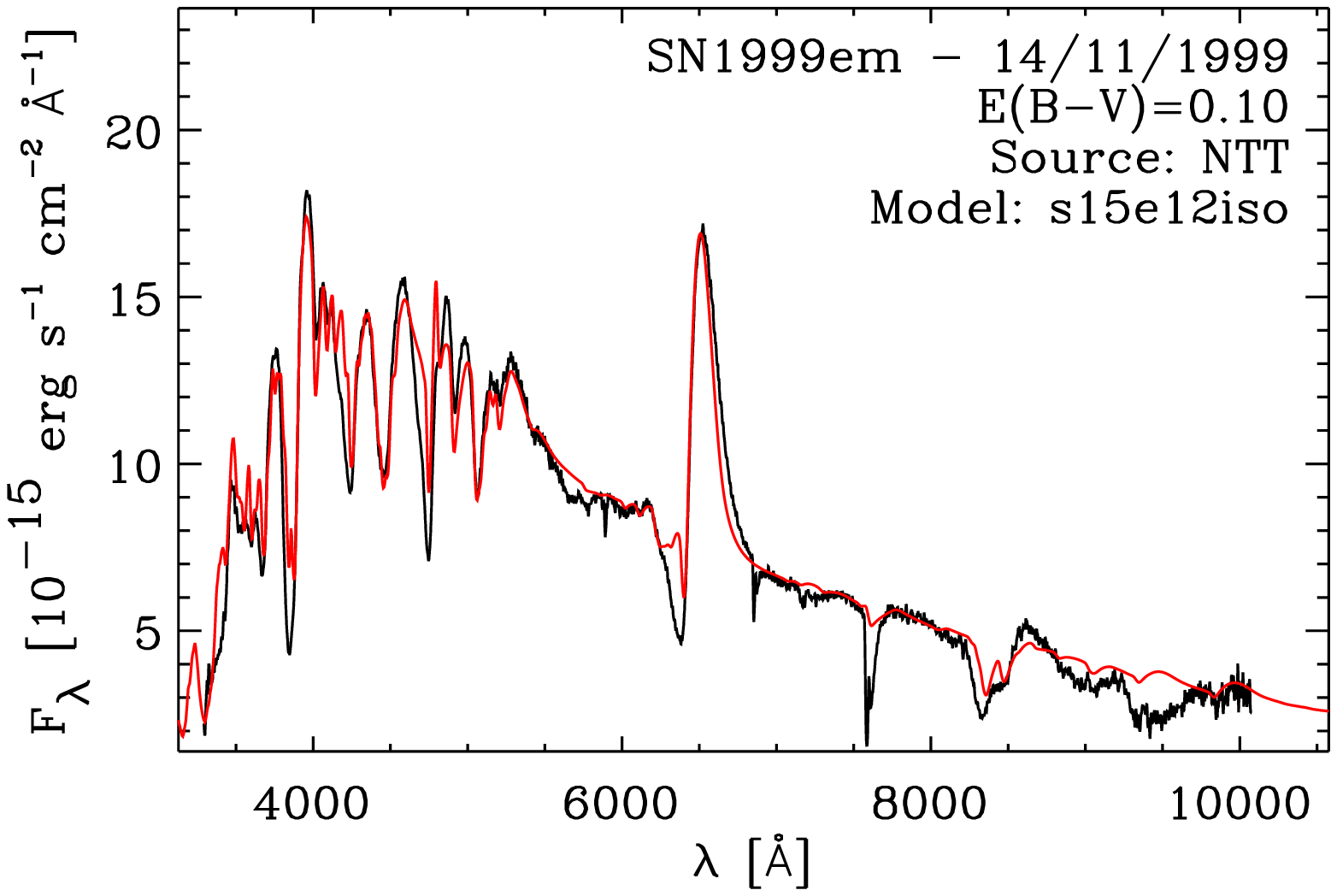,width=8.5cm}
\epsfig{file=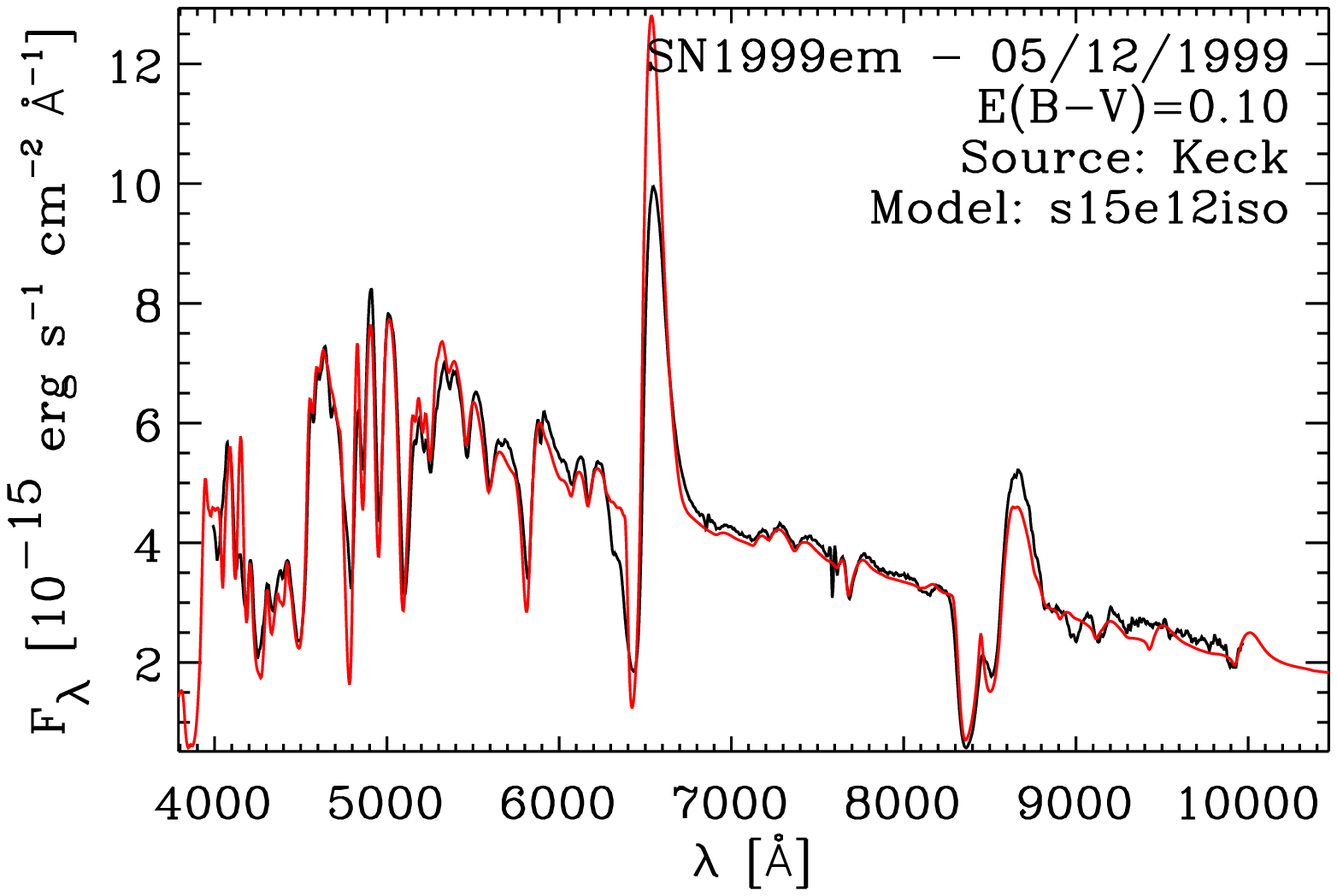,width=8.5cm}
\epsfig{file=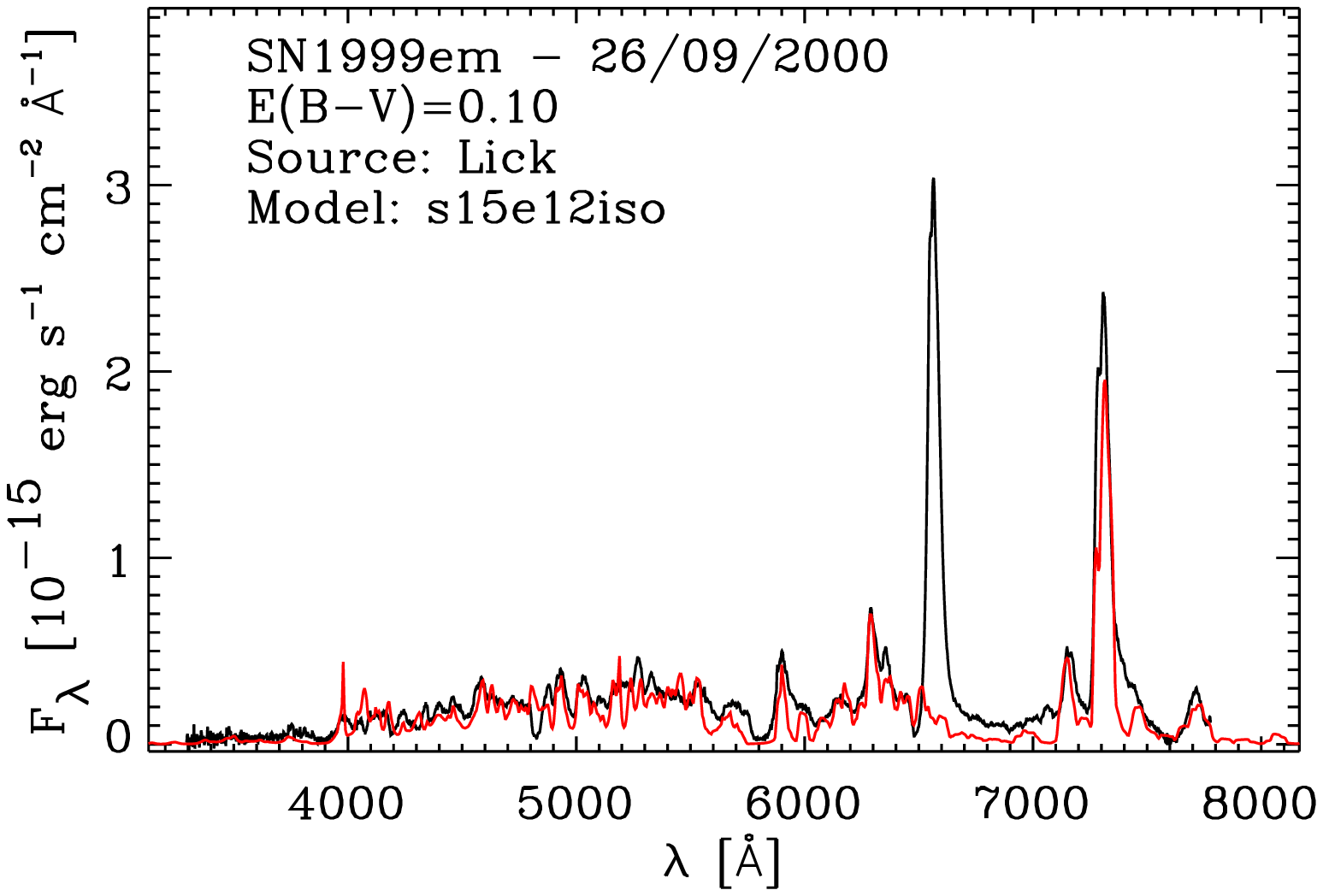,width=8.5cm}
\caption{
{\it Top:} Comparison between the reddened synthetic spectrum from
our simulations s15e12iso at a post-explosion time of 32.1\,d with the NTT observations on 14 Nov. 1999
of SN1999em \citep{hamuy_etal_01}. This observation corresponds to an inferred post-explosion time of $\sim$22\,d
\citep{DH06_SN1999em}. These dates do not match. Instead, our choice is done so that the observations
and models have the same color.
{\it Middle:} Same as top, but now for the s15e12iso at a post-explosion time of 75.8\,d and the Keck observations
on 5 Dec. 1999  \citep{leonard_etal_02a}, corresponding to a post-explosion time of 42\,d.
{\it Bottom:} Same as top, but now for the model s15e12iso at a post-explosion time of 306\,d and the Lick
observations of 26 September 2000 (331\,d after discovery).
In each panel, the synthetic spectra have been scaled to correct for the magnitude mismatch
shown in Fig.~\ref{fig_comp_obs_LC}. \label{fig_spec_99em_2}
}
\end{figure}

   As the ejecta recombine and the photosphere recedes into deeper layers, the line strengths increase.
   At such times, we reproduce the observations of SN1999em extremely well (upper two panels of
   Fig.~\ref{fig_spec_99em_2}). For the observations of SN1999em on 5 Dec 1999, we also illustrate the
   various bound-bound transitions that contribute to the observed line features in Fig.~\ref{fig_ladder_plot}.
   We also include here a comparison of the nebular-phase spectrum of SN1999em taken on 26 Sept. 2000
   \citep{leonard_etal_02a} with the synthetic spectrum at 306\,d from our simulation based on model s15e12iso
   (synthetic spectra are scaled to correct for the magnitude mismatch illustrated in Fig.~\ref{fig_comp_obs_LC}).
   Recall that SN1999em is thought to have produced 0.036\,\msun\
   of $^{56}$Ni \citep{utrobin_07} while model s15e12iso contains initially 0.086\,\msun.
   The agreement is good, especially since no parameter is adjusted in model s15e12iso to match the observations.
   The only, and sizeable discrepancy, is the complete absence of H$\alpha$ in our synthetic spectrum, while
   a strong flux is observed instead. We also note that we predict a sudden disappearance of the continuum flux at the
   end of the plateau phase, while observations suggest a more gradual fading of the continuum flux relative to lines.
   Such discrepancies stem entirely from our neglect of
   non-thermal excitation/ionisation. Remarkably, it may be that this additional process does not affect other species
   much\footnote{Non-thermal processes will likely increase the electron density but this will only have a secondary effect on many of the lines since they are the coolants, and hence their strength is primarily determined by energy deposition from radioactive decay}. In any case, this physical ingredient is essential for quantitative modelling of the nebular phase of SNe II
   and will be incorporated in {\sc cmfgen} in the near future.

\section{Discussion and conclusions}
\label{sect_concl}

We have presented the first non-LTE time-dependent radiative-transfer simulations of SNe II-P
     covering from the early photospheric phase until $\gtrsim$1000\,d after explosion. Our
     simulations start from hydrodynamical models of the explosion of two RSG stars,
     evolved from non-rotating solar-metallicity stars with main-sequence masses of 15 and 25\,\msun.
     The main assets of our approach are 1) the treatment of non-LTE for all level populations
     2) the explicit treatment of line-blanketing, 3) the consideration of all time-dependent terms appearing in the statistical-equilibrium,
     energy, and radiative-transfer equations, 4) the adoption of a hydrodynamical model
     describing the density/temperature/radius/velocity distribution as well as the complex chemical stratification in the SN ejecta, and
     5) the modelling of the {\it entire} ejecta at all times.
     The main approximations of our approach are spherical symmetry, homologous expansion, and the treatment
     of radioactive decay as a local heating source (thus ignoring the effects of non-thermal excitation and ionisation).

From our simulations we find/infer the following:

\begin{enumerate}

\item
The photospheric evolution during the plateau phase shows a continuous
radius increase up to 2.5--3$\times$10$^{15}$\,cm at $\sim$80\,d; a decrease and then a plateau in temperature
at $\sim$5000\,K; and a continuous decrease in velocity as the ejecta optical thickness diminishes with expansion and
recombination. This is in qualitative agreement with that inferred from observation, and is similar to the trend we obtained
for SN1987A \citep{DH10}.

\item
   The synthetic bolometric luminosity shows a very gradual decline (not quite a plateau) during the photospheric phase.
     At 50\,d after explosion our models have bolometric luminosities of $\sim 6 \times10^8$\,\lsun\ for model s15e12 and
   $\sim10^9$\,\lsun\ for model s25e12.  These are a factor of 2 to 3 larger than the standard 50-d luminosity  of
   $\sim$3$\times$10$^8$\,\lsun\ for observed SNe II-P  \citep{bersten_hamuy_09}. Similar overestimates result from the simulations
    of \citet{KW09}, which are based on similar progenitor models.  However the simulation of \citet{eastman_etal_94}, based on a more compact pre-SN RSG star,  is in agreement with observations. This suggests that the pre-SN RSG stars produced by current stellar-evolutionary models
   have surface radii that are too large, on average.

\item
 The density distribution of the photosphere is too steep in our models,  at 10 to 30\,d after core collapse, yielding line strengths that are weaker than observed
 prior to the onset of the recombination phase. This may independently  indicate that the progenitor surface radii are too large in
 our models. Steady state models indicate that density exponents of $\sim$10 (i.e., $\rho \propto r^{-10}$) rather than $\sim$25 (valid at early times for
 the present models) are required to fit the line strengths.

 \item
      At the end of the plateau phase, our computed bolometric luminosities decrease by a factor of a few
     over a timescale of about a month until reaching a rate that corresponds to the instantaneous decay-energy production
     rate. In our simulations, the magnitude of the fading at the end of the photospheric phase is conditioned by two properties
     that are not obviously linked ---  the progenitor radius and the $^{56}$Ni mass produced, and is thus expected to
     occur with a scatter in Nature.

\item
  Our multi-band light curves tend to show  bell shaped curves, with a peak that occurs earlier for filters positioned
     at shorter wavelength, and a width that increases for filters positioned at longer wavelength.
     This morphology is similar to that seen for SN1987A \citep{phillips_etal_88,DH10}, but the larger RSG progenitors
     corresponding to our models causes the effective/color temperature to decline more slowly (and the more so for
     model s25e12 because of its larger-radius). Comparison
     with SN1999em \citep{leonard_etal_02a} confirms that decline in  effective/color temperature is too slow,
     with the $UB$ magnitudes remaining large for about 10-20\,d.

 \item
 As for our simulations of SN1987A, we predict too strong UV and $U$-band fluxes during much of the photosphere phase.
This is likely due to inadequate model atoms, or the H/He abundance ratio in the envelope, and is currently being investigated.

  \item
  In both simulations the photosphere resides in the outer 1\,\msun\ of the SN ejecta for $\sim$ 50\,d after the explosion. Thus spectra taken during this time do not reveal any information about the
  chemical and density structure of the bulk of the mass. The confinement of the photosphere to the outer layers
  also  provides a simple explanation for the low linear-polarisation measures of SNe II-P at early times  \citep{leonard_filippenko_05,leonard_etal_06,chornock_etal_10}, and  the notion that early-time SNe II-P spectra are relatively unaffected
     by the details of the explosion (e.g., geometry and $^{56}$Ni production). Rather, SNe II-P evolve at constant composition through
     changes in ionisation alone --- it is at such early-times that SNe II-P are best used for distance determinations based on
     spectroscopic modelling. Given the issues with the density structure, and the confinement of the photosphere to the outer layers
     at early times, it is clear that higher resolution pre-SN models are needed to allow more accurate results in our present modelling approach --
specifically we require that the outer SN ejecta be resolved and accurately described all the way to large velocities.

  \item
  At the same color,  photospheric  spectra show excellent agreement with those of SN1999em.
  Nebular spectra, with the exception of the H\,{\sc i} spectrum, are also strikingly similar to those observed.
  The problem with the H\,{\sc i} lines is directly attributable to the neglect of excitation/ionisation by non-thermal electrons.
   In our models only forbidden lines of O, Na, Ca, Ti, and Fe, and resonance lines of Ca shine
      at a few 100\,d after explosion (in the optical and/or near-IR ranges). At 1000\,d after explosion, the spectra
      are dominated by the double line of O\,{\sc i} and numerous lines of Fe\,{\sc i}--{\sc ii}; no other species
      included in our calculations seems to contribute at such late times.

 \item
     The s15e12 and s25e12 models show differences in peak luminosity and the rise-time to peak V-band brightness.
     However, at  the same colour, photospheric spectra are very similar.  This suggests that the composition contrast between the two progenitors
     is too small to be observable spectroscopically during the ``plateau''  phase.  It also corroborates the spectral degeneracy of
      SNe II-P spectra and highlights the difficulty of distinguishing between different progenitor masses or evolutionary paths
      based on ``plateau'' phase spectra.

      We find that nebular-phase spectra show differences that unambiguously distinguish a lower-mass progenitor
      from a higher-mass progenitor in a standard SN II-P explosion. These are related to the property that stars with
      higher main-sequence masses have systematically a higher helium-core mass, and
      shows up in late-time [O\,{\sc i}]\,6303--6363\AA\ line profiles, which appear broader and stronger for a higher-mass
      progenitor \citep{DLW10b}.

\end{enumerate}

  In this work, we have demonstrated that there is much to learn from synthetic spectra of SNe II-P. Spectra contain additional
  information that is either absent or degenerate in multi-band light curves alone. Ultimately, it is through non-LTE time-dependent
  radiative-transfer simulations of a large set of non-rotating and rotating pre-SN RSG stars, evolved at various metallicities, that
  an accurate mapping between SNe II-P observations and progenitor/explosions properties will be possible.
  We plan to undertake this task in the near future.
  Securing a large dataset of photospheric- and nebular-phase observations of SNe II-P, covering all
  stages of temporal evolution, is essential for the fulfilment of this task.

\section*{Acknowledgments}

      We thank Stan Woosley for providing the input hydrodynamical models that were used as initial
conditions for our radiative-transfer calculations.
LD acknowledges financial support from the European Community through an
International Re-integration Grant, under grant number PIRG04-GA-2008-239184.
LD also acknowledges financial support in the early stages of this work from the Scientific Discovery
through Advanced Computing (SciDAC) program of the DOE, under grant number DOE-FC02-06ER41452.
DJH acknowledges support from STScI theory grant HST-AR-11756.01.A and NASA theory grant NNX10AC80G.

\label{lastpage}


\begin{thebibliography}{75}
\expandafter\ifx\csname natexlab\endcsname\relax\def\natexlab#1{#1}\fi

\bibitem[{{Ambwani} \& {Sutherland}(1988)}]{AS88}
{Ambwani}, K. \& {Sutherland}, P. 1988, \apj, 325, 820

\bibitem[{{Axelrod}(1980)}]{axelrod_80}
{Axelrod}, T.~S. 1980, PhD thesis, California Univ., Santa Cruz.

\bibitem[{{Baklanov} {et~al.}(2005){Baklanov}, {Blinnikov}, \&
  {Pavlyuk}}]{baklanov_etal_05}
{Baklanov}, P.~V., {Blinnikov}, S.~I., \& {Pavlyuk}, N.~N. 2005, Astronomy
  Letters, 31, 429

\bibitem[{{Baron} {et~al.}(2007){Baron}, {Branch}, \&
  {Hauschildt}}]{baron_etal_07}
{Baron}, E., {Branch}, D., \& {Hauschildt}, P.~H. 2007, \apj, 662, 1148

\bibitem[{{Bersten} \& {Hamuy}(2009)}]{bersten_hamuy_09}
{Bersten}, M.~C. \& {Hamuy}, M. 2009, \apj, 701, 200

\bibitem[{{Blinnikov} {et~al.}(2000){Blinnikov}, {Lundqvist}, {Bartunov},
  {Nomoto}, \& {Iwamoto}}]{blinnikov_etal_00}
{Blinnikov}, S., {Lundqvist}, P., {Bartunov}, O., {Nomoto}, K., \& {Iwamoto},
  K. 2000, \apj, 532, 1132

\bibitem[{{Burrows} {et~al.}(2007{\natexlab{a}}){Burrows}, {Dessart}, {Livne},
  {Ott}, \& {Murphy}}]{burrows_etal_07b}
{Burrows}, A., {Dessart}, L., {Livne}, E., {Ott}, C.~D., \& {Murphy}, J.
  2007{\natexlab{a}}, \apj, 664, 416

\bibitem[{{Burrows} {et~al.}(2007{\natexlab{b}}){Burrows}, {Livne}, {Dessart},
  {Ott}, \& {Murphy}}]{burrows_etal_07a}
{Burrows}, A., {Livne}, E., {Dessart}, L., {Ott}, C.~D., \& {Murphy}, J.
  2007{\natexlab{b}}, \apj, 655, 416

\bibitem[{{Cappellaro} {et~al.}(1997){Cappellaro}, {Turatto}, {Tsvetkov},
  {Bartunov}, {Pollas}, {Evans}, \& {Hamuy}}]{cappellaro_etal_97}
{Cappellaro}, E., {Turatto}, M., {Tsvetkov}, D.~Y., {Bartunov}, O.~S.,
  {Pollas}, C., {Evans}, R., \& {Hamuy}, M. 1997, \aap, 322, 431

\bibitem[{{Castor}(1970)}]{castor_70}
{Castor}, J.~I. 1970, \mnras, 149, 111

\bibitem[{{Chornock} {et~al.}(2010){Chornock}, {Filippenko}, {Li}, \&
  {Silverman}}]{chornock_etal_10}
{Chornock}, R., {Filippenko}, A.~V., {Li}, W., \& {Silverman}, J.~M. 2010,
  \apj, 713, 1363

\bibitem[{{Cohen} {et~al.}(2003){Cohen}, {Wheaton}, \&
  {Megeath}}]{cohen_etal_03}
{Cohen}, M., {Wheaton}, W.~A., \& {Megeath}, S.~T. 2003, \aj, 126, 1090

\bibitem[{{Colgate} \& {White}(1966)}]{colgate_white_66}
{Colgate}, S.~A. \& {White}, R.~H. 1966, \apj, 143, 626

\bibitem[{{Dessart} {et~al.}(2008){Dessart}, {Blondin}, {Brown}, {Hicken},
  {Hillier}, {Holland}, {Immler}, {Kirshner}, {Milne}, {Modjaz}, \&
  {Roming}}]{dessart_etal_08}
{Dessart}, L., {Blondin}, S., {Brown}, P.~J., {Hicken}, M., {Hillier}, D.~J.,
  {Holland}, S.~T., {Immler}, S., {Kirshner}, R.~P., {Milne}, P., {Modjaz}, M.,
  \& {Roming}, P.~W.~A. 2008, \apj, 675, 644

\bibitem[{{Dessart} \& {Hillier}(2005{\natexlab{a}})}]{DH05_epm}
{Dessart}, L. \& {Hillier}, D.~J. 2005{\natexlab{a}}, \aap, 439, 671

\bibitem[{{Dessart} \& {Hillier}(2005{\natexlab{b}})}]{DH05_qs_SN}
---. 2005{\natexlab{b}}, \aap, 437, 667

\bibitem[{{Dessart} \& {Hillier}(2006)}]{DH06_SN1999em}
---. 2006, \aap, 447, 691

\bibitem[{{Dessart} \& {Hillier}(2008)}]{DH08}
---. 2008, \mnras, 383, 57

\bibitem[{{Dessart} \& {Hillier}(2009)}]{DH09_review}
{Dessart}, L. \& {Hillier}, D.~J. 2009, in American Institute of Physics
  Conference Series, Vol. 1111, Probing stellar populations out to the distant
  universe, ed. {G.~Giobbi, A.~Tornambe, G.~Raimondo, M.~Limongi,
  L.~A.~Antonelli, N.~Menci, \& E.~Brocato}, 565--572

\bibitem[{{Dessart} \& {Hillier}(2010)}]{DH10}
---. 2010, \mnras, 405, 2141

\bibitem[{{Dessart} {et~al.}(2010{\natexlab{a}}){Dessart}, {Livne}, \&
  {Waldman}}]{DLW10b}
{Dessart}, L., {Livne}, E., \& {Waldman}, R. 2010{\natexlab{a}},
  arXiv:1006.2268

\bibitem[{{Dessart} {et~al.}(2010{\natexlab{b}}){Dessart}, {Livne}, \&
  {Waldman}}]{DLW10a}
---. 2010{\natexlab{b}}, \mnras, 405, 2113

\bibitem[{{Eastman} {et~al.}(1996){Eastman}, {Schmidt}, \& {Kirshner}}]{E96}
{Eastman}, R.~G., {Schmidt}, B.~P., \& {Kirshner}, R. 1996, \apj, 466, 911

\bibitem[{{Eastman} {et~al.}(1994){Eastman}, {Woosley}, {Weaver}, \&
  {Pinto}}]{eastman_etal_94}
{Eastman}, R.~G., {Woosley}, S.~E., {Weaver}, T.~A., \& {Pinto}, P.~A. 1994,
  \apj, 430, 300

\bibitem[{{Falk} \& {Arnett}(1977)}]{falk_arnett_77}
{Falk}, S.~W. \& {Arnett}, W.~D. 1977, \apjs, 33, 515

\bibitem[{{Gezari} {et~al.}(2008){Gezari}, {Dessart}, {Basa}, {Martin},
  {Neill}, {Woosley}, {Hillier}, {Bazin}, {Forster}, {Friedman}, {Le Du},
  {Mazure}, {Morrissey}, {Neff}, {Schiminovich}, \& {Wyder}}]{gezari_etal_08}
{Gezari}, S., {Dessart}, L., {Basa}, S., {Martin}, D.~C., {Neill}, J.~D.,
  {Woosley}, S.~E., {Hillier}, D.~J., {Bazin}, G., {Forster}, K., {Friedman},
  P.~G., {Le Du}, J., {Mazure}, A., {Morrissey}, P., {Neff}, S.~G.,
  {Schiminovich}, D., \& {Wyder}, T.~K. 2008, \apjl, 683, L131

\bibitem[{{Hamuy}(2003)}]{hamuy_03}
{Hamuy}, M. 2003, \apj, 582, 905

\bibitem[{{Hamuy} \& {Pinto}(2002)}]{hamuy_pinto_02}
{Hamuy}, M. \& {Pinto}, P.~A. 2002, \apjl, 566, L63

\bibitem[{{Hamuy} {et~al.}(2001){Hamuy}, {Pinto}, {Maza}, {Suntzeff},
  {Phillips}, {Eastman}, {Smith}, {Corbally}, {Burstein}, {Li}, {Ivanov},
  {Moro-Martin}, {Strolger}, {de Souza}, {dos Anjos}, {Green}, {Pickering},
  {Gonz{\'a}lez}, {Antezana}, {Wischnjewsky}, {Galaz}, {Roth}, {Persson}, \&
  {Schommer}}]{hamuy_etal_01}
{Hamuy}, M., {Pinto}, P.~A., {Maza}, J., {Suntzeff}, N.~B., {Phillips}, M.~M.,
  {Eastman}, R.~G., {Smith}, R.~C., {Corbally}, C.~J., {Burstein}, D., {Li},
  Y., {Ivanov}, V., {Moro-Martin}, A., {Strolger}, L.~G., {de Souza}, R.~E.,
  {dos Anjos}, S., {Green}, E.~M., {Pickering}, T.~E., {Gonz{\'a}lez}, L.,
  {Antezana}, R., {Wischnjewsky}, M., {Galaz}, G., {Roth}, M., {Persson},
  S.~E., \& {Schommer}, R.~A. 2001, \apj, 558, 615

\bibitem[{{Heger} {et~al.}(2003){Heger}, {Fryer}, {Woosley}, {Langer}, \&
  {Hartmann}}]{heger_etal_03}
{Heger}, A., {Fryer}, C.~L., {Woosley}, S.~E., {Langer}, N., \& {Hartmann},
  D.~H. 2003, \apj, 591, 288

\bibitem[{{Hillier} \& {Miller}(1998)}]{HM98_lb}
{Hillier}, D.~J. \& {Miller}, D.~L. 1998, \apj, 496, 407

\bibitem[{{Kasen} {et~al.}(2006){Kasen}, {Thomas}, \& {Nugent}}]{KTN06_SN_MC}
{Kasen}, D., {Thomas}, R.~C., \& {Nugent}, P. 2006, \apj, 651, 366

\bibitem[{{Kasen} \& {Woosley}(2009)}]{KW09}
{Kasen}, D. \& {Woosley}, S.~E. 2009, \apj, 703, 2205

\bibitem[{{Kirshner} \& {Kwan}(1974)}]{KK74_EPM}
{Kirshner}, R.~P. \& {Kwan}, J. 1974, \apj, 193, 27

\bibitem[{{Kozma} \& {Fransson}(1992)}]{KF92}
{Kozma}, C. \& {Fransson}, C. 1992, \apj, 390, 602

\bibitem[{{Kozma} \& {Fransson}(1998{\natexlab{a}})}]{KF98a}
---. 1998{\natexlab{a}}, \apj, 496, 946

\bibitem[{{Kozma} \& {Fransson}(1998{\natexlab{b}})}]{KF98b}
---. 1998{\natexlab{b}}, \apj, 497, 431

\bibitem[{{Kurucz}(2009)}]{Kur09_ATD}
{Kurucz}, R.~L. 2009, in American Institute of Physics Conference Series, Vol.
  1171, American Institute of Physics Conference Series, ed. {I.~Hubeny,
  J.~M.~Stone, K.~MacGregor, \& K.~Werner}, 43--51

\bibitem[{{Lan{\c c}on} {et~al.}(2007){Lan{\c c}on}, {Hauschildt}, {Ladjal}, \&
  {Mouhcine}}]{lancon_etal_07}
{Lan{\c c}on}, A., {Hauschildt}, P.~H., {Ladjal}, D., \& {Mouhcine}, M. 2007,
  \aap, 468, 205

\bibitem[{{Landolt}(1992)}]{landolt_92}
{Landolt}, A.~U. 1992, \aj, 104, 340

\bibitem[{{Leonard} \& {Filippenko}(2005)}]{leonard_filippenko_05}
{Leonard}, D.~C. \& {Filippenko}, A.~V. 2005, in Astronomical Society of the
  Pacific Conference Series, Vol. 342, 1604-2004: Supernovae as Cosmological
  Lighthouses, ed. {M.~Turatto, S.~Benetti, L.~Zampieri, \& W.~Shea}, 330--+

\bibitem[{{Leonard} {et~al.}(2006){Leonard}, {Filippenko}, {Ganeshalingam},
  {Serduke}, {Li}, {Swift}, {Gal-Yam}, {Foley}, {Fox}, {Park}, {Hoffman}, \&
  {Wong}}]{leonard_etal_06}
{Leonard}, D.~C., {Filippenko}, A.~V., {Ganeshalingam}, M., {Serduke},
  F.~J.~D., {Li}, W., {Swift}, B.~J., {Gal-Yam}, A., {Foley}, R.~J., {Fox},
  D.~B., {Park}, S., {Hoffman}, J.~L., \& {Wong}, D.~S. 2006, \nat, 440, 505

\bibitem[{{Leonard} {et~al.}(2002){Leonard}, {Filippenko}, {Gates}, {Li},
  {Eastman}, {Barth}, {Bus}, {Chornock}, {Coil}, {Frink}, {Grady}, {Harris},
  {Malkan}, {Matheson}, {Quirrenbach}, \& {Treffers}}]{leonard_etal_02a}
{Leonard}, D.~C., {Filippenko}, A.~V., {Gates}, E.~L., {Li}, W., {Eastman},
  R.~G., {Barth}, A.~J., {Bus}, S.~J., {Chornock}, R., {Coil}, A.~L., {Frink},
  S., {Grady}, C.~A., {Harris}, A.~W., {Malkan}, M.~A., {Matheson}, T.,
  {Quirrenbach}, A., \& {Treffers}, R.~R. 2002, \pasp, 114, 35

\bibitem[{{Levesque} {et~al.}(2005){Levesque}, {Massey}, {Olsen}, {Plez},
  {Josselin}, {Maeder}, \& {Meynet}}]{levesque_etal_05}
{Levesque}, E.~M., {Massey}, P., {Olsen}, K.~A.~G., {Plez}, B., {Josselin}, E.,
  {Maeder}, A., \& {Meynet}, G. 2005, \apj, 628, 973

\bibitem[{{Li} \& {McCray}(1992)}]{LM92_SN1987A_OI}
{Li}, H. \& {McCray}, R. 1992, \apj, 387, 309

\bibitem[{{Li} \& {McCray}(1993)}]{LM93_CaII}
---. 1993, \apj, 405, 730

\bibitem[{{Li} \& {McCray}(1995)}]{LM95_SN87A_HeI}
---. 1995, \apj, 441, 821

\bibitem[{{Litvinova} \& {Nadezhin}(1985)}]{litvinova_nadezhin_85}
{Litvinova}, I.~Y. \& {Nadezhin}, D.~K. 1985, Soviet Astronomy Letters, 11, 145

\bibitem[{{Lucy}(1991)}]{L91_HeI}
{Lucy}, L.~B. 1991, \apj, 383, 308

\bibitem[{{Marek} \& {Janka}(2009)}]{marek_janka_09}
{Marek}, A. \& {Janka}, H. 2009, \apj, 694, 664

\bibitem[{{Mazzali} {et~al.}(1992){Mazzali}, {Lucy}, \&
  {Butler}}]{MLB92_SN1987A}
{Mazzali}, P.~A., {Lucy}, L.~B., \& {Butler}, K. 1992, \aap, 258, 399

\bibitem[{{McCray}(1993)}]{McC_SN1987A_rev}
{McCray}, R. 1993, \araa, 31, 175

\bibitem[{{Meynet} {et~al.}(2006){Meynet}, {Ekstr{\"o}m}, \&
  {Maeder}}]{meynet_etal_06}
{Meynet}, G., {Ekstr{\"o}m}, S., \& {Maeder}, A. 2006, \aap, 447, 623

\bibitem[{{Meynet} \& {Maeder}(2000)}]{meynet_maeder_00}
{Meynet}, G. \& {Maeder}, A. 2000, \aap, 361, 101

\bibitem[{{Mitchell} {et~al.}(2002){Mitchell}, {Baron}, {Branch}, {Hauschildt},
  {Nugent}, {Lundqvist}, {Blinnikov}, \& {Pun}}]{MBB02_87A}
{Mitchell}, R.~C., {Baron}, E., {Branch}, D., {Hauschildt}, P.~H., {Nugent},
  P.~E., {Lundqvist}, P., {Blinnikov}, S., \& {Pun}, C.~S.~J. 2002, \apj, 574,
  293

\bibitem[{{Phillips} {et~al.}(1990){Phillips}, {Hamuy}, {Heathcote},
  {Suntzeff}, \& {Kirhakos}}]{PHH90_87A}
{Phillips}, M.~M., {Hamuy}, M., {Heathcote}, S.~R., {Suntzeff}, N.~B., \&
  {Kirhakos}, S. 1990, \aj, 99, 1133

\bibitem[{{Phillips} {et~al.}(1988){Phillips}, {Heathcote}, {Hamuy}, \&
  {Navarrete}}]{phillips_etal_88}
{Phillips}, M.~M., {Heathcote}, S.~R., {Hamuy}, M., \& {Navarrete}, M. 1988,
  \aj, 95, 1087

\bibitem[{{Poznanski} {et~al.}(2009){Poznanski}, {Butler}, {Filippenko},
  {Ganeshalingam}, {Li}, {Bloom}, {Chornock}, {Foley}, {Nugent}, {Silverman},
  {Cenko}, {Gates}, {Leonard}, {Miller}, {Modjaz}, {Serduke}, {Smith}, {Swift},
  \& {Wong}}]{poznanski_etal_09}
{Poznanski}, D., {Butler}, N., {Filippenko}, A.~V., {Ganeshalingam}, M., {Li},
  W., {Bloom}, J.~S., {Chornock}, R., {Foley}, R.~J., {Nugent}, P.~E.,
  {Silverman}, J.~M., {Cenko}, S.~B., {Gates}, E.~L., {Leonard}, D.~C.,
  {Miller}, A.~A., {Modjaz}, M., {Serduke}, F.~J.~D., {Smith}, N., {Swift},
  B.~J., \& {Wong}, D.~S. 2009, \apj, 694, 1067

\bibitem[{{Rabinak} \& {Waxman}(2010)}]{rabinak_waxman_10}
{Rabinak}, I. \& {Waxman}, E. 2010, ArXiv e-prints

\bibitem[{{Smartt}(2009)}]{smartt_09}
{Smartt}, S.~J. 2009, \araa, 47, 63

\bibitem[{{Smartt} {et~al.}(2009){Smartt}, {Eldridge}, {Crockett}, \&
  {Maund}}]{smartt_etal_09}
{Smartt}, S.~J., {Eldridge}, J.~J., {Crockett}, R.~M., \& {Maund}, J.~R. 2009,
  \mnras, 395, 1409

\bibitem[{{Sugerman} {et~al.}(2006){Sugerman}, {Ercolano}, {Barlow}, {Tielens},
  {Clayton}, {Zijlstra}, {Meixner}, {Speck}, {Gledhill}, {Panagia}, {Cohen},
  {Gordon}, {Meyer}, {Fabbri}, {Bowey}, {Welch}, {Regan}, \&
  {Kennicutt}}]{sugerman_etal_06}
{Sugerman}, B.~E.~K., {Ercolano}, B., {Barlow}, M.~J., {Tielens}, A.~G.~G.~M.,
  {Clayton}, G.~C., {Zijlstra}, A.~A., {Meixner}, M., {Speck}, A., {Gledhill},
  T.~M., {Panagia}, N., {Cohen}, M., {Gordon}, K.~D., {Meyer}, M., {Fabbri},
  J., {Bowey}, J.~E., {Welch}, D.~L., {Regan}, M.~W., \& {Kennicutt}, R.~C.
  2006, Science, 313, 196

\bibitem[{{Swartz}(1991)}]{swartz_91}
{Swartz}, D.~A. 1991, \apj, 373, 604

\bibitem[{{Swartz} {et~al.}(1993){Swartz}, {Filippenko}, {Nomoto}, \&
  {Wheeler}}]{swartz_etal_93}
{Swartz}, D.~A., {Filippenko}, A.~V., {Nomoto}, K., \& {Wheeler}, J.~C. 1993,
  \apj, 411, 313

\bibitem[{{Swartz} {et~al.}(1995){Swartz}, {Sutherland}, \&
  {Harkness}}]{swartz_etal_95}
{Swartz}, D.~A., {Sutherland}, P.~G., \& {Harkness}, R.~P. 1995, \apj, 446, 766

\bibitem[{{Tominaga} {et~al.}(2009){Tominaga}, {Blinnikov}, {Baklanov},
  {Morokuma}, {Nomoto}, \& {Suzuki}}]{tominaga_etal_09}
{Tominaga}, N., {Blinnikov}, S., {Baklanov}, P., {Morokuma}, T., {Nomoto}, K.,
  \& {Suzuki}, T. 2009, \apjl, 705, L10

\bibitem[{{Utrobin}(2007)}]{utrobin_07}
{Utrobin}, V.~P. 2007, \aap, 461, 233

\bibitem[{{Utrobin} \& {Chugai}(2005)}]{UC05_time_dep}
{Utrobin}, V.~P. \& {Chugai}, N.~N. 2005, \aap, 441, 271

\bibitem[{{Weaver} {et~al.}(1978){Weaver}, {Zimmerman}, \&
  {Woosley}}]{weaver_etal_78}
{Weaver}, T.~A., {Zimmerman}, G.~B., \& {Woosley}, S.~E. 1978, \apj, 225, 1021

\bibitem[{{Woosley} \& {Heger}(2007)}]{WH07}
{Woosley}, S.~E. \& {Heger}, A. 2007, \physrep, 442, 269

\bibitem[{{Woosley} {et~al.}(2002){Woosley}, {Heger}, \& {Weaver}}]{WHW02}
{Woosley}, S.~E., {Heger}, A., \& {Weaver}, T.~A. 2002, Reviews of Modern
  Physics, 74, 1015

\bibitem[{{Woosley} \& {Weaver}(1995)}]{woosley_weaver_95}
{Woosley}, S.~E. \& {Weaver}, T.~A. 1995, \apjs, 101, 181

\bibitem[{{Xu} \& {McCray}(1991)}]{XM91_87A_energetic}
{Xu}, Y. \& {McCray}, R. 1991, \apj, 375, 190

\bibitem[{{Xu} {et~al.}(1992){Xu}, {McCray}, {Oliva}, \&
  {Randich}}]{XMO92_H_SN1987A}
{Xu}, Y., {McCray}, R., {Oliva}, E., \& {Randich}, S. 1992, \apj, 386, 181

\bibitem[{{Xu} {et~al.}(1991){Xu}, {Ross}, \& {McCray}}]{XRM91_compt}
{Xu}, Y., {Ross}, R.~R., \& {McCray}, R. 1991, \apj, 371, 280

\end{thebibliography}
\end{document}